\documentclass[fleqn,usenatbib]{mnras}

\usepackage{newtxtext,newtxmath}

\usepackage[T1]{fontenc}

\DeclareRobustCommand{\VAN}[3]{#2}
\let\VANthebibliography\thebibliography
\def\thebibliography{\DeclareRobustCommand{\VAN}[3]{##3}\VANthebibliography}

\usepackage{graphicx}	
\usepackage{amsmath}	
\usepackage{multirow}
\usepackage{pdflscape}
\usepackage[flushleft]{threeparttable}
\usepackage{caption}

\title[Cold Dust at $z=7$]{Cold Dust and Low [OIII]/[CII] Ratios: an Evolved Star-forming Population at Redshift 7}

\author[H.S.B. Algera et al.]{
Hiddo S. B. Algera,$^{1,2}$\thanks{E-mail: algera@hiroshima-u.ac.jp}
Hanae Inami,$^{1}$
Laura Sommovigo,$^{3}$
Yoshinobu Fudamoto,$^{4,2}$
Raffaella Schneider,$^{5,6}$
\newauthor
Luca Graziani,$^{5,6}$
Pratika Dayal,$^{7}$
Rychard Bouwens,$^{8}$
Manuel Aravena,$^{9}$
Elisabete da Cunha,$^{10,11}$
\newauthor
Andrea Ferrara,$^{3}$
Alexander P. S. Hygate,$^{8}$
Ivana van Leeuwen,$^{8}$
Ilse De Looze,$^{12,13}$
Marco Palla,$^{14,15,13}$
\newauthor
Andrea Pallottini,$^{3}$
Renske Smit,$^{16}$
Mauro Stefanon,$^{17,18}$
Michael Topping,$^{19}$ and
Paul P. van der Werf$^{8}$ \\
$^{1}$Hiroshima Astrophysical Science Center, Hiroshima University, 1-3-1 Kagamiyama, Higashi-Hiroshima, Hiroshima 739-8526, Japan \\
$^{2}$National Astronomical Observatory of Japan, 2-21-1, Osawa, Mitaka, Tokyo, Japan \\
$^{3}$Scuola Normale Superiore, Piazza dei Cavalieri 7, I-56126 Pisa, Italy \\
$^{4}$Waseda Research Institute for Science and Engineering, Faculty of Science and Engineering, Waseda University, 3-4-1 Okubo, Shinjuku, Tokyo 169-8555, Japan \\
$^{5}$Dipartimento di Fisica, Sapienza, Universita di Roma, Piazzale Aldo Moro 5, I-00185 Roma, Italy \\
$^{6}$INAF/Osservatorio Astronomico di Roma, via Frascati 33, I-00078 Monte Porzio Catone, Roma, Italy \\
$^{7}$Kapteyn Astronomical Institute, University of Groningen, P.O. Box 800, 9700 AV Groningen, The Netherlands \\
$^{8}$Leiden Observatory, Leiden University, NL-2300 RA Leiden, Netherlands \\
$^{9}$Departamento de Astronomia, Universidad de Chile, Camino del Observatorio 1515, Las Condes, Santiago 7591245, Chile \\
$^{10}$International Centre for Radio Astronomy Research, University of Western Australia, 35 Stirling Hwy, Crawley, 26WA 6009, Australia \\
$^{11}$ARC Centre of Excellence for All Sky Astrophysics in 3 Dimensions (ASTRO 3D) \\
$^{12}$Sterrenkundig Observatorium, Ghent University, Krijgslaan 281 - S9, 9000 Gent, Belgium \\
$^{13}$Dept. of Physics \& Astronomy, University College London, Gower Street, London WC1E 6BT, United Kingdom \\
$^{14}$INAF – Osservatorio Astronomico di Trieste, Via G.B. Tiepolo 11, 34131 Trieste, Italy \\
$^{15}$Dipartimento di Fisica, Sezione di Astronomia, Universit\'{a} di Trieste, Via G.B. Tiepolo 11, 34131 Trieste, Italy \\
$^{16}$Astrophysics Research Institute, Liverpool John Moores University, 146 Brownlow Hill, Liverpool L3 5RF, UK \\
$^{17}$Departament d'Astronomia i Astrof\`isica, Universitat de Val\`encia, C. Dr. Moliner 50, E-46100 Burjassot, Val\`encia, Spain \\
$^{18}$Unidad Asociada CSIC "Grupo de Astrof\'isica Extragal\'actica y Cosmolog\'ia" (Instituto de F\'isica de Cantabria - Universitat de Val\`encia) \\
$^{19}$Steward Observatory, University of Arizona, 933 N Cherry Ave, Tucson, AZ 85721, USA
}

\date{Accepted XXX. Received YYY; in original form ZZZ}

\pubyear{2022}

\begin{document}
\label{firstpage}
\pagerange{\pageref{firstpage}--\pageref{lastpage}}
\maketitle

\begin{abstract}
We present new ALMA Band 8 (rest-frame $90\,\mu$m) observations of three massive ($M_\star \approx 10^{10}\,M_\odot$) galaxies at $z\approx7$ previously detected in [CII]$158\,\mu$m and underlying dust continuum emission in the Reionization Era Bright Emission Line Survey (REBELS). We detect the dust continuum emission of two of our targets in Band 8 (REBELS-25 and REBELS-38), while REBELS-12 remains undetected. Through modified blackbody fitting we determine cold dust temperatures ($T_\mathrm{dust} \approx 30 - 35\,$K) in both of the dual-band detected targets, given a fiducial model of optically thin emission with $\beta = 2.0$. Their dust temperatures are lower than most $z\sim7$ galaxies in the literature, and consequently their dust masses are higher ($M_\mathrm{dust} \approx 10^{8}\,M_\odot$). Nevertheless, these large dust masses are still consistent with predictions from models of dust production in the early Universe. In addition, we target and detect [OIII]$88\,\mu$m emission in both REBELS-12 and REBELS-25, and find $L_\mathrm{[OIII]} / L_\mathrm{[CII]}$ ratios of approximately unity, low compared to the $L_\mathrm{[OIII]} / L_\mathrm{[CII]} \gtrsim 2 - 10$ observed in the known $z\gtrsim6$ population thus far. We argue the lower line ratios are due to a comparatively weaker ionizing radiation field resulting from the less starbursty nature of our targets. This low burstiness supports the cold dust temperatures and below average $\mathrm{[OIII]}\lambda\lambda4959,5007 + \mathrm{H}\beta$ equivalent widths of REBELS-25 and REBELS-38, compared to the known high-redshift population. Overall, this provides evidence for the existence of a massive, dust-rich galaxy population at $z\approx7$ which has previously experienced vigorous star formation, but is currently forming stars in a steady, as opposed to bursty, manner.
\end{abstract}






\section{Introduction}
\label{sec:introduction}

Understanding how the evolution of galaxies proceeds across cosmic time is one of the fundamental goals of modern day astronomy. Over the last two decades, ever-increasing samples of high-redshift ($z\gtrsim6$) galaxies are being discovered, largely based on observations probing their rest-frame ultraviolet (UV) and optical emission (e.g., \citealt{mclure2013,bouwens2015,finkelstein2015,stark2016,oesch2018,stefanon2019}). More recently, with the launch of the \textit{James Webb Space Telescope} (JWST), an even clearer window on the rest-frame UV to near-infrared (NIR) emission of high-redshift galaxies has been opened (e.g., \citealt{atek2022,castellano2022,harikane2022,naidu2022,yan2022}). However, observations of galaxies across all redshifts have long demonstrated that the presence of dust severely impacts their detectability at short wavelengths, as well as the information that can be extracted from their UV-to-NIR spectral energy distributions (SEDs; e.g., \citealt{draine1989,draine2003,calzetti2000,blain2002,casey2014,dudzeviciute2020}). Specifically, light emitted at UV and optical wavelengths is readily absorbed by dust, and subsequently re-emitted at longer wavelengths. As a result, UV and optical observations alone provide only an incomplete and biased view of the high redshift galaxy population, necessitating the use of observations at far-infrared wavelengths and beyond.

In the last several years, the Atacama Large Millimeter/submillimeter Array (ALMA) has enabled the detailed study of the dust and interstellar medium (ISM) properties of distant galaxies at (sub-)millimeter wavelengths (see \citealt{hodge2020} for a review). Various emission lines, most significantly the [CII]$158\,\mu$m and [OIII]$88\,\mu$m lines, are now routinely used to probe the ISM conditions of high-redshift sources and to constrain their systemic redshifts \citep{hashimoto2018,hashimoto2019,carniani2020,harikane2020,bouwens2022,schouws2022b,witstok2022}. Given its high ionization potential, [OIII] emission predominantly emanates from dense HII regions close to sites of star formation \citep{cormier2012,vallini2017,arata2020}. [CII], on the other hand, has a variety of origins, but is thought to mostly originate in photo-dissociation regions (PDRs; \citealt{stacey2010,vallini2015,gullberg2015,lagache2018,cormier2019}). While in nearby starburst galaxies the [CII] line is the dominant coolant of the ISM (i.e., $L_\mathrm{[OIII]} / L_\mathrm{[CII]} < 1$; \citealt{delooze2014,diazsantos2017}), at high redshift [OIII] is observed to become more luminous \citep{carniani2020,harikane2020,witstok2022}. This has been attributed to a highly ionized ISM resulting from strong starburst activity \citep{inoue2016,ferrara2019,arata2020,vallini2021,sugahara2022}, possibly in combination with other effects such as a low metallicity and/or a low carbon abundance resulting from a top-heavy initial mass function (IMF; e.g., \citealt{arata2020,lupi2020,katz2022}).

Albeit generally harder to detect at high-redshift than the [OIII] and [CII] lines, the underlying dust continuum emission is also occasionally observed in $z\gtrsim6.5$ galaxies (e.g., \citealt{watson2015,laporte2017,laporte2019,bowler2018,harikane2020,schouws2022a,witstok2022}; see also \citealt{inami2022} for a recent compilation). The Reionization Era Bright Emission Line Survey (REBELS; \citealt{bouwens2022}) in particular has provided the first statistical insights into the dust and ISM properties of UV-selected galaxies at $z\gtrsim6.5$. \citet{inami2022} show that dust is common in these high-redshift sources even for UV-selected galaxies, detecting it in $\geq 40\%$ of REBELS targets. Additionally, \citet{algera2022} find that, even at $z=7$, dust-obscured star formation still accounts for $\sim30\%$ of the overall cosmic star formation rate density. 

Not only may dust conceal an appreciable fraction of star formation at high redshift, it is also thought to be important for studies of reionization, as dust is capable of attenuating ionizing photons (e.g., \citealt{hayes2011,katz2017,glatzle2019}). Furthermore, dust alters the chemical equilibrium in galaxies and can provide an important pathway for the formation of molecular hydrogen, thereby providing the fuel for subsequent star formation \citep{gould1963,hirashita2002}.

Despite the importance of dust in the early Universe, the pathways through which significant dust reservoirs can already be assembled in only a fraction of a gigayear remain actively studied (e.g., \citealt{todini2001,mancini2015,michalowski2015,popping2017,behrens2018,vijayan2019,graziani2020,sommovigo2020,dayal2022,dicesare2022}). However, to date, the bulk of the dust detections at $z\gtrsim7$ remain limited to a single continuum measurement, such that dust masses and infrared luminosities often need to be extrapolated from a single wavelength (e.g., \citealt{bowler2018,schouws2022a,inami2022}). While inventive models have been developed to predict dust parameters (temperature, mass) from single-band continuum data (e.g., \citealt{inoue2020,sommovigo2021,fudamoto2022b}), such methods remain to be tested on larger samples of distant galaxies to fully establish their robustness and predictive power.

Accurately measuring dust properties observationally, however, requires multi-band continuum photometry. One of the key parameters that can be constrained when at least two ALMA bands are available, is the dust temperature $T_\mathrm{dust}$ \citep{hodge2020,bakx2021}. Given that the infrared luminosity, and therefore the obscured star formation rate (SFR; e.g., \citealt{kennicutt2012}), scales as $L_\mathrm{IR} \propto M_\mathrm{dust} T_\mathrm{dust}^{\beta + 4}$ with $\beta\sim1.5 - 2$, small variations in dust temperature imply potentially large variations in dust mass or infrared luminosity. As such, accurately measuring dust temperatures is crucial for a robust census of dust-obscured cosmic star formation and for properly constraining early dust enrichment.

Several observations \citep{schreiber2018,laporte2019,bakx2020,viero2022} and simulations \citep{behrens2018,ma2018,liang2019,pallottini2022} have suggested that dust may be hotter at high redshift. Given that observationally dust mass and temperature are often degenerate, warmer dust lowers the need for massive dust reservoirs. From a physical perspective, hot dust may be expected in high-redshift galaxies, due to their generally compact sizes \citep{vanderwel2014,fudamoto2022} and correspondingly high star formation rate surface densities (e.g., \citealt{schreiber2018}). In addition, when total dust masses are modest, the available energy injected by stars per unit dust mass increases, thereby increasing the overall heating \citep{sommovigo2022b}. However, given that multi-band ALMA observations of galaxies in the epoch of reionization remain rare, are often limited to non-detections or low-S/N measurements, and are potentially biased to warmer and therefore more luminous sources, larger samples of high-redshift galaxies with robust observational constraints on their dust temperatures are essential.

In this paper, we investigate the dust properties of three $z\approx7$ galaxies from the REBELS survey using combined ALMA Band 6 and Band 8 observations. In Section \ref{sec:observations} we introduce the REBELS survey and the newly acquired Band 8 observations, followed by the identification of dust continuum and [OIII]$88\,\mu$m emission in our targets in Section \ref{sec:source_detection}. Section \ref{sec:mbb_fitting} details our method of fitting far-infrared SEDs, while in Section \ref{sec:results} we describe our results. In Section \ref{sec:discussion} we discuss our findings in detail, and finally we summarize them in Section \ref{sec:summary}. Throughout this work, we assume a standard $\Lambda$CDM cosmology, with $H_0=70\,\text{km\,s}^{-1}\text{\,Mpc}^{-1}$, $\Omega_m=0.30$ and $\Omega_\Lambda=0.70$. We further adopt a \citet{chabrier2003} IMF across a mass range of $0.1 - 300\,M_\odot$.

\section{Data}
\label{sec:observations}

\subsection{REBELS}
REBELS is a Cycle 7 ALMA Large Program targeting 40 UV-bright galaxies with robustly measured photometric redshifts in the range $6.5 \lesssim z_\mathrm{phot} \lesssim 9.5$. Galaxies were targeted either in the [CII]$158\,\mu$m (36 sources) or [OIII]$88\,\mu$m (4 sources) emission line through a spectral scanning technique, designed to cover $\sim90\%$ of the photometric redshift probability distribution. In tandem, sensitive observations of the dust continuum are therefore obtained. The observing strategy as well as a summary of initial results of the REBELS program are outlined in detail in \citet{bouwens2022}. In the observations taken during Cycle 7, $23$ galaxies were detected in [CII] emission, and $16$ were detected in dust emission at rest-frame $158\,\mu$m. For a full analysis of the [CII]-detected sources, we refer the reader to Schouws et al. (in prep), while the dust continuum detections are presented in \citet{inami2022}. 

The UV luminosities of the full REBELS sample have been measured by Stefanon et al. (in prep), while stellar masses are presented in \citet{topping2022}. The latter study makes use of SED-fitting code {\tt{Prospector}} \citep{johnson2021} under the assumption of a non-parametric star formation history (SFH). Such SFHs are particularly well-suited to model any older stellar populations that may be present, even if outshone by more recent bursts of star formation (e.g., \citealt{leja2019,leja2020,topping2022,whitler2022}). The REBELS sample spans a stellar mass range of $\log_{10}(M_\star/M_\odot) = 8.8 - 10.4$; in this work, we focus on three select galaxies at the massive end.

\subsection{ALMA Band 8 Observations}

\begin{table}
	\centering
	\caption{Summary of the ALMA Band 8 continuum image properties of our three REBELS targets.}
	\label{tab:imaging}
	\begin{tabular}{ccccccc}
		\hline
		\textbf{ID} & $\nu_\mathrm{cen}$ & $t_\mathrm{int}^\mathrm{a}$ & RMS & $\theta_\mathrm{major}^\mathrm{b}$ & $\theta_\mathrm{minor}^\mathrm{b}$ & $\mathrm{PA}^\mathrm{b}$ \\
		\hline
		 & [GHz] & min & [$\mu\mathrm{Jy\,bm}^{-1}$] & [asec] & [asec] & [deg] \\
		\hline
        REBELS-12 & 401.4 & 83.2 & 48.9 & 0.61 & 0.45 & 58.5 \\
        REBELS-25 & 403.4 & 12.6 & 111.7 & 1.09 & 0.64 & -63.4 \\
        REBELS-38 & 405.0 & 38.3 & 41.9 & 0.61 & 0.48 & 56.9 \\
		\noalign{\smallskip} \hline
		\multicolumn{4}{l}{$^a$\footnotesize{Integration time (on source) in minutes}} \\
		\multicolumn{4}{l}{$^b$\footnotesize{Size and orientation of the synthesized beam}}
	\end{tabular}
\end{table}

Three REBELS sources detected in both [CII] and Band 6 continuum emission ($\lambda_\mathrm{obs} \approx 1.3\,$mm) were followed up with Band 8 ($\lambda_\mathrm{obs} \approx 750\,\mu$m) in ALMA Cycle 8. This sample includes the two brightest Band 6 continuum sources, REBELS-25 at $z_{\mathrm{[CII]}} = 7.3065$ and REBELS-38 at $z_{\mathrm{[CII]}} = 6.5770$ (2021.1.00318.S, PI: Inami), as well as REBELS-12 at $z_{\mathrm{[CII]}} = 7.3459$ (2021.1.01297.S, PI: Fudamoto). The Band 6 continuum observations of these REBELS targets are described in detail in \citet{inami2022}, while in this work we focus on the newly obtained Band 8 data. 

REBELS-25 and REBELS-38 are in the phase center of their respective Band 8 observations, while the primary beam sensitivity of the Band 8 map at the location of REBELS-12 is $\approx0.70$. This is due to the simultaneous observation of a neighboring source roughly $11.5\,$arcsec from the main target, serendipitously detected in the original REBELS observations \citep{fudamoto2021}. The observations of this neighbor will be presented in Fudamoto et al. (in prep), while this paper focuses solely on the main REBELS targets (REBELS-12, REBELS-25 and REBELS-38).

The Band 8 data were calibrated using the ALMA pipeline incorporated in {\tt{CASA}} (version 6.2.1). Continuum imaging was performed using {\tt{TCLEAN}} with natural weighting to optimize sensitivity, excluding channels contaminated by the [OIII] emission line (see below). Details of the final images, including the on-source time, RMS noise, central frequency and resolution, are listed in Table \ref{tab:imaging}. 

For REBELS-12 and REBELS-25, the spectral setup also covers the [OIII] emission line. We therefore additionally create datacubes of both targets in order to identify possible [OIII] emission. For this, we first run CASA task {\sc{uvcontsub}} to subtract any continuum emission, assuming a zeroth-order polynomial. We use natural weighting to produce the cubes, and create initial moment zero maps by collapsing the channels where the emission line is expected based on the known [CII] redshift and full width at half maximum (FWHM). We fit the line center and FWHM iteratively until the fit converges to a stable solution, and then create a moment-0 map by collapsing channels across the FWHM to maximize its S/N. In the continuum imaging of REBELS-12 and REBELS-25, we ensured that the spectral range contaminated by the [OIII] emission line was excluded by removing all channels within $3\times$ the [OIII] FWHM around the line center.

\section{Identification of Continuum and [OIII] Emission}
\label{sec:source_detection}

\begin{figure*}
    \centering
    \includegraphics[width=0.95\textwidth]{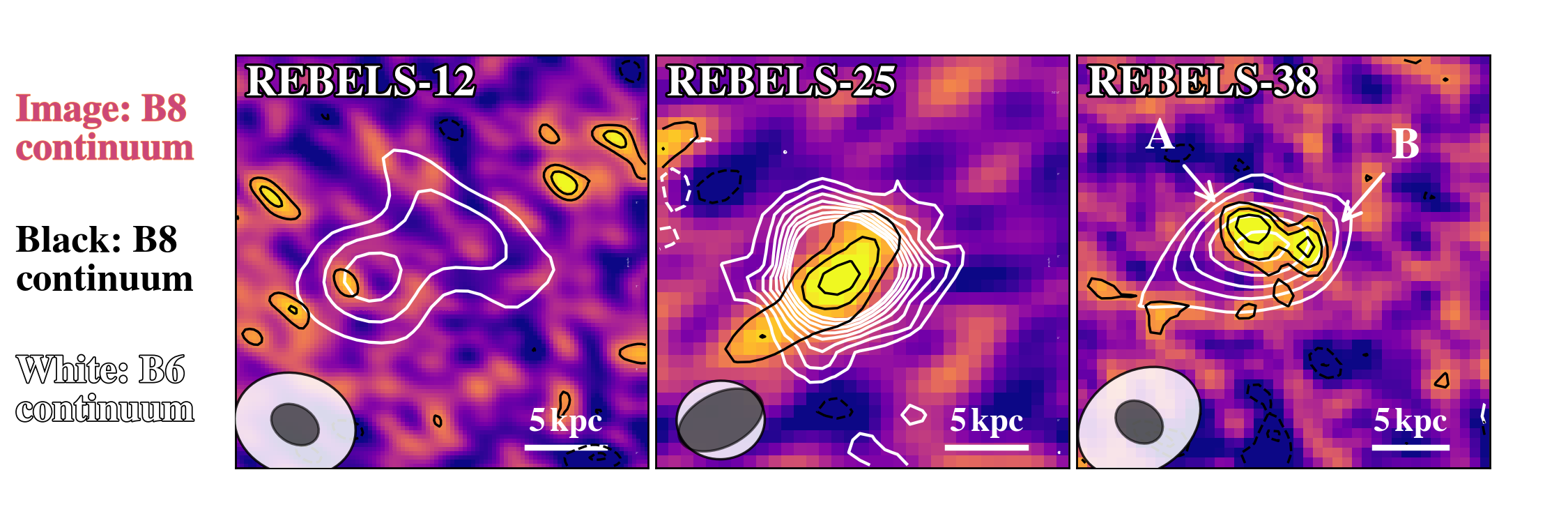}\vspace*{-0.6cm}
    \includegraphics[width=0.95\textwidth]{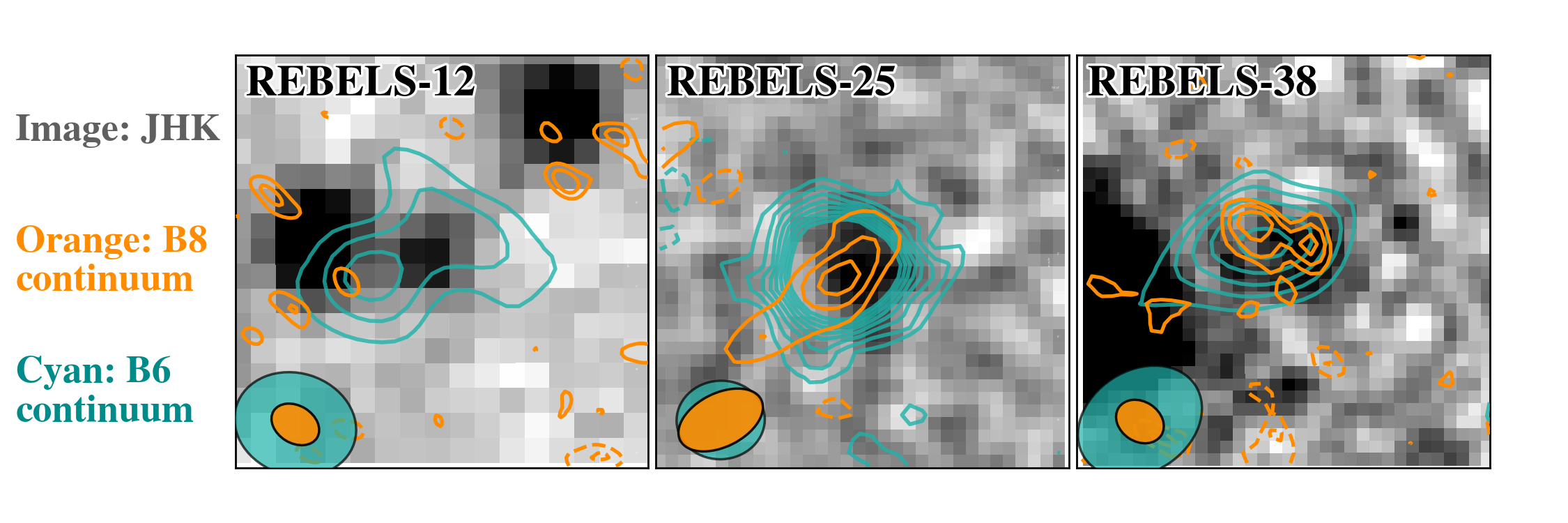}
    \caption{The three sources analyzed in this work -- from left to right, REBELS-12, REBELS-25 and REBELS-38. All cutouts span an area of $5''\times5''$. The \textbf{top row} shows the Band 8 continuum maps at their native resolution in the background, with black contours outlining the continuum emission at the $2,3$ and $4\sigma$ levels where $\sigma$ is the local RMS in the image. REBELS-12 is undetected in the Band 8 continuum data, while REBELS-25 and REBELS-38 are detected at a peak S/N in the untapered maps of $4.4$ and $4.3\sigma$, respectively. The overlaid white contours show the Band 6 continuum emission from $2 - 10\sigma$ in steps of $1\sigma$. The black (white) beam in the corner represents the FWHM of the Band 8 (Band 6) observations. The \textbf{bottom row} shows the same Band 6 (cyan) and Band 8 (orange) contours on top of stacked ground-based JHK-imaging. In the higher resolution Band 8 observations, REBELS-38 appears to separate into two dusty components, labeled A and B in the top right image.}
    \label{fig:B8_maps}
\end{figure*}

To identify continuum emission in the ALMA Band 8 images, we use {\sc{PyBDSF}} \citep{mohanrafferty2015}, following the source detection procedure in the original REBELS survey \citep{inami2022}. {\sc{PyBDSF}} finds islands of contiguous emission and fits these with two-dimensional Gaussians to extract their flux densities. We detect Band 8 continuum emission in REBELS-25 at a peak S/N of $4.4\sigma$, while the continuum remains undetected in REBELS-12 ($<3\sigma$). The new Band 8 observations of REBELS-38 highlight a two-component dust morphology that was not visible in the original Band 6 data. The two dust peaks are individually detected at a S/N of $4.3\sigma$ and $4.1\sigma$. This interesting morphology of REBELS-38 is discussed further in Appendix \ref{app:rebels_38morphology}.

We show the Band 8 continuum images of our targets in the top row of Figure \ref{fig:B8_maps}, with the Band 6 continuum contours overlaid. In addition, in the bottom row, we show the Band 8 continuum contours on top of stacked ground-based JHK images. For REBELS-12, the ground-based imaging is from the VIDEO survey \citep{jarvis2013}, while for REBELS-25 and REBELS-38 the images are from COSMOS/UltraVISTA DR4 \citep{scoville2007,mccracken2012}. The rest-frame optical images for REBELS-12 were aligned to the Gaia DR3 catalog by \citet{inami2022}, while the UltraVISTA JHK images were already aligned to Gaia DR1 as part of the fourth data release, ensuring an astrometric accuracy of $0\farcs03 - 0\farcs 12$. \\

We additionally create tapered maps of REBELS-25 and REBELS-38 in order to match the Band 6 resolution and more fairly compare the continuum flux densities across wavelength. The Band 6 continuum images of REBELS-25 and REBELS-38, created using natural weighting, have a resolution of $1\farcs1 \times 0\farcs94$  and $1\farcs6\times1\farcs2$, respectively \citep{inami2022}. The Band 8 image of REBELS-25 has an asymmetric native beam (Table \ref{tab:imaging}), and we therefore taper the beam to a circular Gaussian of $1\farcs1$. For REBELS-38, we apply tapering to obtain a circular beam of $1\farcs2$ in order to better match the Band 6 resolution. This tapered beam is sufficiently coarse that the two dust peaks are blended together, while ensuring the continuum sensitivity remains adequate.

We re-run {\tt{PyBDSF}} to extract the continuum flux densities from these tapered maps, and compile these -- as well as the untapered flux densities and other relevant physical properties of our targets -- in Table \ref{tab:source_parameters}. The significance of the continuum detection in the tapered map of REBELS-25 is comparable to that at the native resolution. For REBELS-38, on the other hand, the two dust components are blended together in the tapered image, such that the detection significance of the peak flux density is enhanced to $5.3\sigma$. The flux in the tapered map is consistent with the sum of the two individual components as measured in the higher resolution image. In what follows, we utilize the continuum flux densities obtained from the tapered images for both REBELS-25 and REBELS-38, unless specified otherwise. \\

\begin{figure}
    \includegraphics[width=0.49\textwidth]{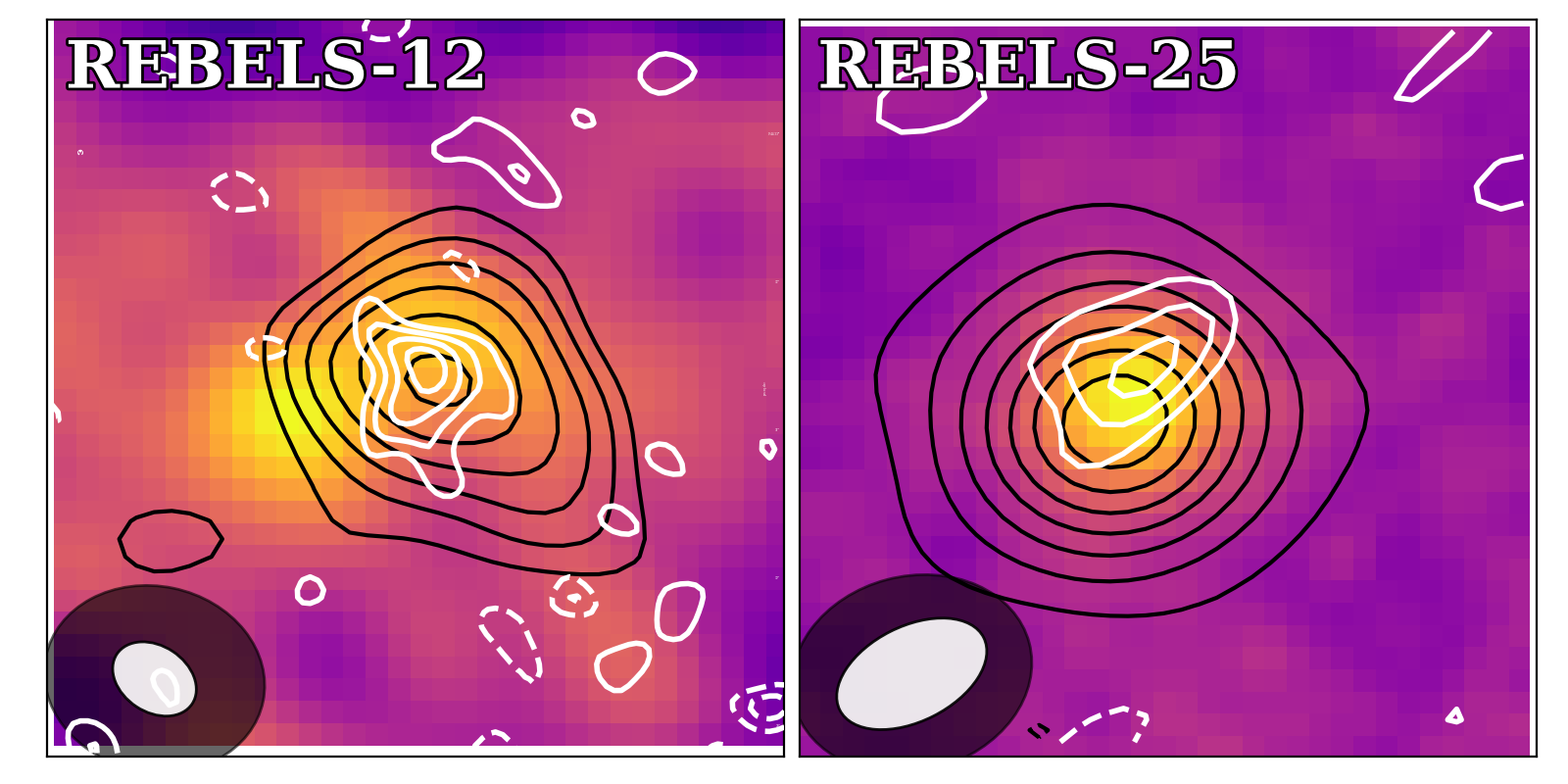}
    \includegraphics[width=0.49\textwidth]{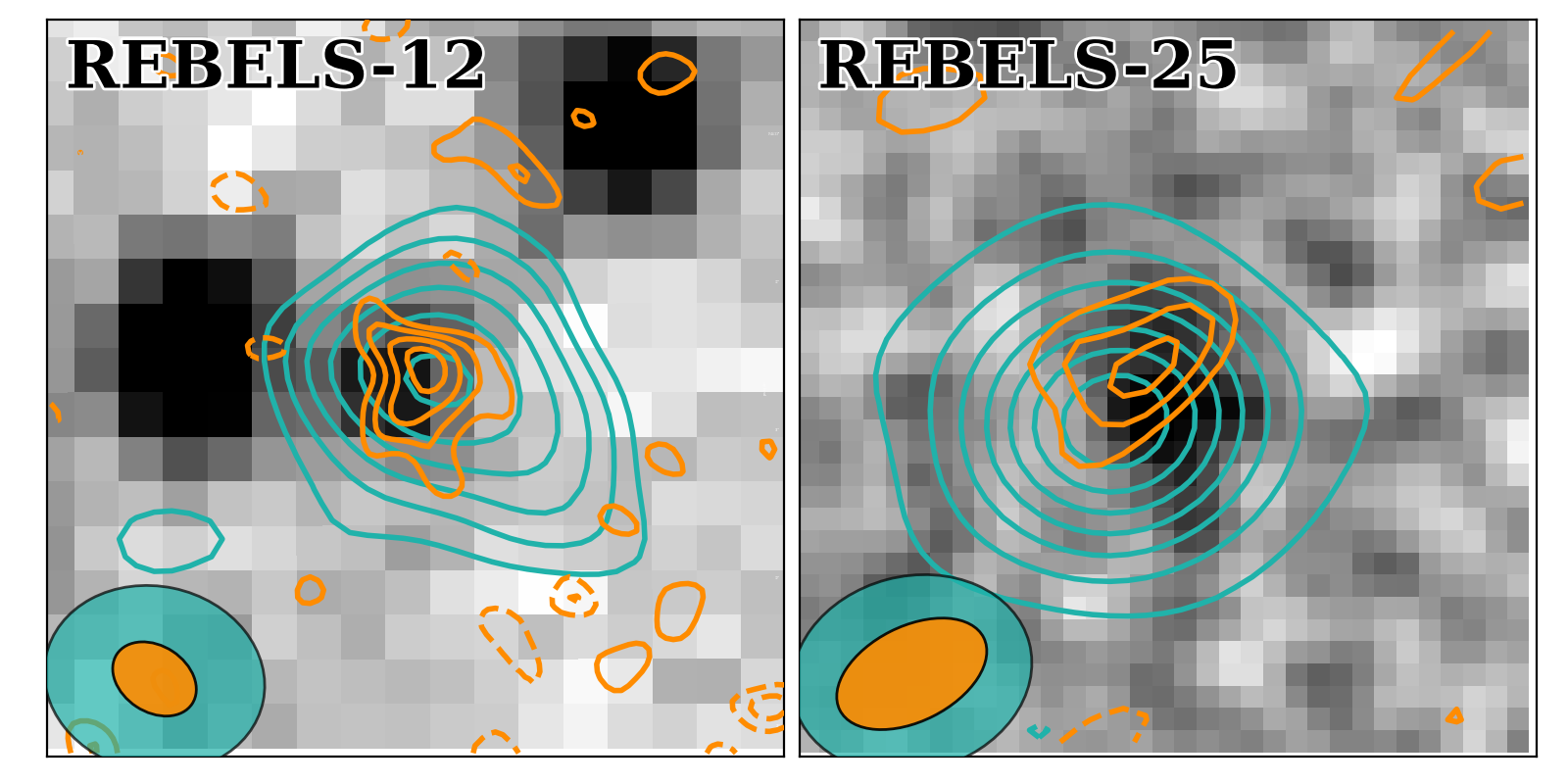}
    \caption{\textbf{Top:} Moment-0 maps of the [OIII] emission in REBELS-12 and REBELS-25 (white contours), on top of [CII] emission (black contours) and Band 6 dust continuum emission (background; $5''\times5''$). \textbf{Bottom:} The moment-0 maps on top of stacked JHK images. Here, [OIII] and [CII] are shown in orange and cyan, respectively, for visual clarity. In all panels, [OIII] contours range from $2-5\sigma$, and the [CII] contours start at $3\sigma$ and increase in steps of $1\sigma$ (steps of $4\sigma$) for REBELS-12 (REBELS-25). The beam sizes corresponding to the emission line data are shown in the bottom left corner. In REBELS-25, the [CII], [OIII] and rest-frame $158\,\mu$m dust emission are roughly co-spatial, while the peak of the dust continuum in REBELS-12 appears offset from the emission lines.}
    \label{fig:moment0}
\end{figure}

For REBELS-12 and REBELS-25, the spectral setup also covers the [OIII] emission line. We show the corresponding moment-0 maps in Figure \ref{fig:moment0}, comparing to the spatial distribution of the dust continuum emission (top row) and rest-frame UV emission (bottom). We detect the [OIII] line in both REBELS-12 and REBELS-25, at a significance of $5.6\sigma$ and $4.5\sigma$, respectively, measured as the peak S/N in the moment-0 map. To extract the 1D spectra, we sum all pixels in the moment-0 map with significance $\geq2\sigma$ within a radius of $2\farcs0$ around the peak pixel. We discuss the 1D spectra, as well as the morphology of our targets, in further detail in Section \ref{sec:results_OIII}.

\begin{table*}
	
	\caption{Physical and dust continuum properties of the three REBELS targets analyzed in this work.}
	\label{tab:source_parameters}
	
	\def\arraystretch{1.2}
	
	\hspace*{-1.1cm}
	\begin{tabular}{cccccccc}
	    \hline
	    ID$^\mathrm{a}$ & RA$^\mathrm{b}$ & DEC$^\mathrm{b}$ & $z_\mathrm{[CII]}^\mathrm{c}$ & $\log_{10}(M_\star)^\mathrm{d}$ & $S_\mathrm{Band\,6}^\mathrm{e}$ & $S_\mathrm{Band\,8}$ & $S_\mathrm{Band\,8,tapered}$ \\
	    \hline 
	    - & - & - & - & $\log_{10}(M_\odot)$ & $\mu$Jy & $\mu$Jy & $\mu$Jy \\
	    \hline 
        REBELS-12 & 02:25:07.94 & -05:06:40.70 & $7.3459 \pm 0.0010$ & $9.94_{-0.42}^{+0.32}$ & $87\pm24$ & $ < 210^\mathrm{f}$ &  \\
        REBELS-25 & 10:00:32.34 & +01:44:31.11 & $7.3065 \pm 0.0001$ & $10.27_{-0.23}^{+0.1}$ & $260\pm22$ & $493\pm112$ & $580 \pm 130$ \\
        REBELS-38 & 10:02:54.06 & +02:42:12.12 & $6.5770 \pm 0.0001$ & $10.37_{-0.36}^{+0.15}$ & $163\pm23$ & - & $334 \pm 63$ \\
        \emph{REBELS-38 A} & 10:02:54.08 & +02:42:12.31 & - & - & - & $196\pm45$ & - \\
        \emph{REBELS-38 B} & 10:02:54.03 & +02:42:12.11 & - & - & - & $165\pm40$ & - \\
        \noalign{\smallskip} \hline
        
        \multicolumn{8}{l}{$^\mathrm{a}${\footnotesize{Source IDs listed in italic (with suffices \emph{A}, \emph{B}) denote the individual dust components labeled in Figure \ref{fig:B8_maps}.}}} \\
        \multicolumn{8}{l}{$^\mathrm{b}${\footnotesize{Band 8 dust continuum coordinates, except for the $88\,\mu$m-undetected REBELS-12, for which we quote Band 6 coordinates instead.}}} \\
        \multicolumn{8}{l}{$^\mathrm{c}$\footnotesize{[CII]-based spectroscopic redshifts from Schouws et al. (in prep).}} \\
        \multicolumn{8}{l}{$^\mathrm{d}$\footnotesize{Stellar masses from \citet{topping2022}, under the assumption of a non-parametric star formation history.}} \\
        \multicolumn{8}{l}{$^\mathrm{e}$\footnotesize{ALMA Band 6 (rest-frame $160\,\mu$m) continuum flux densities taken from \citet{inami2022}.}} \\
        \multicolumn{8}{l}{$^\mathrm{f}$\footnotesize{The quoted upper limit is $3\sigma$.}}\\
	\end{tabular}
\end{table*}

\section{Dust SED Fitting}
\label{sec:mbb_fitting}

\subsection{Modified Blackbody Formalism}

When multi-frequency observations of a galaxy's dust continuum emission are available, it is possible to fit a physical model to the dust emission to constrain its properties (e.g., dust mass and temperature, infrared luminosity). Generally, a modified blackbody (MBB) is used (e.g., \citealt{greve2012,liang2019,jones2020}), which depends on three main fitting parameters: the dust temperature ($T_\mathrm{dust}$), the dust mass ($M_\mathrm{dust}$) and the dust emissivity ($\beta$). In addition, the dust may be optically thick at short wavelengths ($\lambda \lesssim 100 - 200\,\mu$m; e.g., \citealt{casey2012}), introducing an additional parameter $\lambda_\mathrm{thick} = c / \nu_\mathrm{thick}$ where the optical depth equals unity. The most general functional form of an optically thick modified blackbody may therefore be written as 

\begin{align}
    S_{\nu_\mathrm{obs}} = \left(\frac{1+z}{d_\mathrm{L}^2}\right) \left(\frac{1 - e^{-\tau_\nu}}{\tau_\nu}\right) M_\mathrm{dust} \kappa_\nu B_\nu(T_\mathrm{dust}) \ .
    \label{eq:mbb_thick}
\end{align}

\noindent Here $\nu$ and $\nu_\mathrm{obs}$ represent frequencies in the source rest- and observed frame, respectively,\footnote{In what follows, flux densities are implied to be at observer-frame frequencies, such that the subscript ``obs'' is omitted.} $d_\mathrm{L}$ is the luminosity distance at redshift $z$ and $\kappa_\nu = \kappa_0 (\nu / \nu_0)^\beta$ is the dust mass absorption coefficient at normalization frequency $\nu_0$. We adopt Milky Way dust with $\kappa_0 = \kappa(\nu_0=1900\,\mathrm{GHz}) = 10.41\,\mathrm{cm}^2\mathrm{g}^{-1}$, which appears best suited to match the dust in high-redshift sources (see e.g., \citealt{behrens2018,schouws2022a,ferrara2022}) and has been used throughout the series of REBELS papers to model dust properties (e.g., \citealt{inami2022,sommovigo2022}). The optical depth is defined as the integral over the line of sight via $\tau_\nu = \int \alpha_\nu ds = \int \kappa_\nu \rho_\mathrm{dust} ds$, where $\alpha_\nu$ is the absorption coefficient and $\rho_\mathrm{dust}$ is the dust mass density. As such, the optical depth can be written as $\tau_\nu = \kappa_\nu \Sigma_\mathrm{dust}$, where $\Sigma_\mathrm{dust}$ is the dust mass surface density. Equivalently, we may write $\tau_\nu = (\nu / \nu_\mathrm{thick})^\beta = (\lambda / \lambda_\mathrm{thick})^{-\beta}$. In the limit where the dust is optically thin across all frequencies, that is $\tau_\nu \ll 1$, the term $(1-e^{-\tau_\nu}) / \tau_\nu \approx 1$ and we recover the general functional form of an optically thin MBB. 

We must further consider that at $z\sim7$ the cosmic microwave background (CMB) is sufficiently warm [$T_\mathrm{CMB}(z=7) = T_\mathrm{CMB,0} \times (1 + z) = 21.8\,$K] that it may contribute to the heating of the dust. As shown by \citet{dacunha2013}, the ``effective'' dust temperature of a high-redshift galaxy, $T_{\mathrm{dust},z}$, is related to the dust temperature it would have at $z=0$ via

\begin{align}
    T_{\mathrm{dust},z} = \left(T_\mathrm{dust,0}^{4+\beta} + T_\mathrm{CMB,0}^{4+\beta}\left[(1+z)^{4+\beta} - 1\right]\right)^{1/(4+\beta)} \ .
    \label{eq:tdust_z}
\end{align}

\noindent If not accounted for, a modified blackbody with intrinsic dust temperature $T_\mathrm{dust,0}$ at $z=0$ is therefore less luminous than its higher redshift counterpart by a factor of $\left(T_{\mathrm{dust},z} / T_\mathrm{dust,0}\right)^{4+\beta}$. 

However, not only is the CMB causing dust to be warmer at high redshift, it also acts as a background against which the submillimeter emission from a galaxy is observed, since interferometers are insensitive to large-scale uniform emission \citep{dacunha2013}. The magnitude of this effect depends on the observed (i.e., CMB-heated) temperature of the dust and the temperature of the CMB itself, and is demonstrated by \citet{dacunha2013} to be

\begin{align}
    \frac{S_\nu^\mathrm{no\,contrast}}{S_\nu^\mathrm{against\,CMB}} = \left(1 - \frac{B_\nu(T_{\mathrm{CMB},z})}{B_\nu(T_{\mathrm{dust},z})} \right)^{-1} \ .
\end{align}

\noindent Combining the above equation with Equation \ref{eq:mbb_thick}, we may write the general functional form of a modified blackbody affected by the CMB as

{\footnotesize
\begin{align}
    S_\nu = \left(\frac{1+z}{d_\mathrm{L}^2}\right) \left(\frac{1 - e^{-\tau_\nu}}{\tau_\nu}\right) M_\mathrm{dust} \kappa_0 \left(\frac{\nu}{\nu_0}\right)^\beta \left[B_\nu(T_{\mathrm{dust},z}) - B_\nu(T_{\mathrm{CMB},z}) \right] \ .
    \label{eq:mbb_fit}
\end{align}
}

\noindent This equation, which depends on $M_\mathrm{dust}, T_{\mathrm{dust},z}, \beta$ and potentially $\lambda_\mathrm{thick}$ in the optically thick scenario, will be used to fit the dust continuum emission of our high-redshift targets, as outlined in the following Section.

\subsection{Modified Blackbody Fitting}

In this work, we aim to simultaneously constrain the dust temperature and mass of three $z\approx7$ galaxies through MBB fitting. Whilst we cannot robustly constrain $\beta$ due to a lack of sampling of the dust SED, we explore various fixed values for $\beta$ in the fits, and will additionally marginalize across $\beta$ after choosing a suitable prior (see below).

We adopt the Monte Carlo Markov Chain (MCMC) algorithm implemented in Python library {\sc{emcee}} \citep{foreman-mackey2013} to fit the dust SEDs of the REBELS targets. We fit to the measured flux densities where available, and implement upper limits following the procedure described in \citet{sawicki2012} (see also \citealt{bakx2020,witstok2022}). In practice, we fit the logarithm of the dust mass $\log_{10}\left(M_\mathrm{dust}/M_\odot\right)$, for which we introduce a simple uniform prior of $\log_{10}\left(M_\mathrm{dust}/M_\odot\right) \in [4, 10]$. We have verified that adopting a wider range does not affect our results. For $T_{\mathrm{dust},z}$, we adopt a flat prior with a lower bound of $T_{\mathrm{CMB},z}$, as we expect the dust to be at least as warm as the CMB. Given the wide variety observed in the dust temperatures of high-redshift galaxies (Section \ref{sec:discussion_Tdust}), we adopt a uniform prior between $T_{\mathrm{CMB},z}$ and $T_\mathrm{dust,max}=150\,$K, while beyond $T_\mathrm{dust,max}$ we smoothly decrease the prior probability by a Gaussian with a width of $\sigma=30\,$K. This ensures that high-temperature solutions are not excluded a priori.

For $\beta$ we adopt a Gaussian prior centered around a mean of $\langle\beta\rangle = 1.8$ with a standard deviation of $\sigma_\beta = 0.5$, following the analysis of \citet{faisst2020} for four $z\sim5.5$ galaxies. This value is consistent with that of local galaxies \citep{hildebrand1983} and submillimeter galaxies \citep{dacunha2021}. However, we also use MBB fitting with a fixed value of $\beta = (1.5, 2.0)$. The latter value is similar to that of Milky Way dust ($\beta=2.03$; \citealt{weingartner2001,draine2003}), which was assumed by \citet{sommovigo2022} to model the dust temperatures of REBELS targets with both a dust and a [CII] detection.

Two further remarks about the modified blackbody fitting routine are warranted, beginning with the implementation of upper limits. As per the \citet{sawicki2012} formalism, it is favourable for a model (that is, a MBB given some sampled $T_{\mathrm{dust},z}, M_\mathrm{dust}$ and $\beta$) to be significantly below an upper limit rather than just below it. The reason is that, given a model with flux density $f_\mathrm{M}$, an error of $\sigma$ and an upper limit of $3\sigma$, the log-likelihood $\ln\mathcal{L}$ is penalized by a factor of 

\begin{align}
    \Delta \ln\mathcal{L} = \ln \left[ \frac{1}{2} \left(1 + \mathrm{erf}\left(\frac{3\sigma - f_\mathrm{M}}{\sqrt{2}\sigma}\right)\right)\right] \ ,
\label{eq:logllh}
\end{align}

\noindent where $\mathrm{erf}(x)$ denotes the error function. Intuitively this can be understood as follows: when a model is just below the limiting flux, i.e., $f_\mathrm{M} = 0.99 \times 3\sigma$, the probability of exceeding the flux limit given Gaussian noise is nearly 50\%. As such, a model with $f_\mathrm{M} \ll 3\sigma$ is preferred, given the decreased probability of exceeding the upper limit in the presence of noise. Nevertheless, this prescription does not exclude solutions which exceed the upper limit a priori, as any imposed limit may be exceeded with non-zero probability.

The second point is in reference to the fitting of modified blackbodies with a single detection and an upper limit. As per Equation \ref{eq:logllh}, the solution that maximizes the log-likelihood is an MBB that matches the observed flux density, while simultaneously staying sufficiently far from the upper limit. In the case of REBELS-12, with a detection at rest-frame 158$\,\mu$m and an upper limit at $88\,\mu$m, one such optimal solution will be a very cold MBB ($T_{\mathrm{dust},z}\sim T_{\mathrm{CMB},z}\approx 22\,$K). However, MCMC ensures that (nearly) the full parameter space is explored, and as such allows for a variety of warmer MBB solutions that also accurately represent the data. This demonstrates that it is important to consider the full posterior distributions of the parameters of interest, and not only the maximum a posteriori solution (e.g., \citealt{hogg2018}). In this work, we therefore adopt the median of the posterior distributions as the fitted dust temperature, mass and emissivity, while the quoted errors represent the 16th and 84th percentiles of the corresponding posterior. We further remark that the posteriors of the fitted parameters are single-peaked in all cases (see also Appendix \ref{app:posteriors}).

We execute our MCMC routine with 32 walkers for a total of $10^5\,$steps each, thereby conservatively discarding the first $10^4$ steps as the burn-in phase. We confirm that the number of samples utilized in the MCMC fitting is greater than $50\times$ the autocorrelation time to ensure sufficient sampling of the parameter space (e.g., \citealt{foreman-mackey2013}). The robustness of our fitting routine is further supported by the acceptance fractions obtained, averaging $0.33$ across the three REBELS sources when $\beta$ is fixed ($1.5, 2.0$), and $0.22$ when beta is included in the fitting (c.f., an optimal range of $0.2 - 0.5$; \citealt{foreman-mackey2013}), and through the visual inspection of trace plots for $M_\mathrm{dust}, T_\mathrm{dust}$ and $\beta$.

\section{Results}
\label{sec:results}

\begin{figure*}
    \centering
    \includegraphics[width=0.95\textwidth]{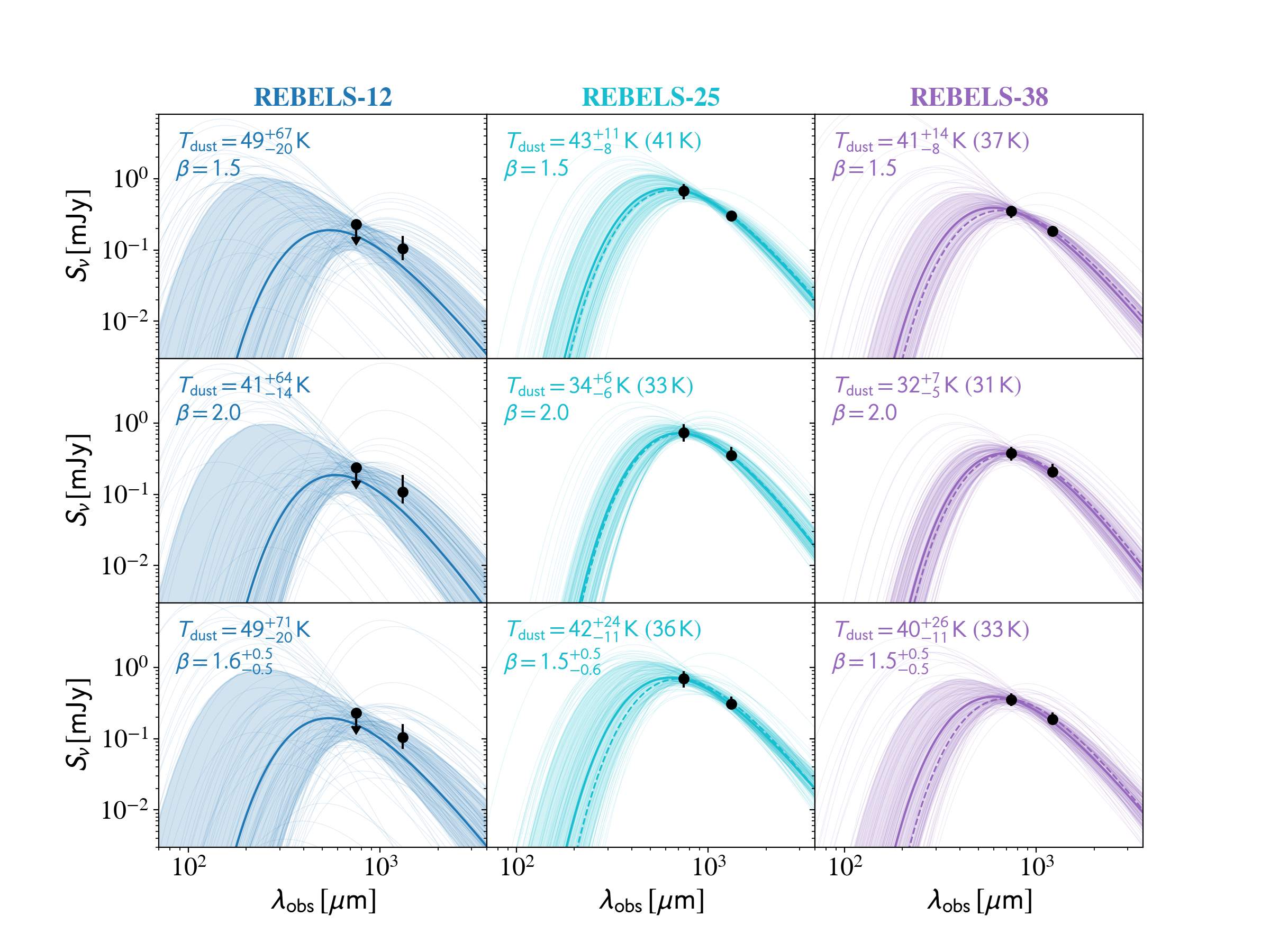}
    \caption{Optically thin MBB fits to REBELS-12, REBELS-25 and REBELS-38 using $\beta=1.5$ (upper row), $\beta=2.0$ (middle row) and a Gaussian prior on $\beta$ (bottom row). The thick solid line shows the ``best fit'' using the median values from the posterior distributions of $M_\mathrm{dust}$ and $T_{\mathrm{dust},z}$ (and $\beta$ for the lower row only), while the shaded region indicates the corresponding $1\sigma$ confidence interval. The dashed line shows the model using the fitting parameters that maximize the log-likelihood, and is omitted from the first column given that REBELS-12 is detected at only a single distinct frequency (Section \ref{sec:mbb_fitting}). The median dust temperature and its corresponding uncertainty are indicated on the panel, with the maximum a posteriori value in parentheses. Individual semi-transparent lines represent randomly drawn Monte Carlo samples. Both sources with two continuum detections, REBELS-25 and REBELS-38, appear to have cold dust temperatures of $T_{\mathrm{dust},z} \approx30 - 35\,$K. given a fiducial $\beta=2.0$. For REBELS-12, which is not detected in Band 8 and only at modest S/N in Band 6 ($\sim3.6\sigma$ at $\lambda_\mathrm{obs}\approx1300\,\mu$m), we cannot place strong constraints on $T_{\mathrm{dust},z}$. However, the non-detection in Band 8 hints towards a similarly modest dust temperature.}
    \label{fig:mbb_fit}
\end{figure*}

\begin{figure*}
    \centering
    \includegraphics[width=0.95\textwidth]{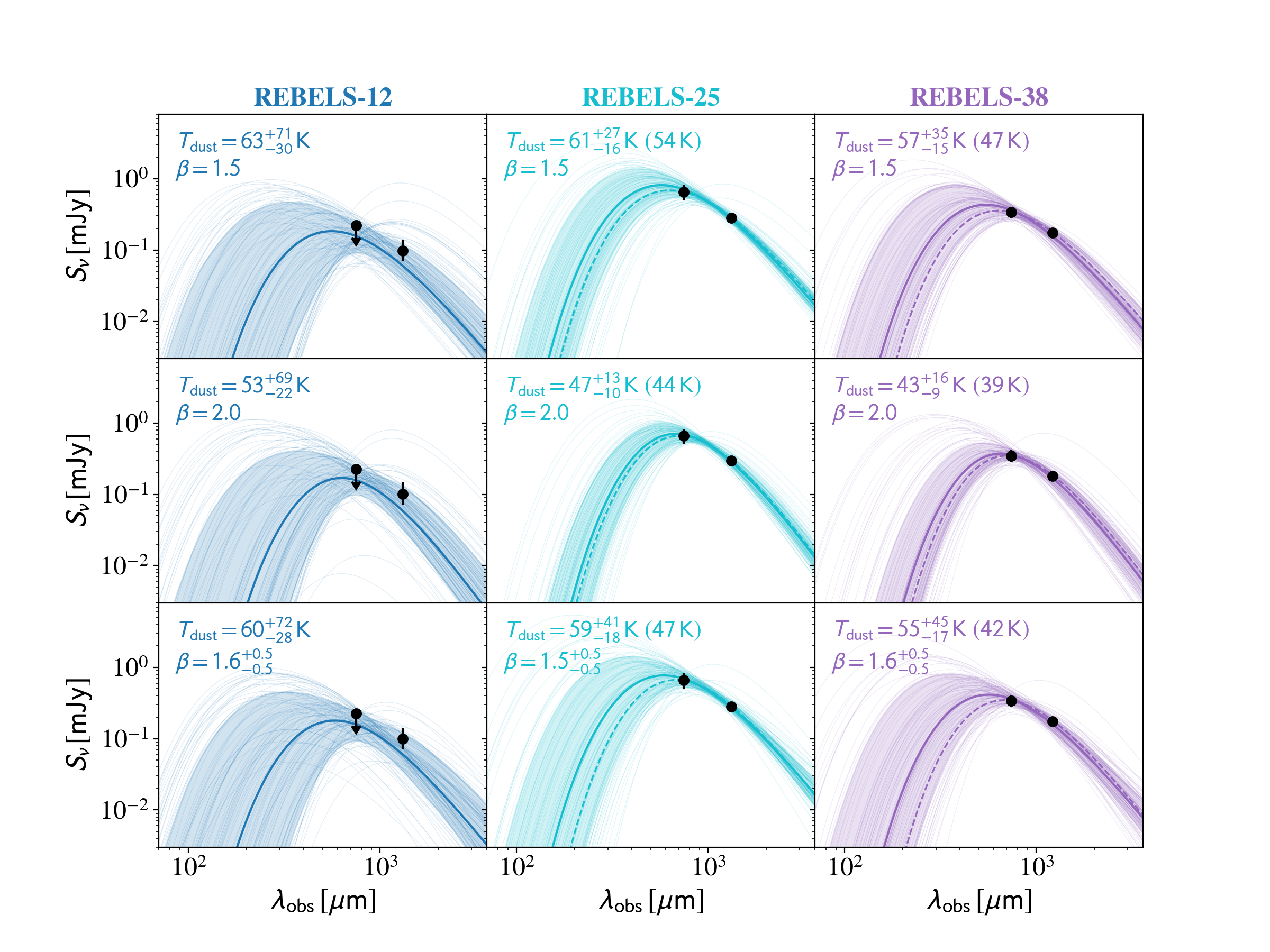}
    \caption{Same as Figure \ref{fig:mbb_fit}, now assuming an optically thick modified blackbody with a fixed $\lambda_\mathrm{thick} = 100\,\mu$m to investigate how this affects the inferred temperatures. While we expect our sources to be optically thin beyond $\lambda \gtrsim 40\,\mu$m (Section \ref{sec:results_thick}), an optically thick scenario would result in higher inferred dust temperatures by $\sim10-15\,$K.}
    \label{fig:mbb_fit_thick}
\end{figure*}

\subsection{Dust Emission at $z\sim7$}

As outlined in Section \ref{sec:source_detection}, we detect Band 8 continuum emission (rest-frame $90\,\mu$m) in two targets (REBELS-25 and REBELS-38), while the continuum is not detected in REBELS-12. In addition, all three targets were previously detected in Band 6 continuum (rest-frame $160\,\mu$m; \citealt{inami2022}). In what follows, we use our modified blackbody fitting routine to model the dust continuum SEDs of our targets in order to constrain their dust temperatures and masses. We first adopt an optically thin MBB, before considering optically thick dust.

\begin{table*}
    \begin{threeparttable}
    \caption{Dust parameters derived from modified blackbody fitting. Optically thick MBB fitting assumes a fixed $\lambda_\mathrm{thick}=100\,\mu$m. The dust emissivity is kept fixed to either $\beta=1.5, 2.0$ (top and middle rows), or is marginalized across using a Gaussian prior (bottom row; see text).}
    \label{tab:MBB_parameters}
    \def\arraystretch{1.2}
    \begin{tabular}{lllllll}
    \hline 
    \multicolumn{1}{c}{} & \multicolumn{2}{c}{\textbf{REBELS-12}} & \multicolumn{2}{c}{\textbf{REBELS-25}} & \multicolumn{2}{c}{\textbf{REBELS-38}}\\
    \hline
    \textbf{Optical Depth:} & \textbf{Thin} & \textbf{Thick} & \textbf{Thin} & \textbf{Thick} & \textbf{Thin} & \textbf{Thick}  \\
    \hline
    $\beta$ & $1.5$ & $1.5$ & $1.5$ & $1.5$ & $1.5$ & $1.5$ \\
    $\log_{10}(\mathrm{M}_\mathrm{dust}/M_\odot)$ &$6.95_{-0.94}^{+0.93}$ & $6.85_{-0.68}^{+0.89}$ & $7.79_{-0.26}^{+0.31}$ & $7.54_{-0.31}^{+0.30}$ & $7.56_{-0.37}^{+0.35}$ & $7.30_{-0.44}^{+0.37}$ \\
    $T_\mathrm{dust}\,[\mathrm{K}]$ &$49_{-20}^{+67}$ & $63_{-30}^{+71}$ & $43_{-8}^{+11}$ & $61_{-16}^{+27}$ & $41_{-8}^{+14}$ & $57_{-15}^{+35}$ \\
    $\log_{10}(\mathrm{L}_\mathrm{IR}/L_\odot)$ &$11.56_{-0.38}^{+1.04}$ & $11.51_{-0.40}^{+0.72}$ & $12.02_{-0.20}^{+0.28}$ & $12.12_{-0.28}^{+0.38}$ & $11.66_{-0.20}^{+0.33}$ & $11.76_{-0.26}^{+0.46}$ \\
    \hline
    $\beta$ & $2.0$ & $2.0$ & $2.0$ & $2.0$ & $2.0$ & $2.0$ \\
    $\log_{10}(\mathrm{M}_\mathrm{dust}/M_\odot)$ &$7.12_{-1.24}^{+0.94}$ & $6.98_{-0.87}^{+0.91}$ & $8.13_{-0.27}^{+0.38}$ & $7.79_{-0.27}^{+0.30}$ & $7.90_{-0.35}^{+0.39}$ & $7.58_{-0.39}^{+0.35}$ \\
    $T_\mathrm{dust}\,[\mathrm{K}]$ &$41_{-14}^{+64}$ & $53_{-22}^{+69}$ & $34_{-6}^{+6}$ & $47_{-10}^{+13}$ & $32_{-5}^{+7}$ & $43_{-9}^{+16}$ \\
    $\log_{10}(\mathrm{L}_\mathrm{IR}/L_\odot)$ &$11.44_{-0.31}^{+1.06}$ & $11.33_{-0.32}^{+0.68}$ & $11.85_{-0.10}^{+0.18}$ & $11.84_{-0.18}^{+0.23}$ & $11.47_{-0.10}^{+0.20}$ & $11.47_{-0.16}^{+0.26}$ \\
    \hline
    $\beta$ & $1.6_{-0.5}^{+0.5}$ & $1.6_{-0.5}^{+0.5}$ & $1.5_{-0.6}^{+0.5}$ & $1.5_{-0.5}^{+0.5}$ & $1.5_{-0.5}^{+0.5}$ & $1.6_{-0.5}^{+0.5}$ \\
    $\log_{10}(\mathrm{M}_\mathrm{dust}/M_\odot)$ &$6.93_{-0.96}^{+0.96}$ & $6.89_{-0.73}^{+0.91}$ & $7.83_{-0.47}^{+0.48}$ & $7.57_{-0.41}^{+0.39}$ & $7.58_{-0.55}^{+0.50}$ & $7.35_{-0.51}^{+0.45}$ \\
    $T_\mathrm{dust}\,[\mathrm{K}]$ &$49_{-20}^{+71}$ & $60_{-28}^{+72}$ & $42_{-11}^{+24}$ & $59_{-18}^{+41}$ & $40_{-11}^{+26}$ & $55_{-17}^{+45}$ \\
    $\log_{10}(\mathrm{L}_\mathrm{IR}/L_\odot)$ &$11.55_{-0.37}^{+1.06}$ & $11.45_{-0.39}^{+0.73}$ & $12.00_{-0.22}^{+0.46}$ & $12.07_{-0.34}^{+0.58}$ & $11.65_{-0.23}^{+0.51}$ & $11.71_{-0.33}^{+0.62}$ \\
    \hline
    \end{tabular}
    \def\arraystretch{1.0}
    \end{threeparttable}
\end{table*}

\subsubsection{Dust SED: Optically Thin Scenario}

Given that high-redshift galaxies generally have a sparsely sampled dust SED, it is common in the literature to assume the dust is optically thin across the wavelength range of interest (e.g., \citealt{harikane2020,bakx2021,sugahara2021}). As such, we proceed by fitting an optically thin MBB to our three REBELS targets, as shown in Figure \ref{fig:mbb_fit}. The top and middle rows show the fits for a fixed $\beta=1.5$ and $\beta=2.0$, respectively, while the bottom row shows the MBB fit using a Gaussian prior on $\beta$. In what follows, we adopt $\beta=2.0$ as our fiducial model, to ensure consistency with the value assumed by \citet{sommovigo2022}, and the series of REBELS papers in general. 

We robustly constrain the dust temperatures of REBELS-25 and REBELS-38 using the high-S/N detection in Band 6 and the new Band 8 data. For REBELS-25, we find $T_{\mathrm{dust},z} = 34 \pm 6\,$K, while for REBELS-38 we determine $T_{\mathrm{dust},z} = 32_{-5}^{+7}\,$K. In addition, we determine dust masses of $M_\mathrm{dust} \approx 10^8\,M_\odot$ for both targets (Table \ref{tab:MBB_parameters}). We discuss these dust masses in the context of dust production mechanisms in Section \ref{sec:discussion_Mdust}. 

Given the non-detection at Band 8 for REBELS-12 and the modest-S/N detection at Band 6, we cannot place robust constraints on its dust temperature, as the fitting allows for a long tail towards high dust temperatures (e.g., for $\beta=2.0$ we find $T_{\mathrm{dust},z} = 41_{-14}^{+64}\,$K, though we note that the exact value depends on the assumed prior on the temperature). While the Band 8 upper limit is relatively constraining, the main uncertainty in our analysis is at present the S/N of the Band 6 data. For example, if we were to artificially reduce the uncertainty of its Band 6 flux density to obtain an $\mathrm{S/N}=10$, we would determine a low dust temperature of $T_{\mathrm{dust},z} = 30_{-5}^{+8}\,$K for REBELS-12. However, with the current observations a higher temperature solution cannot be ruled out, as can be seen from the posterior distributions of $T_{\mathrm{dust},z}$ in Appendix \ref{app:posteriors}. 

By propagating the full posterior distributions of the various MBB parameters determined through MCMC, we compute the infrared luminosities of our targets by integrating the sampled MBBs between rest-frame $8-1000\,\mu$m. For REBELS-25, we determine $\log_{10}\left(L_\mathrm{IR}/L_\odot\right) = 11.85_{-0.10}^{+0.18}$, again with our fiducial $\beta=2.0$. As such, we find that REBELS-25 may not be a ULIRG ($L_\mathrm{IR} \geq 10^{12}\,L_\odot$), as was previously predicted from its [CII]-based dust temperature (\citealt{sommovigo2022}; see also \citealt{hygate2022}). The continuum-detected REBELS sample therefore appears to consist solely of LIRGs, with $10^{11} \leq L_\mathrm{IR}/L_\odot < 10^{12}$. For REBELS-38, we determine $\log_{10}\left(L_\mathrm{IR}/L_\odot\right) = 11.47_{-0.10}^{+0.20}$. Adopting a different dust emissivity and/or optically thick dust does not affect the infrared luminosities of our sources by more than $0.3\,$dex (Table \ref{tab:MBB_parameters}; see also the next section).

If we assume a lower $\beta=1.5$, the inferred dust temperatures for the two sources detected in Band 8 continuum increase to $T_{\mathrm{dust},z} \approx 40 - 45\,$K, with slightly larger uncertainties given that we now probe further redwards of the dust peak. Nevertheless, the inferred infrared luminosities do not exceed the values determined with the fiducial $\beta=2.0$ by more than $0.2\,$dex due to the anticorrelation between dust temperature and dust mass. Indeed, the dust masses inferred when adopting $\beta=1.5$ are lower compared to the fiducial model by $\sim0.35\,$dex.\footnote{Note, however, that while we here adopt a different value for $\beta$ in the fit, we do not alter the normalization of the dust opacity $\kappa_0 = \kappa_\nu(1900\,\mathrm{GHz})$. In practice, $\kappa_0$ and $\beta$ are not independent, as both depend on the dust grain composition, as well as the grain size distribution (e.g., \citealt{bianchi2013}).}

When allowing for variation in $\beta$ in the form of a Gaussian prior, we find similar temperatures of $\sim40\,$K, though with $1\sigma$ uncertainties spanning a relatively wide range of $T_{\mathrm{dust},z}\sim30 - 65\,$K. The value of $\beta$ itself is not constrained, and is correlated (anti-correlated) with the dust mass (temperature) as expected. However, $\beta \approx 1.5$ is preferred by the fit, mainly because very high values of $\beta \gtrsim2.5$ would require dust temperatures below $T_{\mathrm{CMB},z}$, which are ruled out by our adopted prior. Nevertheless, we suggest the reader treat these results with caution, as we are effectively marginalizing across $\beta$. An increased sampling of the continuum SED, in particular at longer wavelengths, is required to constrain the dust emissivity more precisely.

\subsubsection{Dust SED: Optically Thick Scenario}
\label{sec:results_thick}

Given that the dust optical depth scales linearly with the dust surface density (Section \ref{sec:mbb_fitting}), dust-rich galaxies may become optically thick at short wavelengths. Using a sample of 23 submillimeter galaxies (SMGs), \citet{simpson2017} find a typical value for $\lambda_\mathrm{thick} = 75_{-20}^{+15}\,\mu$m given $\beta=1.8$, though they caution this should be taken as a lower limit given that their results are based on observed galaxy sizes at longer wavelengths. Indeed, some particularly dusty sources are found to be optically thick to $\lambda_\mathrm{thick} \approx200\,\mu$m \citep{blain2003,conley2011,casey2019}. However, making use of \emph{Herschel} observations of local galaxies, \citet{lutz2016} find that the majority of LIRGs are on average optically thin, while infrared-brighter sources may indeed be optically thick at far-infrared wavelengths. 

In their analysis of the dust emission in $z\sim4$ starburst GN-20, \citet{cortzen2020} find the galaxy to be optically thick out to $\lambda_\mathrm{thick} = 170\pm23\,\mu$m and determine its dust temperature to be $T_\mathrm{dust} = 52 \pm 5\,$K. However, they show that, had an optically thin MBB been assumed, the recovered dust temperature is substantially lower ($T_\mathrm{dust} = 33 \pm 2\,$K). As such, they clearly demonstrate that wrongfully adopting optically thin dust can significantly decrease the inferred dust temperatures.

For the REBELS targets studied in this work, we cannot fit $\lambda_\mathrm{thick}$ directly given that the dust SED is sampled at only two distinct wavelengths. However, we can determine a plausible value for $\lambda_\mathrm{thick}$ by assuming galaxies can be characterized by a spherical dust distribution of size $R$. In this case, the source is expected to become optically thick at a wavelength of 

\begin{align}
    \lambda_\mathrm{thick} = \lambda_0 \left(\frac{\pi R^2}{\kappa_0 M_\mathrm{dust}}\right)^{-\frac{1}{\beta}} \approx 42\,\mu\mathrm{m}\times\left(\dfrac{R}{1\mathrm{kpc}}\right)^{-1}\left(\dfrac{M_\mathrm{dust}}{10^8M_\odot}\right)^{1/2} ,
    \label{eq:lambda_thick}
\end{align}

\noindent where the numerical value assumes $\beta=2.0$. From our optically thin modified blackbody fitting, we infer a maximum dust mass of $M_\mathrm{dust} \approx 10^{8}\,M_\odot$ for our brightest target, REBELS-25. To obtain an estimate of $\lambda_\mathrm{thick}$, we adopt this as a conservative upper limit on the true dust mass, given that, if the source is in fact optically thick, its true dust mass would be lower than inferred by optically thin models. For the size $R$, we adopt the median stacked continuum size of the dust-detected REBELS sample of $R = 1.1 \pm 0.3\,$kpc \citep{fudamoto2022}. With these estimates, we expect $\lambda_\mathrm{thick} \lesssim 40\,\mu$m to be a sensible upper limit on the wavelength where the dust transitions to optically thick. These calculations indicate that the REBELS sample is unlikely to be optically thick at the shortest wavelengths probed with ALMA of $\lambda_\mathrm{rest} \approx88\,\mu$m. 

Nevertheless, we investigate the effect of allowing for optically thick dust by adopting a fixed $\lambda_\mathrm{thick} = 100\,\mu$m (following e.g., \citealt{faisst2020,dacunha2021}). We caveat that, since $\tau_\nu \propto \Sigma_\mathrm{dust}$, this necessarily implies significant dust surface densities due to large overall dust masses and/or compact dust-obscured star-forming regions (Equation \ref{eq:lambda_thick}). In these scenarios, nearly all of the UV emission associated with star formation will be attenuated by dust. However, since the REBELS targets were explicitly selected to be UV-bright galaxies, reconciling this with a high dust optical depth requires spatial offsets between the dust and UV emission, or clumpy dust substructures (e.g., \citealt{behrens2018,ferrara2022}).

The optically thick MBB fits assuming $\lambda_\mathrm{thick} = 100\,\mu$m are shown in Figure \ref{fig:mbb_fit_thick}. The inferred temperatures exceed the ones determined using optically thin models by $\Delta T_\mathrm{dust} \approx 10 - 15\,$K, while dust masses are found to be lower than in our fiducial analysis by $\sim0.3\,$dex (Table \ref{tab:MBB_parameters}). Nevertheless, the inferred infrared luminosities, and therefore the corresponding dust-obscured star formation rates, are consistent with the ones derived from optically thin MBB fitting within the uncertainties. We note that, had we assumed $\lambda_\mathrm{thick} = 40\,\mu$m based on the aforementioned estimate for REBELS (Equation \ref{eq:lambda_thick}), the inferred temperatures agree with the optically thin ones to within $\lesssim 2\,$K ($\lesssim 4\,$K) for $\beta=2.0$ ($\beta=1.5$).


\begin{figure*}
    \centering
    \includegraphics[width=\textwidth]{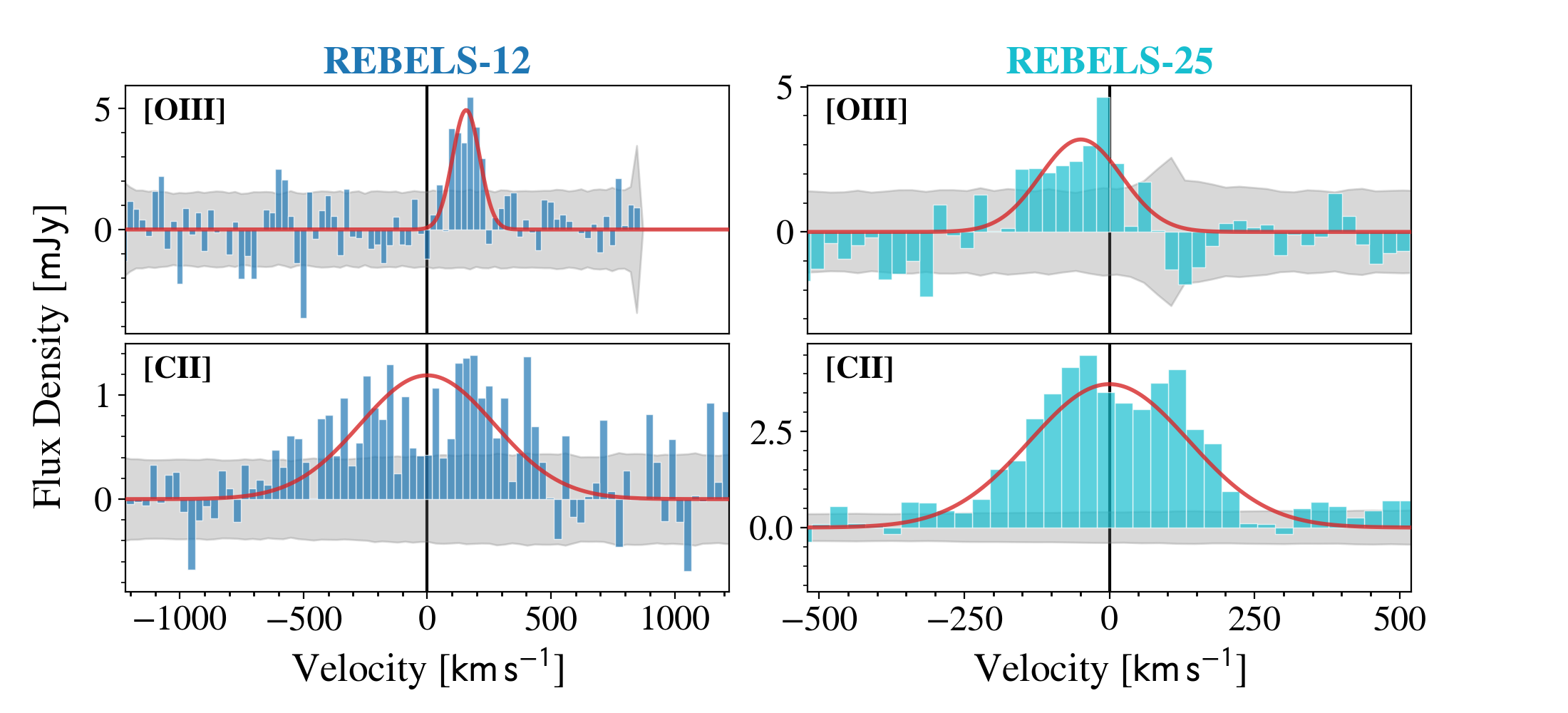}
    \caption{The extracted [OIII] (top) and [CII] emission (bottom) for REBELS-12 (left) and REBELS-25 (right). The grey band shows the noise level per channel and the red line represents a Gaussian fit to the spectrum. The black vertical line corresponds to the center of the [CII] line, defined to be at $v=0$. In both sources, the [CII] line is notably broader than the [OIII] emission, although there is tentative evidence that the [CII] emission in REBELS-12 consists of two components, one of which is blueshifted by $\sim500\,\mathrm{km}\,\mathrm{s}^{-1}$ compared to the observed [OIII] emission (Appendix \ref{app:rebels12_cii}). Adopting the total line fluxes for both emission lines, we find $\mathrm{[OIII]/[CII]} \approx 1$ in both galaxies.}
    \label{fig:OIII}
\end{figure*}

\subsection{[OIII] Emission}
\label{sec:results_OIII}

\begin{table*}
	
	\caption{Emission line properties of REBELS-12 and REBELS-25.}
	\label{tab:emission_line_parameters}
	
	\def\arraystretch{1.2}
	
	\hspace*{-1.1cm}
	\begin{tabular}{ccccccccc}
	    \hline
	    ID & $z_\mathrm{[OIII]}$ & $S_\mathrm{[OIII]}$ & $\mathrm{FWHM}_\mathrm{[OIII]}$ & $L_\mathrm{[OIII]}$ & $S_\mathrm{[CII]}^\mathrm{a}$ & $\mathrm{FWHM}_\mathrm{[CII]}^\mathrm{a}$ & $L_\mathrm{[CII]}^\mathrm{a}$ & $L_\mathrm{[OIII]} / L_\mathrm{[CII]}$ \\
	\hline 
	& & $\mathrm{mJy\,km\,s}^{-1}$ & $\mathrm{km\,s}^{-1}$ & $10^8 L_\odot$ & $\mathrm{mJy\,km\,s}^{-1}$ & $\mathrm{km\,s}^{-1}$ & $10^8 L_\odot$ & \\
    \hline 
    REBELS-12 & $7.3491\pm0.0002$ & $670_{-136}^{+152}$ & $129\pm21$ & $15.0_{-3.0}^{+3.5}$ & $807\pm155$ & $637\pm82$ & $10.1\pm4.1$ & $1.5_{-0.5}^{+1.0}$ \\
    REBELS-25 & $7.3052\pm0.0004$ & $552_{-135}^{+148}$ & $165\pm33$ & $12.3_{-3.0}^{+3.3}$ & $1280\pm36$ & $322\pm12$ & $15.9\pm1.0$ & $0.8 \pm 0.2$ \\
        \noalign{\smallskip} \hline
        \multicolumn{8}{l}{$^\mathrm{a}$\footnotesize{[CII] fluxes, FWHMs and luminosities from Schouws et al. (in prep).}} \\
	\end{tabular}
\end{table*}

Given the known [CII]-based spectroscopic redshifts of REBELS-12 and REBELS-25, we search for and detect the [OIII] line in the Band 8 datacubes. The extracted 1D spectra of the [OIII] emission, including all pixels with $\mathrm{S/N} > 2$ in the moment-0 map (c.f., Figure \ref{fig:moment0}), are shown in the top row of Figure \ref{fig:OIII}. We fit the spectra with a Gaussian profile, and determine the total line flux by integrating the area under the Gaussian curve. The line fluxes, luminosities and FWHMs are listed in Table \ref{tab:emission_line_parameters}.

In the bottom row of Figure \ref{fig:OIII}, we compare to the [CII] emission lines for our sources (Schouws et al. in prep), adopting $v=0\,\mathrm{km}\,\mathrm{s}^{-1}$ as the center of the [CII] line, given its higher S/N compared to [OIII]. Interestingly, the [CII] line in REBELS-12 is very broad, with a $\mathrm{FWHM}\approx640\,\mathrm{km\,s}^{-1}$, while the [OIII] emission is comparatively narrow, $\mathrm{FWHM} \approx 120\,\mathrm{km\,s}^{-1}$. Moreover, the [CII] emission appears to consist of two peaks separated by $\Delta v \sim 500\,\mathrm{km}\mathrm{s}^{-1}$, indicating that REBELS-12 may be a merger, although it could also be a signature of disk rotation (e.g., \citealt{kohandel2019}). The velocity of [OIII] emission coincides with the redder peak, while no counterpart to the bluer emission is visible in the [OIII] spectrum. 

In this work, we adopt the total flux density in the [CII] line as given by Schouws et al. (in prep), integrated over the full linewidth. However, if REBELS-12 is indeed a merger, the bluer component is detected only in [CII] emission, while being undetected in UV, dust and [OIII] emission. We discuss this interesting possibility in detail in Appendix \ref{app:rebels12_cii}, and comment on how this affects a comparison of the [OIII] / [CII] flux ratio. In our fiducial analysis, however, we determine a line ratio of $\mathrm{[OIII]/[CII]} = L_\mathrm{[OIII]} / L_\mathrm{[CII]} = 1.5_{-0.5}^{+1.0}$ for REBELS-12.

Further complicating this picture is the dust emission in REBELS-12: as detailed in \citet{inami2022}, the dust in this source shows two separate components, the brightest of which is spatially offset from the [CII] emission (Figure \ref{fig:moment0}). The [OIII] line, on the other hand, is co-spatial with both the [CII], UV and the fainter dust component. Clearly, deeper and higher resolution ALMA observations of REBELS-12 are required to better understand the nature of its [CII] and dust emission.

For REBELS-25, we find that the [OIII] emission is slightly blueshifted compared to the [CII] emission by $\Delta v = - 46\pm12\,\mathrm{km\,s}^{-1}$. Similar to what we observe for REBELS-12, the [OIII] line is narrower than the [CII] emission. However, for REBELS-25 we do not find evidence for significant spatial offsets between the dust continuum and the emission lines. The [CII] line in REBELS-25 consists of a double-peaked component and shows a velocity field consistent with that of a rotating disk (\citealt{hygate2022}). The larger linewidth of the [CII] line therefore likely implies it traces an extended gas reservoir at larger circular velocities than the [OIII] emission.  We adopt the [CII] line luminosity for REBELS-25 determined by Schouws et al. (in prep) and determine its line ratio to be $\mathrm{[OIII]}/\mathrm{[CII]} = 0.8 \pm 0.2$.

While [CII] emission in distant galaxies has been observed to be more spatially extended than the [OIII] emission (e.g., \citealt{carniani2020,akins2022}; Section \ref{sec:discussion_OIII}), we note that our results remain unchanged if we re-extract the [CII] fluxes following the procedure adopted for the [OIII] line described in Section \ref{sec:source_detection}. This is as expected, given that the fluxes are extracted in an aperture with a radius of $\sim10\,$kpc at $z=7$, which is much larger than the typical [CII] sizes of galaxies of $r_\mathrm{[CII]}\sim2\,$kpc \citep{fujimoto2020,fudamoto2022}.


\section{Discussion}
\label{sec:discussion}

We proceed by first discussing the physical properties of our targets in terms of the [OIII]/[CII] emission line ratio, and subsequently link these to their dust continuum properties in the sections that follow.

\subsection{[OIII]/[CII] as a proxy for Burstiness}
\label{sec:discussion_OIII}

The ratio between the [OIII] and [CII] luminosities provides a global characterization of the ISM. This line ratio is observed to have typical values of $\lesssim1$ in starbursts at $z\sim0$ \citep{diazsantos2017}, while generally being $\gtrsim 2 - 10$ in high-redshift galaxies ($z\gtrsim6$; \citealt{harikane2020}). This is likely indicative of an increasingly ionized ISM at early cosmic times, which is further supported by observations at intermediate redshifts ($z\sim2$; \citealt{steidel2016,kashino2017}).

Combining a sample of three newly observed $z\approx6$ Lyman Break Galaxies with six $6 \lesssim z \lesssim 9$ galaxies from the literature, \citet{harikane2020} find the line ratios to span a broad range of $\mathrm{[OIII]/[CII]} = 2 - 20$.\footnote{Their literature sample comprises MACS1149-JD1 \citep{hashimoto2018}, A2744-YD4 \citep{laporte2017}, MACS0416-Y1 \citep{tamura2019,bakx2020}, SXDF-NB1006-2 \citep{inoue2016}, The Big Three Dragons \citep{hashimoto2019} and BDF-3299 \citep{carniani2017}.} While \citet{carniani2020} show that some [CII] emission may be missed due to surface brightness dimming, given the typically larger extent of [CII] compared to [OIII], even upon accounting for this, they still find line ratios greater than unity ($\mathrm{[OIII]/[CII]} = 2 - 8$) for the same nine $z > 6$ galaxies analyzed by \citet{harikane2020}. This is in agreement with recent measurements from \citet{witstok2022} for five additional galaxies at $z\approx7$ detected in both [CII] and [OIII] emission, for which they similarly find the line ratios to span $\mathrm{[OIII]} / \mathrm{[CII]} \sim 2 - 8$. In addition, \citet{akins2022} find a ratio of $\mathrm{[OIII]} / \mathrm{[CII]} \sim 2$ for A1689-zD1 at $z=7.13$, when adopting an aperture large enough to include the [CII] emission from the extended halo around the source.

Through detailed {\sc{Cloudy}} \citep{ferland1998,ferland2017} modelling, \citet{harikane2020} find that the large [OIII]/[CII] ratios observed in high-redshift galaxies are likely due to A) a high ionization parameter $U_\mathrm{ion}$, and/or B) a low PDR covering fraction, while other effects, such as metallicity, C/O abundance, and ISM density are subdominant. Both of these scenarios involve suppressing the [CII] luminosity: at fixed metallicity and density, the [CII] luminosity decreases for a larger $U_\mathrm{ion}$ as C$^+$ becomes increasingly ionized (\citealt{harikane2020}; their Figure 11). Similarly, as most of the [CII] emission is believed to emanate from PDRs (e.g., \citealt{cormier2019}), a lower PDR covering fraction also decreases the overall [CII] luminosity.

\begin{figure}
    \centering
    \includegraphics[width=0.49\textwidth]{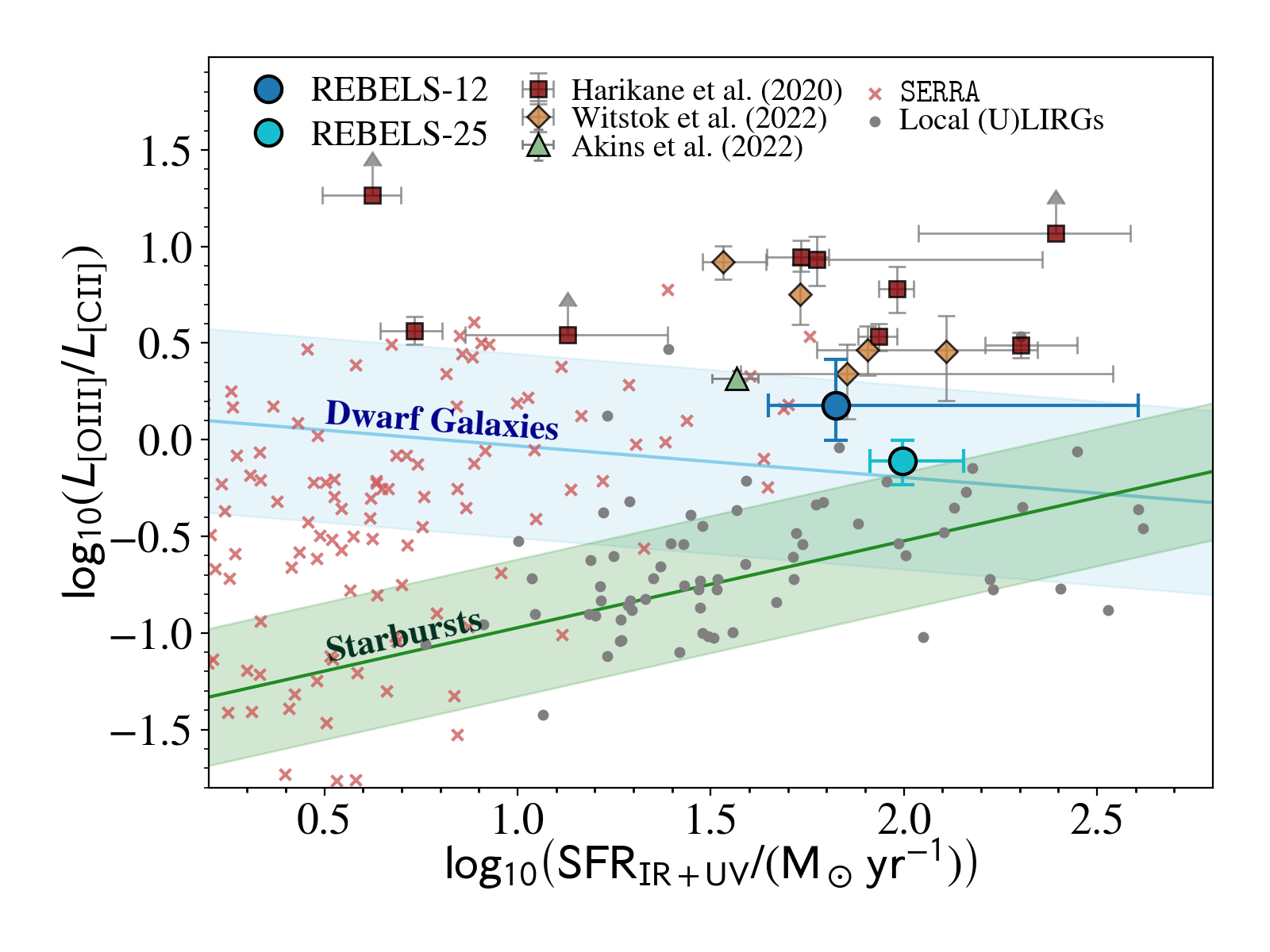}
    \caption{The [OIII]/[CII] luminosity ratio versus total (IR + UV) star formation rate for REBELS-12, REBELS-25 and a variety of galaxies from the literature. The compilation of nine $6 \lesssim z \lesssim 9$ galaxies from \citet{harikane2020} is shown as maroon squares, while the five $z\sim7$ galaxies from \citet{witstok2022} are plotted as orange diamonds. We additionally show the $z=7.13$ galaxy A1689-zD1 recently analyzed by \citet{akins2022}. The $z\sim0$ GOALS (U)LIRGs from \citet{diazsantos2017} are shown as grey dots, and the local relations for starburst and dwarf galaxies from \citet{delooze2014} are indicated through the green and blue bands, respectively. In addition, we show the $z=7.7$ galaxies from the {\tt{SERRA}} simulation suite \citep{pallottini2022} through the pink crosses. Compared to the $z\gtrsim6$ population, REBELS-12 and in particular REBELS-25 show low [OIII]/[CII] ratios.}
    \label{fig:OIII_CII}
\end{figure}

Our sample of three $z\approx7$ galaxies, on the other hand, shows [CII] line luminosities consistent with the local relation between [CII] luminosity and SFR (Schouws et al. in prep). While [OIII] observations exist only for REBELS-12 and REBELS-25, the inferred line ratios ($\mathrm{[OIII]/[CII]}\approx0.8 - 1.5$; Figure \ref{fig:OIII_CII}) are indicative of less extreme physical conditions in our targets compared to the known $z\gtrsim6$ population. Instead, our targets show line ratios consistent with the upper range inferred for local (U)LIRGs \citep{diazsantos2017}.

The line ratios determined for REBELS-12 and REBELS-25 are additionally consistent with the sample of $\sim200$ simulated galaxies analyzed by \citet{pallottini2022}. These galaxies are drawn from the cosmological zoom-in simulations {\tt{SERRA}} at $z=7.7$, which are post-processed in order to obtain galaxy UV, FIR and line luminosities. While the {\tt{SERRA}} sample mostly probes galaxies with $\mathrm{SFR} \sim1 - 10\,M_\odot\,\mathrm{yr}^{-1}$ -- lower than currently accessible with observations of unlensed galaxies such as REBELS -- the agreement between the simulations and observations is nonetheless encouraging. \\

In what follows, we investigate the origin of the different [OIII]/[CII] ratios seen in our REBELS targets and the previously observed $z\gtrsim6$ population quantitatively in the context of the ionization parameter, which \citet{harikane2020} show is the dominant ingredient governing the [OIII]/[CII] emission line ratio. While in practice there is a degeneracy between a lower/higher ionization parameter and higher/lower PDR covering fraction, we here focus on the former scenario. As discussed in \citet{harikane2020}, a low PDR covering fraction would allow for the efficient transmission of ionizing photons, and can therefore be independently investigated through Lyman-$\alpha$ spectroscopy. Keck/MOSFIRE observations of REBELS-12 and REBELS-25, however, do not reveal a Lyman-$\alpha$ detection for these two $z\sim7.3$ targets (Algera et al. in prep). 

As shown by \citet{ferrara2019}, the ionization parameter is related to the star formation rate and molecular gas surface densities via $U_\mathrm{ion} \propto \Sigma_\mathrm{SFR} / \Sigma_\mathrm{gas}^2$. Defining the parameter $\kappa_\mathrm{s}$ as the offset from the local Schmidt-Kennicutt relation via $\Sigma_\mathrm{SFR} = \kappa_\mathrm{s} \Sigma_\mathrm{gas}^{1.4}$, \citet{ferrara2019} rewrite the ionization parameter as $U_\mathrm{ion} \propto \kappa^{1.4}_\mathrm{s}\Sigma_\mathrm{SFR}^{-0.4}$. As a result, large values of the [OIII]/[CII] ratio may be related to high values of the ``burstiness'' $\kappa_s$, corresponding to large upward deviations from the local Schmidt-Kennicutt relation.

We next adopt the models introduced by \citet{vallini2021} to relate the observed emission line ratios to the burstiness of our sources. Their framework, dubbed {\tt{GLAM}},\footnote{\url{https://lvallini.github.io/MCMC_galaxyline_analyzer/}} takes the [OIII], [CII] and star formation rate surface densities as inputs and uses these to constrain the burstiness, metallicity and ISM density through a Monte Carlo-based approach. In order to determine the required line and IR surface densities, we define $\Sigma_\mathrm{line} = L_\mathrm{line} / (2\pi r_\mathrm{line}^2)$, where $r$ is the source radius measured in the image plane and ``line'' refers to either [CII] or [OIII], and $\Sigma_\mathrm{IR} = L_\mathrm{IR} / (2\pi r_\mathrm{IR}^2)$. We determine the galaxy sizes of the different components through image plane fitting with a 2D Gaussian, and quote circularized radii, defined as $r = \sqrt{ab}/2$ where $a,b$ are the FWHM of the major and minor axes, respectively. Given the moderate resolution of our data, we further assume a fiducial $30\%$ uncertainty on all of the size measurements when utilizing them in the \citet{vallini2021} models.

For REBELS-12, both emission lines are spatially resolved, and we adopt $r_\mathrm{[CII]} = 3.8 \pm 1.1\,$kpc and $r_\mathrm{[OIII]} = 1.5 \pm 0.4\,$kpc. The Band 6 continuum emission is additionally resolved, but shows a complex morphology (Section \ref{sec:results_OIII} and Figure \ref{fig:moment0}). In particular, at present it remains unclear if both dust components are physically associated to REBELS-12, or whether the brighter one corresponds to a foreground object. We here conservatively adopt the combined flux density of the two dust components, while adopting a compact size of $r_\mathrm{IR} = r_\mathrm{[OIII]}$ for REBELS-12. This ensures that we effectively determine an upper limit on its true burstiness, which will strengthen our subsequent claims that the REBELS sources are less bursty than the known $z\gtrsim6$ population (Section \ref{sec:discussion_evolved}).

For REBELS-25, we only find the Band 6 continuum to be resolved \citep{inami2022}. Based on a 2D Gaussian fit, we determine $r_\mathrm{IR} = 1.8 \pm 0.5\,$kpc, and we adopt an identical size for the unresolved [OIII] emission. The [CII] emission is similarly not resolved, such that the beam size translates to a robust upper limit on its [CII] size of $r_\mathrm{[CII]} < 3.7\,$kpc. Given the general observation among high-redshift galaxies that [CII] is more extended than the dust continuum (e.g., \citealt{carniani2020}), we therefore adopt a fiducial size of $r_\mathrm{[CII]} = 3.0 \pm 0.9\,$kpc for REBELS-25, in agreement with the findings by \citet{fudamoto2022} that the [CII] sizes of the REBELS sources are $\sim2\times$ larger than their dust continuum sizes.

Based on the inferred line luminosities and galaxy sizes, we determine the corresponding surface densities which we provide as inputs to {\tt{GLAM}}. We recover burstiness parameters of $\kappa_\mathrm{s} = 13.5_{-5.4}^{+7.9}$ and $\kappa_\mathrm{s} = 14.5_{-4.6}^{+9.5}$ for REBELS-12 and REBELS-25, respectively. This is on the low end of the burstiness of $\kappa_\mathrm{s} = 10 - 100$ (average of $\kappa_\mathrm{s} \sim40$) inferred among the sample of nine high-redshift galaxies analyzed by \citet{vallini2021}. However, the results are consistent with the median burstiness of the simulated {\tt{SERRA}} galaxies at $z=7.7$ of $\kappa_\mathrm{s} = 6.5_{-4.0}^{+11.5}$ \citep{pallottini2022}. 

We relate the inferred burstiness to the global physical properties of our REBELS targets in Section \ref{sec:discussion_Tdust}, and we utilize them in the context of single-band dust temperature modelling in Appendix \ref{app:singleband_Tdust}. There we show that the temperatures predicted from the \citet{sommovigo2022} single-band dust models are in agreement with those obtained from MBB-fitting, when adopting appropriate priors on the burstiness. In addition, in Appendix \ref{app:singleband_Tdust} we show that the fitted temperatures agree with those predicted by the single-band models from \citet{inoue2020} and \citet{fudamoto2022b}.

\subsection{Dust Temperatures at High Redshift}
\label{sec:discussion_Tdust}

\begin{figure*}
    \centering
    \includegraphics[width=0.9\textwidth]{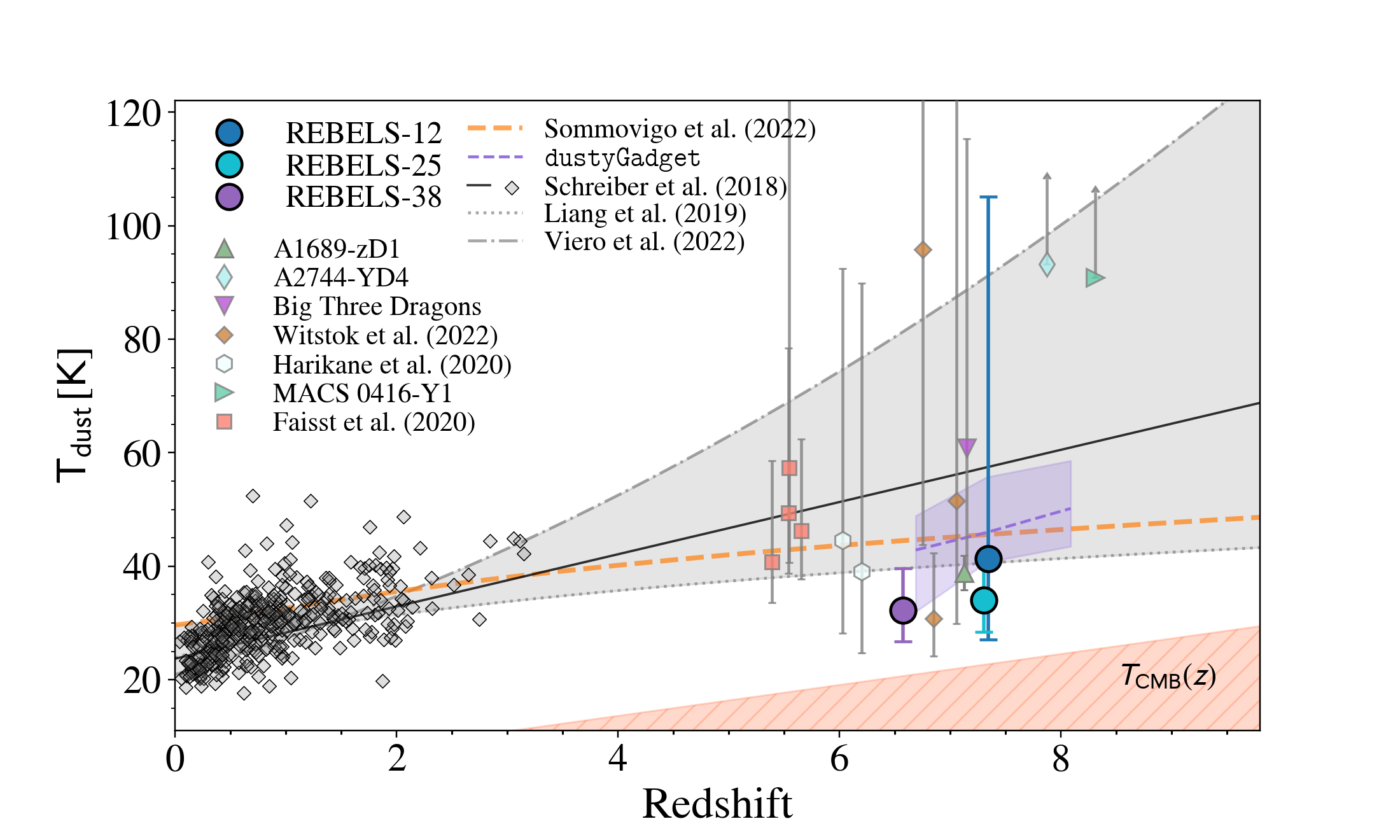}
    \caption{Dust temperature as a function of redshift for the three REBELS targets analyzed in this work (large circles). In addition, we consistently analyze all sources at $z > 6$ with multi-frequency ALMA continuum observations, at least one of which constitutes a detection, using $\beta=2.0$. At $z\sim5.5$ we overplot the four sources studied by \citet{faisst2020}, while at $z\lesssim4$ we show individual galaxies from \citet{schreiber2018}. We additionally include their linear fit as the solid black line. Moreover, we overplot the $T_\mathrm{dust}-z$ relations from \citet{liang2019}, \citet{sommovigo2022}, \citet{viero2022} and {\tt{dustyGadget}} (Schneider et al.\ in prep) as indicated in the legend. The grey shading highlights the large spread in these relations, which extend up to $T_\mathrm{dust}\sim100\,$K at $z\sim7$. In contrast, REBELS-25 and REBELS-38 are both among the coldest known sources at this redshift.}
    \label{fig:Tdust_redshift}
\end{figure*}

Since the advent of ALMA, considerable effort has been dedicated to constraining the dust temperatures of individual galaxies in the epoch of reionization, in order to get an accurate census of their dust-obscured star formation rates, and to constrain models of dust build-up in the early Universe. While it remains unclear if, and to what extent, average dust temperatures evolve with redshift, what \emph{is} clear from observations of individual sources at $z\gtrsim6$ is that the scatter in $T_\mathrm{dust}$ is large. This is illustrated by the low temperature of $T_\mathrm{dust}\sim30\,$K found for COS-3018555981 at $z\sim7$ by \citet{witstok2022}, and the existence of two sources at $z\sim8$ with dust temperatures in excess of 80\,K \citep{laporte2017,behrens2018,bakx2020}. However, the set of high-redshift galaxies with multi-band ALMA photometry is very inhomogeneous, and any comparisons across the literature sample are further complicated by the adoption of different fitting techniques and priors.\footnote{For example, different studies adopt different values of $\beta$, different priors on $T_\mathrm{dust}$, and different definitions of `best fit' (e.g., maximum a posteriori likelihood vs. median of the posteriors).} In order to fairly compare the REBELS targets analyzed in this work to the literature, we therefore compile the observed continuum flux densities for all $z\gtrsim6$ galaxies targeted at a minimum of two distinct ALMA bands with at least one of those being a continuum detection at $>3\sigma$. This results in a sample of nine galaxies (listed in Table \ref{tab:literature} in the Appendix), four of which are detected in continuum emission in two or more bands (at $>3\sigma$). This literature sample includes two galaxies at $z\sim6$ from \citet{harikane2020}, three sources at $z\sim7$ from \citet{witstok2022}, the Big Three Dragons at $z=7.15$ \citep{hashimoto2019,sugahara2021}, A1689-zD1 at $z=7.13$ \citep{bakx2021,akins2022}, A2744-YD4 at $z=7.88$ \citep{laporte2017,laporte2019,morishita2022} and MACS0416-Y1 at $z=8.31$ \citep{bakx2020}.\footnote{A redshift of $z=8.38$ was previously adopted for A2744-YD4 by \citet{laporte2017}, but recent \emph{JWST}/NIRSpec observations from \citet{morishita2022} indicate the source is likely to be at $z=7.88$ instead.} We note that the last three sources are gravitationally lensed galaxies. In what follows, we adopt our fiducial model of optically thin dust with $\beta=2.0$ for the REBELS and $z\gtrsim6$ literature samples. However, we note that our results are qualitatively consistent when a different fixed value of $\beta$ is adopted. \\ 

We show the dust temperatures of the REBELS and re-fitted $z\gtrsim6$ literature samples as a function of redshift in Figure \ref{fig:Tdust_redshift}. We additionally overplot the $z\sim5.5$ galaxies from \citet{faisst2020} and several sources at lower redshift from \citet{schreiber2018}, adopting the dust temperatures directly from these respective works (see also \citealt{sommovigo2022}).

Compared to the previously observed population of $z>6$ galaxies, the three REBELS sources analyzed in this work are characterized by colder dust of $T_\mathrm{dust}\sim30-35\,$K. By concatenating the posterior $T_\mathrm{dust}$ probability density functions of our REBELS targets (c.f., Figure \ref{fig:Tdust_posterior}), we determine a median dust temperature of $\langle T_{\mathrm{dust},z} \rangle = 34_{-7}^{+14}\,$K across the three sources. The unconstrained dust temperature for REBELS-12 results in a relatively large tail towards higher $T_\mathrm{dust}$ in the stacked posterior distribution. However, the combination of a low [OIII]/[CII] ratio and a Band 8 non-detection likely implies that REBELS-12, too, hosts relatively cold dust (see also Section \ref{sec:discussion_evolved}). Instead, the long tail towards higher temperatures is likely due to the low-S/N detection ($3.6\sigma$) at rest-frame $160\,\mu$m in combination with the upper limit at rest-frame $90\,\mu$m. In the case of REBELS-12 -- and sources with similarly poorly-constraining data -- fitted dust temperatures will be quite sensitive to the adopted prior on the dust temperature. If a wider prior is adopted, hotter solutions are generally allowed, resulting in a large uncertainty on $T_\mathrm{dust}$. For example, if we widen our prior on the dust temperature to allow values up to $300\,$K, the upper error on dust temperature of REBELS-12 increases from $64\,$K (c.f., Table \ref{tab:source_parameters}) to $129\,$K, while the fits for REBELS-25 and REBELS-38 are unaffected. As such, one has to be careful when interpreting dust temperatures obtained from low-S/N photometry and upper limits.

For the literature sample, we determine a median dust temperature of $\langle T_{\mathrm{dust},z} \rangle = 55_{-24}^{+88}\,$K by concatenating the posteriors obtained from fitting our fiducial MBB model. The long tail of this distribution towards high temperatures is the result of the stringent continuum non-detection at $158\,\mu$m for both lensed galaxies at $z\sim8$, MACS0416-Y1 and A2744-YD4. Upon leaving out these sources, we determine a median dust temperature of $\langle T_{\mathrm{dust},z}\rangle = 42_{-13}^{+63}\,$K, still slightly higher than the median temperature determined for the three REBELS sources presented in this work. This could be due to the low-S/N detections for part of the sample, or due to an intrinsically higher dust temperature. In the latter case, Band 8 observations do not probe the peak of the dust emission, such that the temperature cannot accurately be constrained even in the case of a high-S/N detection. This, too, manifests as a long tail of plausible high-temperature solutions in the $T_\mathrm{dust}$ posterior distribution (see also \citealt{bakx2020}). \\

We proceed by comparing to various redshift-dependent parameterizations of the dust temperature, obtained from stacking either observed \citep{schreiber2018,viero2022} or simulated (\citealt{liang2019}; Schneider et al. in prep) galaxy samples, and the physical model from \citet{sommovigo2022}.

Through stacking UV-selected galaxies in {\emph{Herschel}} maps, \citet{schreiber2018} obtain a linear trend between $T_\mathrm{dust}$ and redshift ranging from 20\,K locally to 40\,K at $z\sim4$ (see also \citealt{bethermin2015}). However, several recent studies have shown widely contrasting results in terms of dust temperature evolution. For instance, \citet{drew2022} find no evidence for an increase in $T_\mathrm{dust}$ out to $z\approx2$, while, in contrast, \citet{viero2022} prefer a rapidly rising dust temperature that reaches a typical $T_\mathrm{dust}\sim100\,$K at $z\approx8$. The high temperatures measured by the latter are consistent with the two $z\sim8$ lensed galaxies observed by \citet{laporte2017} and \citet{bakx2020}, but appear inconsistent with the majority of the known $z\gtrsim6$ population, including our REBELS targets. As \citet{viero2022} admit, the high-redshift sample they use for stacking is likely to show significant contamination from low-redshift interlopers, thereby potentially biasing the temperature upwards (see also the discussion in \citealt{sommovigo2022b}).

Instead, better agreement is found with the trends from \citet{liang2019}, \citet{sommovigo2022} and the {\tt{dustyGadget}} simulations (\citealt{graziani2020,dicesare2022}, Schneider et al. in prep), which all predict a slightly increasing dust temperature towards higher redshift. \citet{liang2019} focus on the dust temperatures of $\sim30$ galaxies identified in the {\tt{MassiveFIRE}} cosmological zoom-in simulations \citep{feldmann2016}, with stellar masses spanning $10^9 \lesssim M_\star / M_\odot \lesssim 10^{11}$ at $z=6$. They use radiative transfer code {\tt{SKIRT}} \citep{baes2011} to generate dust SEDs and self-consistently calculate galaxy dust temperatures. While they do not find evidence for a strong correlation between stellar mass and (peak) dust temperature, \citet{liang2019} find temperatures of $T_\mathrm{dust}(z=6)\approx 35 - 40\,$K for $M_\star\sim10^{10}\,M_\odot$ galaxies, in agreement with the values derived for our similarly massive REBELS targets.\footnote{We note that \citet{liang2019} fit a trend to the peak dust temperature, $T_\mathrm{dust}^\mathrm{peak} \propto \lambda_\mathrm{peak}^{-1}$, where $\lambda_\mathrm{peak}$ is the wavelength where the dust SED peaks. However, for our fiducial model of optically thin dust with $\beta=2.0$, the modified blackbody temperature and peak temperature are equal to one another.} In addition, by considering the dust temperatures of their full sample between $2 \lesssim z \lesssim 6$, \citet{liang2019} fit a trend between $T_\mathrm{dust}$ and redshift to all galaxies in their sample with $L_\mathrm{IR} > 10^{11}\,L_\odot$, which predicts $T_\mathrm{dust}(z=7)\approx 40\,$K, consistent with our measurements.

In addition, our results are in reasonable agreement with the models from \citet{sommovigo2022}. They argue that the dust temperatures of massive, relatively metal-rich galaxies depend predominantly on their depletion timescales, whereby sources with a shorter $t_\mathrm{depl}$ are expected to be hotter. Using the redshift-dependent parameterization of the depletion timescale from \citet{tacconi2020}, their models predict a typical $T_\mathrm{dust}(z=7)\approx45\,$K.\footnote{When instead assuming the depletion timescale traces the cosmic halo accretion rate as predicted by numerical simulations, \citet{sommovigo2022} estimate a higher temperature of $T_\mathrm{dust}(z=7)\approx55\,$K.} While slightly hotter (by $\sim10\,$K) than the temperatures determined for our Band 8 detected REBELS targets, their model can physically explain the scatter towards lower dust temperatures by supposing colder galaxies have longer depletion timescales (see also Section \ref{sec:discussion_evolved}).

Finally, we compare our results to the dust temperatures computed at $6.7 \lesssim z \lesssim 8.1$ from the {\tt{dustyGadget}} simulations (Schneider et al. in prep), obtained from post-processing the simulations with {\tt{SKIRT}}. Galaxy-averaged dust temperatures are computed by modelling the absorption and re-emission of individual dust patches within the galaxy, assuming each emits as a modified blackbody. In Figure \ref{fig:Tdust_redshift}, we show the typical peak dust temperature obtained from averaging the temperatures across the simulated galaxy sample at three distinct redshifts. The temperatures obtained from {\tt{dustyGadget}} show a slight redshift dependence, and predict a typical temperature of $T_\mathrm{dust}\sim40 - 45\,$K at $z\approx7$, in good agreement with the measurements for REBELS-25 and REBELS-38.

In summary, the dust temperatures we measure for our REBELS sample are among the lowest observed thus far at $z\approx7$. However, the temperatures are in agreement with those predicted from simulations (\citealt{liang2019}, Schneider et al. in prep), and the physical model from \citet{sommovigo2022}. This may indicate that galaxies with $T_\mathrm{dust} \gg 50\,$K are outliers at $z \gtrsim 6$, and that typical dust temperatures exhibit only a moderate increase towards high redshift.

\subsection{The Dust Properties of UV-Bright Galaxies at $z=7$}
\label{sec:discussion_Mdust}

\begin{figure*}
    \centering
    \includegraphics[width=0.9\textwidth]{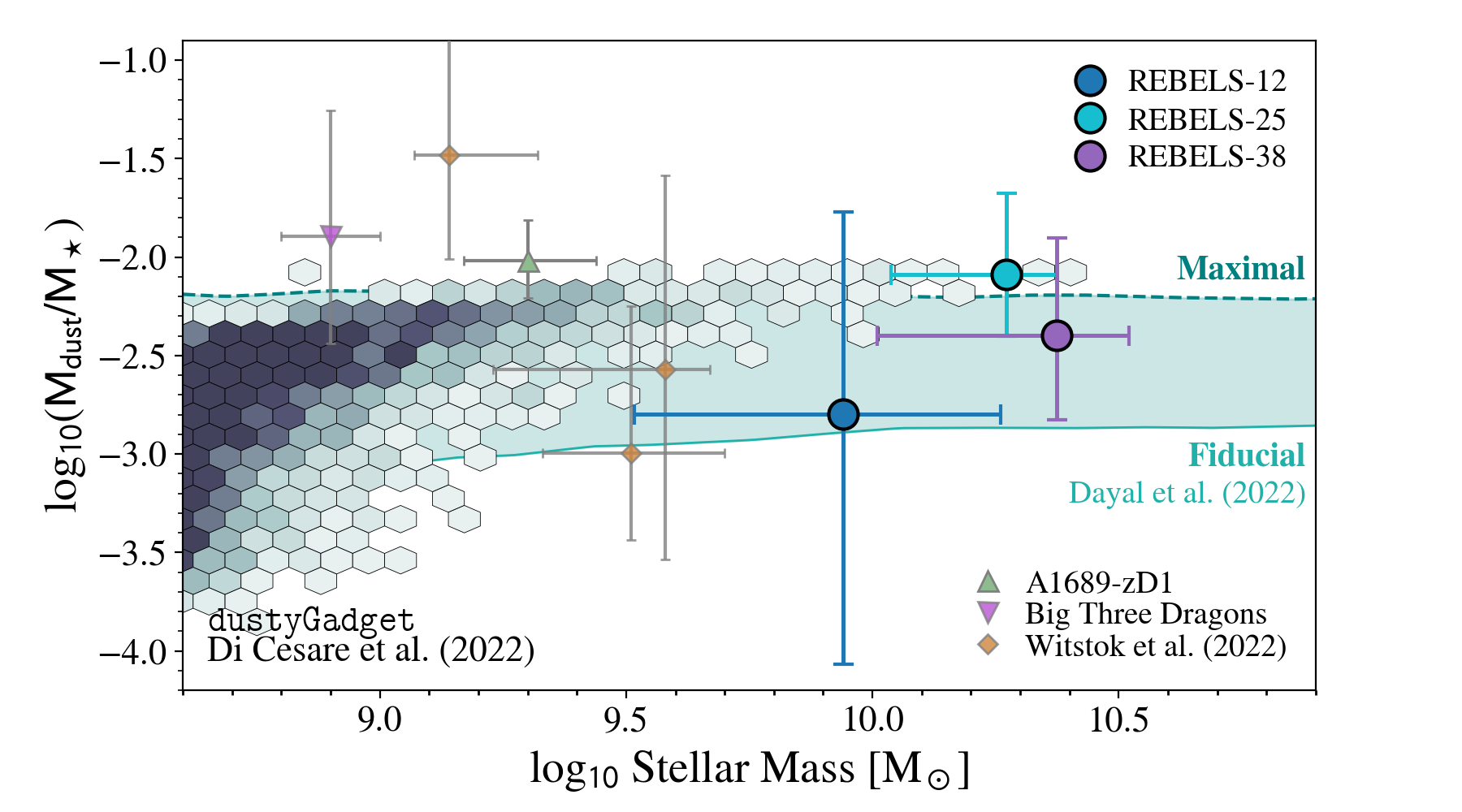}
    \caption{A comparison of the inferred REBELS dust masses, expressed as the ratio of dust-to-stellar mass, with observations and simulations of high-redshift galaxies. We show a compilation of $z\approx7$ sources with robust temperature measurements, as well as the models from \citet{dayal2022} and \citet{dicesare2022}. The dust masses inferred for REBELS agree with the massive end of the simulated sample from {\tt{dustyGadget}} at $z=6.7$, which predict $\log_{10}(M_\mathrm{dust}/M_\star) \approx -2$. In addition, agreement is found with the models from \citet{dayal2022}, though REBELS-25 would require rapid dust build-up and little destruction as in their ``maximal'' model. Overall, we conclude that while high, the dust masses observed in our massive $z\sim7$ sample can be reproduced by current state-of-the-art models and simulations.}
    \label{fig:Mdust_Mstar}
\end{figure*}

We proceed by discussing the dust properties of the three REBELS targets analyzed in this work. As before, we adopt the analysis using an optically thin modified blackbody with a fixed $\beta = 2.0$ as our fiducial model. We convert the measured IR luminosities to dust-obscured star formation rates via $\mathrm{SFR}_\mathrm{IR}/(M_\odot\,\mathrm{yr}^{-1}) = 1.2\times10^{-10} \times (L_\mathrm{IR}/L_\odot)$ \citep{bouwens2022}. For REBELS-25 and REBELS-38, which are continuum-detected at both rest-frame $90\,\mu$m and $160\,\mu$m, we determine obscured SFRs of $84_{-17}^{+45}\,M_\odot\,\mathrm{yr}^{-1}$ and $36_{-8}^{+21}\,M_\odot\,\mathrm{yr}^{-1}$, respectively. These values indicate that the bulk of the star formation in these galaxies is obscured, with $f_\mathrm{obs} = \mathrm{SFR}_\mathrm{IR} / (\mathrm{SFR}_\mathrm{IR} + \mathrm{SFR}_\mathrm{UV}) = 0.86 \pm 0.05$ and $0.68_{-0.08}^{+0.10}$ for REBELS-25 and REBELS-38, respectively. Despite our finding that these sources have a lower infrared luminosity than previously expected based on a Band 6 detection and the \citet{sommovigo2022} models, their obscured fraction remains larger than that of the average REBELS sample based on Band 6 observations alone ($\langle f_\mathrm{obs} \rangle \approx 0.45$; \citealt{algera2022}), reinforcing the highly obscured nature of these luminous UV-selected sources.

The inferred dust masses for REBELS-25 and REBELS-38 are similarly significant. REBELS-25, in particular, has an inferred dust mass of $\log_{10}\left(M_\mathrm{dust}/M_\odot\right) = 8.13_{-0.27}^{+0.38}$ given $\beta=2.0$. Making use of the stellar masses determined by \citet{topping2022}, we find its dust-to-stellar mass ratio to be $M_\mathrm{dust} / M_\star = 8.2_{-4.2}^{+12.9}\times10^{-3}$. For REBELS-38, we infer a slightly lower dust mass of $\log_{10}\left(M_\mathrm{dust}/M_\odot\right) = 7.90_{-0.35}^{+0.39}$, and similarly a slightly lower $M_\mathrm{dust} / M_\star = 4.0_{-2.5}^{+8.5}\times10^{-3}$. 

We show these dust-to-stellar mass ratios in the context of models of dust production and destruction in Figure \ref{fig:Mdust_Mstar}. In addition, we overplot five sources (three from \citealt{witstok2022}, A1689-zD1 and the Big Three Dragons) for which stellar mass measurements are available in the literature, and which provided robust temperature measurements in Section \ref{sec:discussion_Tdust}. We note that the literature stellar masses are based on parametric SED fitting, and may therefore be underestimated. For REBELS, for example, non-parametric mass measurements are larger than those obtained for an assumed constant star formation history by a typical $\sim0.4\,$dex \citep{topping2022}.

Utilizing the dust masses estimated from the single-band REBELS observations at rest-frame $158\,\mu$m, \citet{dayal2022} investigate the dust build-up of the REBELS sample through the {\tt{Delphi}} semi-analytical model. Their fiducial model, which predicts a typical $M_\mathrm{dust} / M_\star \sim 10^{-3}$, can readily explain the dust contents for 11/13 of the dust- and [CII]-detected REBELS galaxies, which they predict is predominantly due to dust produced in supernovae. For the remaining two sources -- the low-mass REBELS-19 and REBELS-39 -- they require minimal dust destruction/ejection, and rapid grain growth in the ISM (the ``maximal dust'' model in \citealt{dayal2022}, which predicts $M_\mathrm{dust} / M_\star \sim 6 - 7 \times 10^{-3}$). For REBELS-25 and REBELS-38, we now infer $\sim0.4 - 0.5\,$dex larger dust masses compared to the earlier estimates based on the [CII] and underlying continuum emission in these sources from \citet{sommovigo2022}.\footnote{For consistency, we adopt the dust masses predicted by \citet{sommovigo2022} adapted to the stellar masses from \citet{topping2022} used in this work; these are provided in \citet{sommovigo2022b}.} REBELS-38 now falls between the fiducial and maximal dust models, while REBELS-25 requires dust build-up nearly as rapid as in the maximal dust model.

We next compare the measured dust masses for REBELS with those found across a large sample of $z=6.7$ galaxies in the {\tt{dustyGadget}} simulations by \citet{dicesare2022}. Their hydrodynamical simulations include a two-phase ISM, which facilitates the modelling of various dust production and destruction mechanisms based on the local physical conditions in the galaxy. In particular, dust is produced through a combination of supernovae, AGB stars and grain growth in the ISM, while it may be destroyed in a variety of processes such as shocks, astration and sputtering \citep{graziani2020,dicesare2022}. The {\tt{dustyGadget}} simulations are capable of producing large dust reservoirs in massive ($M_\star \gtrsim 10^{10}\,M_\odot$) galaxies already at $z\approx7$, reaching up to $M_\mathrm{dust} / M_\star \sim 10^{-2}$. As such, their models are in excellent agreement with the inferred dust properties of our REBELS targets.

Overall, we conclude that REBELS-25 and REBELS-38 have massive dust reservoirs of $M_\mathrm{dust}\sim10^{8}\,M_\odot$, corresponding to a significant fraction of their stellar mass ($\sim 0.4 - 0.8\%$). Nevertheless, the inferred dust masses are consistent with those predicted by state-of-the art models of dust production in the early Universe, and demonstrate that dust build-up can proceed in a rapid manner at high redshift. In the future, larger samples of massive, dusty galaxies will allow for a detailed investigation of the relative contributions of the various avenues of dust production in the earliest galaxies.

\subsection{An Evolved Star-forming Population at Redshift 7}
\label{sec:discussion_evolved}

In the previous sections, we have outlined how our REBELS targets are characterized by 1) low [OIII]/[CII] ratios, and 2) low dust temperatures compared to the known $z\gtrsim6$ population. In addition, our targets have relatively high dust masses ($M_\mathrm{dust} \sim 10^8\,M_\odot$) as well as high stellar masses ($M_\star \sim 10^{10}\,M_\odot$). All of these characteristics indicate that the three REBELS sources analyzed in this work have previously experienced one (or multiple) burst(s) of star formation, while currently residing in a less bursty state. Indeed, while their inferred burstiness parameters ($\kappa_\mathrm{s}\sim10 - 15$; Section \ref{sec:discussion_OIII}) place these galaxies $\sim1\,$dex above the \emph{local} Schmidt-Kennicutt relation, they appear less bursty than the previously observed $z\gtrsim6$ population.

As an independent tracer of burstiness, we consider the combined equivalent width (EW) of the $\mathrm{[OIII]}\lambda\lambda4959,5007$ doublet and $\mathrm{H}\beta$ line [denoted $\mathrm{EW}(\mathrm{[OIII]}+\mathrm{H}\beta)$], which are redshifted into the \emph{Spitzer}/IRAC channels for galaxies at $z\gtrsim6.5$ and are capable of significantly contributing to the observed broadband fluxes (e.g., \citealt{smit2014,smit2015,roberts-borsani2016,schouws2022b}). Given that these are nebular emission lines, they trace the recent formation of massive stars, and hence form a proxy for galaxy burstiness.

By fitting non-parametric star formation histories to available rest-UV/optical observations of the REBELS sample, \citet{topping2022} find a mean equivalent width of $\mathrm{EW}(\mathrm{[OIII]}+\mathrm{H}\beta) = 735_{-93}^{+97}\,$\AA\ at $z\sim7$, in agreement with previous results from \citet{endsley2021}. REBELS-25 and REBELS-38, however, have an inferred $\mathrm{EW}(\mathrm{[OIII]}+\mathrm{H}\beta) = 127_{-92}^{+327}\,$\AA\ and $143_{-91}^{+617}\,$\AA, significantly below the REBELS average, while REBELS-12, on the other hand, is characterized by an equivalent width of $\mathrm{EW}(\mathrm{[OIII]}+\mathrm{H}\beta) = 2217_{-981}^{+710}\,$\AA, among the largest in REBELS.

The low expected burstiness of REBELS-25 and REBELS-38 is therefore supported by their relatively small $\mathrm{[OIII]}+\mathrm{H}\beta$ equivalent widths, while the EW of REBELS-12 implies significant bursty star formation, in apparent contrast with its observed [OIII]/[CII] line ratio. It is possible that the complex kinematics inferred from its broad [CII] line lie at the basis of this discrepancy in REBELS-12. The broad [CII] profile may indicate that REBELS-12 is an ongoing merger, whereby the [OIII]$88\,\mu$m emitting system dominates the $\mathrm{EW}(\mathrm{[OIII]}+\mathrm{H}\beta)$, while both components emit in [CII]. In this scenario, we would infer a line ratio of $\mathrm{[OIII]/[CII]} = 3.4_{-1.1}^{+3.6}$ for the bursty component, and an upper limit of $\mathrm{[OIII]/[CII]} < 1.1$ for the component not detected in [OIII] emission (Appendix \ref{app:rebels12_cii}). This would bring (the bursty component of) REBELS-12 into closer agreement with the line ratios of $\mathrm{[OIII]/[CII]} \sim 2 - 8$ inferred for the five $z\sim7$ galaxies from \citet{witstok2022}, which are also characterized by a relatively high $\mathrm{EW}(\mathrm{[OIII]}+\mathrm{H}\beta) \approx 600 - 1400\,$\AA. We further note that, while the non-detection of REBELS-12 at rest-frame $90\,\mu$m likely implies a low dust temperature, we cannot currently rule out warmer dust, which in turn might be indicative of a larger burstiness.

In what follows, we therefore focus on REBELS-25 and REBELS-38. We can interpret their lack of bursty star formation as both sources having a long depletion timescale ($t_\mathrm{depl} = M_\mathrm{gas} / \mathrm{SFR}$), for which we can provide a rough estimate by converting their [CII] luminosities into a molecular gas mass following \citet{zanella2018}, resulting in $\log_{10}(M_\mathrm{gas} / M_\odot) \approx 10.7 \pm 0.3$. Utilizing the total (UV + IR) SFRs, we determine a depletion timescale of $t_\mathrm{depl} = 470_{-250}^{+520}\,$Myr and $t_\mathrm{depl} = 900_{-490}^{+1010}\,$Myr for REBELS-25 and REBELS-38, respectively. While the uncertainties are large, these depletion timescales appear higher than the typical depletion time found for the ALPINE sample at $z\sim5.5$ by \citet{dessauges-zavadsky2020} of $t_\mathrm{depl} = 230 - 460\,$Myr. In addition, the estimated depletion times exceed an extrapolation of the redshift-dependent trend from \citet{tacconi2018}, which predicts $t_\mathrm{depl}\sim300\,$Myr at $z=7$. As such, REBELS-25 and REBELS-38 may indeed show longer depletion timescales than expected for the typical galaxy population at this epoch. For a detailed discussion of the [CII]-based molecular gas masses and depletion timescales of the full REBELS sample, we refer the reader to Aravena et al. (in prep).

In other words, compared to the typical $z\gtrsim6$ population, our targets may have higher gas masses given their SFRs, and therefore also larger dust masses, assuming a simple (metallicity-dependent) scaling between dust and gas mass known as the dust-to-gas ratio (e.g., \citealt{remyruyer2014}). This is indeed corroborated by our observations, which indicate large dust masses of $M_\mathrm{dust} \sim 10^8\,M_\odot$ for REBELS-25 and REBELS-38. Such elevated dust masses could then provide a natural explanation for the low observed temperatures, as a larger dust content implies there is, on average, less energy available per unit dust mass to heat the dust to higher temperatures. 

We caveat, however, that the spatial distribution of dust within the galaxy is expected to additionally affect the measured dust temperatures. For example, if dust is preferentially located directly around sites of star formation, a higher dust temperature would be inferred observationally given that a hot dust component can dominate the infrared luminosity of a galaxy even if it contributes little to the overall dust mass (e.g., \citealt{liang2019}). A low dust temperature, on the other hand, could therefore also be the result of spatial segregation between the sites of star formation and the dust. We emphasize that high-resolution dust continuum observations are required to test this scenario in detail. However, given the independent constraints on the burstiness of our sources from the [OIII]/[CII] ratio, long depletion timescales and low $\mathrm{EW}(\mathrm{[OIII]}+\mathrm{H}\beta)$, our preferred interpretation is a scenario where the low dust temperatures are the result of a lack of bursty star formation resulting in the inefficient heating of a massive dust reservoir.



\section{Summary}
\label{sec:summary}
We present new ALMA Band 8 ($\lambda_\mathrm{rest} \approx 90\,\mu$m) observations of three massive UV-selected galaxies ($M_\star \approx 10^{9.9 - 10.4}\,M_\odot$) at $z\sim7$, REBELS-12, REBELS-25 and REBELS-38, in order to constrain their far-infrared spectral energy distributions. All three targets were previously detected in $\mathrm{[CII]}158\,\mu$m emission and underlying Band 6 continuum (Schouws et al. in prep and \citealt{inami2022}, respectively). The new Band 8 observations additionally target the $\mathrm{[OIII]}88\,\mu$m line in REBELS-12 and REBELS-25, allowing insight into their global ISM properties through the [OIII]/[CII] line ratio. Our main conclusions are as follows:

\begin{itemize}
    \item We detect Band 8 continuum emission in REBELS-25 ($4.4\sigma$) and REBELS-38 ($5.3\sigma$), and determine an upper limit for REBELS-12 ($<3\sigma$). This allows us to measure dust temperatures for our sample through modified blackbody fitting. We adopt a fiducial model of an optically thin MBB with $\beta=2.0$, but also investigate both optically thick dust and other possible values of $\beta$. Using our fiducial parameterization, we measure relatively cold dust temperatures for REBELS-25 and REBELS-38 of $T_\mathrm{dust}\sim 30 - 35\,$K, while the dust temperature of REBELS-12 cannot be accurately constrained due to its non-detection in Band 8. However, the upper limit in combination with its low [OIII]/[CII] ratio likely implies a similarly cold dust temperature. 
    \item We compare the dust temperatures of our targets to literature sources at $z\gtrsim6$ by fitting them in a consistent manner, and find that the REBELS galaxies are among the coldest known sources at this epoch. Nevertheless, the measured temperatures are in good agreement with those predicted by simulations (e.g., \citealt{liang2019}; Schneider et al. in prep) and with the models from \citet{sommovigo2022}, while being significantly lower than (extrapolations of) stacked samples at $z\sim7$ \citep{schreiber2018,viero2022}.
    \item Given that REBELS-25 and REBELS-38 have low inferred dust temperatures, yet are robustly detected in both ALMA Bands 6 and 8, they require large dust masses of $M_\mathrm{dust}\sim10^{8}\,M_\odot$. Their dust-to-stellar mass ratios are $\sim 0.4 - 0.8\%$, which necessitate rapid dust build-up. However, the dust contents of REBELS-25 and REBELS-38 are consistent with those predicted for massive galaxies by simulations \citep{dayal2022,dicesare2022}.
    \item We detect [OIII] emission in both REBELS-12 and REBELS-25, and find [OIII]/[CII] ratios of approximately unity, lower by a factor of $2 - 8\times$ compared to the previously observed population of $6 \lesssim z \lesssim 9$ galaxies. Through the models of \citet{vallini2021}, we link the [OIII]/[CII] ratios to the physical properties of our targets. In particular, the low line ratios indicate that our targets are less bursty than the known $z\gtrsim6$ galaxy population. Instead, the three REBELS sources are likely characterized by weaker ionizing radiation fields due to their lower star formation rate surface densities. This is consistent with the low $\mathrm{EW}(\mathrm{[OIII]}+\mathrm{H}\beta)\approx140\,$\AA\ inferred from broadband \emph{Spitzer}/IRAC fluxes for REBELS-25 and REBELS-38 \citep{topping2022}.
\end{itemize}

The REBELS sources analyzed in this work are characterized by low dust temperatures, high dust and stellar masses and low [OIII]/[CII] ratios. Taken together, this implies that they may be relatively evolved star-forming galaxies already at $z\approx7$. The low emission line ratios likely indicate that our targets are currently steadily forming stars, as opposed to in a bursty manner, translating into longer depletion timescales and a low heating efficiency per unit dust mass, resulting in lower overall dust temperatures. 

Further dual/multi-band ALMA observations targeting a larger number of $z\sim7$ galaxies are required to elucidate whether this relatively evolved evolutionary state may indeed be common in high-redshift sources, or whether the galaxies analyzed in this work constitute a less common yet highly interesting population at this epoch.

\section*{Acknowledgements}

This paper made use of the following software packages: {\tt{spectral-cube}} \citep{ginsburg2019}, {\tt{radio-beam}} \citep{koch2021}, {\tt{interferopy}} \citep{boogaard2021} and {\tt{GLAM}} \citep{vallini2021}.

The authors thank Livia Vallini for help with the {\tt{GLAM}} software package, Monica Rela\~{n}o for providing feedback on the draft and Claudia di Cesare for sharing data from the {\tt{dustyGadget}} simulations.

This work was supported by NAOJ ALMA Scientific Research Grant Code 2021-19A (HSBA and HI). PD acknowledges support from the NWO grant 016.VIDI.189.162 (``ODIN") and from the European Commission's and University of Groningen's CO-FUND Rosalind Franklin program. MA acknowledges support from FONDECYT grant 1211951, CONICYT + PCI + INSTITUTO MAX PLANCK DE ASTRONOMIA MPG190030, CONICYT + PCI + REDES 190194, and ANID BASAL project FB210003. IDL and MP acknowledge support from ERC starting grant 851622 DustOrigin. RS acknowledges support from a STFC Ernest Rutherford Fellowship (ST/S004831/1). MS acknowledges support from the CIDEGENT/2021/059 grant, from project PID2019-109592GB-I00/AEI/10.13039/501100011033 from the Spanish Ministerio de Ciencia e Innovaci\'on - Agencia Estatal de Investigaci\'on, and from Proyecto ASFAE/2022/025 del Ministerio de Ciencia y Innovaci\'on en el marco del Plan de Recuperaci\'on, Transformaci\'on y Resiliencia del Gobierno de Espa\~na.

\section*{Data Availability}
The data underlying this article will be made available upon reasonable request to the corresponding author.



\bibliographystyle{mnras}
\bibliography{main} 

\begin{thebibliography}{}
\makeatletter
\relax
\def\mn@urlcharsother{\let\do\@makeother \do\$\do\&\do\#\do\^\do\_\do\%\do\~}
\def\mn@doi{\begingroup\mn@urlcharsother \@ifnextchar [ {\mn@doi@}
  {\mn@doi@[]}}
\def\mn@doi@[#1]#2{\def\@tempa{#1}\ifx\@tempa\@empty \href
  {http://dx.doi.org/#2} {doi:#2}\else \href {http://dx.doi.org/#2} {#1}\fi
  \endgroup}
\def\mn@eprint#1#2{\mn@eprint@#1:#2::\@nil}
\def\mn@eprint@arXiv#1{\href {http://arxiv.org/abs/#1} {{\tt arXiv:#1}}}
\def\mn@eprint@dblp#1{\href {http://dblp.uni-trier.de/rec/bibtex/#1.xml}
  {dblp:#1}}
\def\mn@eprint@#1:#2:#3:#4\@nil{\def\@tempa {#1}\def\@tempb {#2}\def\@tempc
  {#3}\ifx \@tempc \@empty \let \@tempc \@tempb \let \@tempb \@tempa \fi \ifx
  \@tempb \@empty \def\@tempb {arXiv}\fi \@ifundefined
  {mn@eprint@\@tempb}{\@tempb:\@tempc}{\expandafter \expandafter \csname
  mn@eprint@\@tempb\endcsname \expandafter{\@tempc}}}

\bibitem[\protect\citeauthoryear{{Akins} et~al.,}{{Akins}
  et~al.}{2022}]{akins2022}
{Akins} H.~B.,  et~al., 2022, \mn@doi [\apj] {10.3847/1538-4357/ac795b}, \href
  {https://ui.adsabs.harvard.edu/abs/2022ApJ...934...64A} {934, 64}

\bibitem[\protect\citeauthoryear{{Algera} et~al.,}{{Algera}
  et~al.}{2023}]{algera2022}
{Algera} H. S.~B.,  et~al., 2023, \mn@doi [\mnras] {10.1093/mnras/stac3195},
  \href {https://ui.adsabs.harvard.edu/abs/2023MNRAS.518.6142A} {518, 6142}

\bibitem[\protect\citeauthoryear{{Arata}, {Yajima}, {Nagamine}, {Abe}  \&
  {Khochfar}}{{Arata} et~al.}{2020}]{arata2020}
{Arata} S.,  {Yajima} H.,  {Nagamine} K.,  {Abe} M.,   {Khochfar} S.,  2020,
  \mn@doi [\mnras] {10.1093/mnras/staa2809}, \href
  {https://ui.adsabs.harvard.edu/abs/2020MNRAS.498.5541A} {498, 5541}

\bibitem[\protect\citeauthoryear{{Atek} et~al.,}{{Atek}
  et~al.}{2022}]{atek2022}
{Atek} H.,  et~al., 2022, arXiv e-prints, \href
  {https://ui.adsabs.harvard.edu/abs/2022arXiv220712338A} {p. arXiv:2207.12338}

\bibitem[\protect\citeauthoryear{{Baes}, {Verstappen}, {De Looze}, {Fritz},
  {Saftly}, {Vidal P{\'e}rez}, {Stalevski}  \& {Valcke}}{{Baes}
  et~al.}{2011}]{baes2011}
{Baes} M.,  {Verstappen} J.,  {De Looze} I.,  {Fritz} J.,  {Saftly} W.,  {Vidal
  P{\'e}rez} E.,  {Stalevski} M.,   {Valcke} S.,  2011, \mn@doi [\apjs]
  {10.1088/0067-0049/196/2/22}, \href
  {https://ui.adsabs.harvard.edu/abs/2011ApJS..196...22B} {196, 22}

\bibitem[\protect\citeauthoryear{{Bakx} et~al.,}{{Bakx}
  et~al.}{2020}]{bakx2020}
{Bakx} T. J.~L.~C.,  et~al., 2020, \mn@doi [\mnras] {10.1093/mnras/staa509},
  \href {https://ui.adsabs.harvard.edu/abs/2020MNRAS.493.4294B} {493, 4294}

\bibitem[\protect\citeauthoryear{{Bakx} et~al.,}{{Bakx}
  et~al.}{2021}]{bakx2021}
{Bakx} T. J.~L.~C.,  et~al., 2021, \mn@doi [\mnras] {10.1093/mnrasl/slab104},
  \href {https://ui.adsabs.harvard.edu/abs/2021MNRAS.508L..58B} {508, L58}

\bibitem[\protect\citeauthoryear{{Behrens}, {Pallottini}, {Ferrara},
  {Gallerani}  \& {Vallini}}{{Behrens} et~al.}{2018}]{behrens2018}
{Behrens} C.,  {Pallottini} A.,  {Ferrara} A.,  {Gallerani} S.,   {Vallini} L.,
   2018, \mn@doi [\mnras] {10.1093/mnras/sty552}, \href
  {https://ui.adsabs.harvard.edu/abs/2018MNRAS.477..552B} {477, 552}

\bibitem[\protect\citeauthoryear{{B{\'e}thermin} et~al.,}{{B{\'e}thermin}
  et~al.}{2015}]{bethermin2015}
{B{\'e}thermin} M.,  et~al., 2015, \mn@doi [\aap]
  {10.1051/0004-6361/201425031}, \href
  {https://ui.adsabs.harvard.edu/abs/2015A&A...573A.113B} {573, A113}

\bibitem[\protect\citeauthoryear{{Bianchi}}{{Bianchi}}{2013}]{bianchi2013}
{Bianchi} S.,  2013, \mn@doi [\aap] {10.1051/0004-6361/201220866}, \href
  {https://ui.adsabs.harvard.edu/abs/2013A&A...552A..89B} {552, A89}

\bibitem[\protect\citeauthoryear{{Blain}, {Smail}, {Ivison}, {Kneib}  \&
  {Frayer}}{{Blain} et~al.}{2002}]{blain2002}
{Blain} A.~W.,  {Smail} I.,  {Ivison} R.~J.,  {Kneib} J.~P.,   {Frayer} D.~T.,
  2002, \mn@doi [\physrep] {10.1016/S0370-1573(02)00134-5}, \href
  {https://ui.adsabs.harvard.edu/abs/2002PhR...369..111B} {369, 111}

\bibitem[\protect\citeauthoryear{{Blain}, {Barnard}  \& {Chapman}}{{Blain}
  et~al.}{2003}]{blain2003}
{Blain} A.~W.,  {Barnard} V.~E.,   {Chapman} S.~C.,  2003, \mn@doi [\mnras]
  {10.1046/j.1365-8711.2003.06086.x}, \href
  {https://ui.adsabs.harvard.edu/abs/2003MNRAS.338..733B} {338, 733}

\bibitem[\protect\citeauthoryear{{Boogaard}, {Meyer}  \& {Novak}}{{Boogaard}
  et~al.}{2021}]{boogaard2021}
{Boogaard} L.,  {Meyer} R.~A.,   {Novak} M.,  2021, {Interferopy: analysing
  datacubes from radio-to-submm observations}, Zenodo,
  \mn@doi{10.5281/zenodo.5775604}

\bibitem[\protect\citeauthoryear{{Bouwens} et~al.,}{{Bouwens}
  et~al.}{2015}]{bouwens2015}
{Bouwens} R.~J.,  et~al., 2015, \mn@doi [\apj] {10.1088/0004-637X/803/1/34},
  \href {https://ui.adsabs.harvard.edu/abs/2015ApJ...803...34B} {803, 34}

\bibitem[\protect\citeauthoryear{{Bouwens} et~al.,}{{Bouwens}
  et~al.}{2022}]{bouwens2022}
{Bouwens} R.~J.,  et~al., 2022, \mn@doi [\apj] {10.3847/1538-4357/ac5a4a},
  \href {https://ui.adsabs.harvard.edu/abs/2022ApJ...931..160B} {931, 160}

\bibitem[\protect\citeauthoryear{{Bowler}, {Bourne}, {Dunlop}, {McLure}  \&
  {McLeod}}{{Bowler} et~al.}{2018}]{bowler2018}
{Bowler} R.~A.~A.,  {Bourne} N.,  {Dunlop} J.~S.,  {McLure} R.~J.,   {McLeod}
  D.~J.,  2018, \mn@doi [\mnras] {10.1093/mnras/sty2368}, \href
  {https://ui.adsabs.harvard.edu/abs/2018MNRAS.481.1631B} {481, 1631}

\bibitem[\protect\citeauthoryear{{Calzetti}, {Armus}, {Bohlin}, {Kinney},
  {Koornneef}  \& {Storchi-Bergmann}}{{Calzetti} et~al.}{2000}]{calzetti2000}
{Calzetti} D.,  {Armus} L.,  {Bohlin} R.~C.,  {Kinney} A.~L.,  {Koornneef} J.,
   {Storchi-Bergmann} T.,  2000, \mn@doi [\apj] {10.1086/308692}, \href
  {https://ui.adsabs.harvard.edu/abs/2000ApJ...533..682C} {533, 682}

\bibitem[\protect\citeauthoryear{{Carniani} et~al.,}{{Carniani}
  et~al.}{2017}]{carniani2017}
{Carniani} S.,  et~al., 2017, \mn@doi [\aap] {10.1051/0004-6361/201630366},
  \href {https://ui.adsabs.harvard.edu/abs/2017A&A...605A..42C} {605, A42}

\bibitem[\protect\citeauthoryear{{Carniani} et~al.,}{{Carniani}
  et~al.}{2020}]{carniani2020}
{Carniani} S.,  et~al., 2020, \mn@doi [\mnras] {10.1093/mnras/staa3178}, \href
  {https://ui.adsabs.harvard.edu/abs/2020MNRAS.499.5136C} {499, 5136}

\bibitem[\protect\citeauthoryear{{Casey}}{{Casey}}{2012}]{casey2012}
{Casey} C.~M.,  2012, \mn@doi [\mnras] {10.1111/j.1365-2966.2012.21455.x},
  \href {https://ui.adsabs.harvard.edu/abs/2012MNRAS.425.3094C} {425, 3094}

\bibitem[\protect\citeauthoryear{{Casey}, {Narayanan}  \& {Cooray}}{{Casey}
  et~al.}{2014}]{casey2014}
{Casey} C.~M.,  {Narayanan} D.,   {Cooray} A.,  2014, \mn@doi [\physrep]
  {10.1016/j.physrep.2014.02.009}, \href
  {https://ui.adsabs.harvard.edu/abs/2014PhR...541...45C} {541, 45}

\bibitem[\protect\citeauthoryear{{Casey} et~al.,}{{Casey}
  et~al.}{2019}]{casey2019}
{Casey} C.~M.,  et~al., 2019, \mn@doi [\apj] {10.3847/1538-4357/ab52ff}, \href
  {https://ui.adsabs.harvard.edu/abs/2019ApJ...887...55C} {887, 55}

\bibitem[\protect\citeauthoryear{{Castellano} et~al.,}{{Castellano}
  et~al.}{2022}]{castellano2022}
{Castellano} M.,  et~al., 2022, arXiv e-prints, \href
  {https://ui.adsabs.harvard.edu/abs/2022arXiv220709436C} {p. arXiv:2207.09436}

\bibitem[\protect\citeauthoryear{{Chabrier}}{{Chabrier}}{2003}]{chabrier2003}
{Chabrier} G.,  2003, \mn@doi [\pasp] {10.1086/376392}, \href
  {https://ui.adsabs.harvard.edu/abs/2003PASP..115..763C} {115, 763}

\bibitem[\protect\citeauthoryear{{Conley} et~al.,}{{Conley}
  et~al.}{2011}]{conley2011}
{Conley} A.,  et~al., 2011, \mn@doi [\apjl] {10.1088/2041-8205/732/2/L35},
  \href {https://ui.adsabs.harvard.edu/abs/2011ApJ...732L..35C} {732, L35}

\bibitem[\protect\citeauthoryear{{Cormier} et~al.,}{{Cormier}
  et~al.}{2012}]{cormier2012}
{Cormier} D.,  et~al., 2012, \mn@doi [\aap] {10.1051/0004-6361/201219818},
  \href {https://ui.adsabs.harvard.edu/abs/2012A&A...548A..20C} {548, A20}

\bibitem[\protect\citeauthoryear{{Cormier} et~al.,}{{Cormier}
  et~al.}{2019}]{cormier2019}
{Cormier} D.,  et~al., 2019, \mn@doi [\aap] {10.1051/0004-6361/201834457},
  \href {https://ui.adsabs.harvard.edu/abs/2019A&A...626A..23C} {626, A23}

\bibitem[\protect\citeauthoryear{{Cortzen} et~al.,}{{Cortzen}
  et~al.}{2020}]{cortzen2020}
{Cortzen} I.,  et~al., 2020, \mn@doi [\aap] {10.1051/0004-6361/201937217},
  \href {https://ui.adsabs.harvard.edu/abs/2020A&A...634L..14C} {634, L14}

\bibitem[\protect\citeauthoryear{{Da Cunha} et~al.,}{{Da Cunha}
  et~al.}{2013}]{dacunha2013}
{Da Cunha} E.,  et~al., 2013, \mn@doi [\apj] {10.1088/0004-637X/766/1/13},
  \href {https://ui.adsabs.harvard.edu/abs/2013ApJ...766...13D} {766, 13}

\bibitem[\protect\citeauthoryear{{Da Cunha} et~al.,}{{Da Cunha}
  et~al.}{2021}]{dacunha2021}
{Da Cunha} E.,  et~al., 2021, \mn@doi [\apj] {10.3847/1538-4357/ac0ae0}, \href
  {https://ui.adsabs.harvard.edu/abs/2021ApJ...919...30D} {919, 30}

\bibitem[\protect\citeauthoryear{{Dayal} et~al.,}{{Dayal}
  et~al.}{2022}]{dayal2022}
{Dayal} P.,  et~al., 2022, \mn@doi [\mnras] {10.1093/mnras/stac537}, \href
  {https://ui.adsabs.harvard.edu/abs/2022MNRAS.512..989D} {512, 989}

\bibitem[\protect\citeauthoryear{{De Looze} et~al.,}{{De Looze}
  et~al.}{2014}]{delooze2014}
{De Looze} I.,  et~al., 2014, \mn@doi [\aap] {10.1051/0004-6361/201322489},
  \href {https://ui.adsabs.harvard.edu/abs/2014A&A...568A..62D} {568, A62}

\bibitem[\protect\citeauthoryear{{Dessauges-Zavadsky}
  et~al.,}{{Dessauges-Zavadsky} et~al.}{2020}]{dessauges-zavadsky2020}
{Dessauges-Zavadsky} M.,  et~al., 2020, \mn@doi [\aap]
  {10.1051/0004-6361/202038231}, \href
  {https://ui.adsabs.harvard.edu/abs/2020A&A...643A...5D} {643, A5}

\bibitem[\protect\citeauthoryear{{Di Cesare}, {Graziani}, {Schneider},
  {Ginolfi}, {Venditti}, {Santini}  \& {Hunt}}{{Di Cesare}
  et~al.}{2022}]{dicesare2022}
{Di Cesare} C.,  {Graziani} L.,  {Schneider} R.,  {Ginolfi} M.,  {Venditti} A.,
   {Santini} P.,   {Hunt} L.~K.,  2022, arXiv e-prints, \href
  {https://ui.adsabs.harvard.edu/abs/2022arXiv220905496D} {p. arXiv:2209.05496}

\bibitem[\protect\citeauthoryear{{D{\'\i}az-Santos} et~al.,}{{D{\'\i}az-Santos}
  et~al.}{2017}]{diazsantos2017}
{D{\'\i}az-Santos} T.,  et~al., 2017, \mn@doi [\apj]
  {10.3847/1538-4357/aa81d7}, \href
  {https://ui.adsabs.harvard.edu/abs/2017ApJ...846...32D} {846, 32}

\bibitem[\protect\citeauthoryear{{Draine}}{{Draine}}{1989}]{draine1989}
{Draine} B.~T.,  1989, in {B{\"o}hm-Vitense} E.,  ed., Infrared Spectroscopy in
  Astronomy. p.~93

\bibitem[\protect\citeauthoryear{{Draine}}{{Draine}}{2003}]{draine2003}
{Draine} B.~T.,  2003, \mn@doi [\araa]
  {10.1146/annurev.astro.41.011802.094840}, \href
  {https://ui.adsabs.harvard.edu/abs/2003ARA&A..41..241D} {41, 241}

\bibitem[\protect\citeauthoryear{{Drew} \& {Casey}}{{Drew} \&
  {Casey}}{2022}]{drew2022}
{Drew} P.~M.,  {Casey} C.~M.,  2022, \mn@doi [\apj] {10.3847/1538-4357/ac6270},
  \href {https://ui.adsabs.harvard.edu/abs/2022ApJ...930..142D} {930, 142}

\bibitem[\protect\citeauthoryear{{Dudzevi{\v{c}}i{\={u}}t{\.{e}}}
  et~al.,}{{Dudzevi{\v{c}}i{\={u}}t{\.{e}}} et~al.}{2020}]{dudzeviciute2020}
{Dudzevi{\v{c}}i{\={u}}t{\.{e}}} U.,  et~al., 2020, \mn@doi [\mnras]
  {10.1093/mnras/staa769}, \href
  {https://ui.adsabs.harvard.edu/abs/2020MNRAS.494.3828D} {494, 3828}

\bibitem[\protect\citeauthoryear{{Endsley}, {Stark}, {Chevallard}  \&
  {Charlot}}{{Endsley} et~al.}{2021}]{endsley2021}
{Endsley} R.,  {Stark} D.~P.,  {Chevallard} J.,   {Charlot} S.,  2021, \mn@doi
  [\mnras] {10.1093/mnras/staa3370}, \href
  {https://ui.adsabs.harvard.edu/abs/2021MNRAS.500.5229E} {500, 5229}

\bibitem[\protect\citeauthoryear{{Faisst} et~al.,}{{Faisst}
  et~al.}{2020}]{faisst2020}
{Faisst} A.~L.,  et~al., 2020, \mn@doi [\apjs] {10.3847/1538-4365/ab7ccd},
  \href {https://ui.adsabs.harvard.edu/abs/2020ApJS..247...61F} {247, 61}

\bibitem[\protect\citeauthoryear{{Feldmann}, {Hopkins}, {Quataert},
  {Faucher-Gigu{\`e}re}  \& {Kere{\v{s}}}}{{Feldmann}
  et~al.}{2016}]{feldmann2016}
{Feldmann} R.,  {Hopkins} P.~F.,  {Quataert} E.,  {Faucher-Gigu{\`e}re} C.-A.,
   {Kere{\v{s}}} D.,  2016, \mn@doi [\mnras] {10.1093/mnrasl/slw014}, \href
  {https://ui.adsabs.harvard.edu/abs/2016MNRAS.458L..14F} {458, L14}

\bibitem[\protect\citeauthoryear{{Ferland}, {Korista}, {Verner}, {Ferguson},
  {Kingdon}  \& {Verner}}{{Ferland} et~al.}{1998}]{ferland1998}
{Ferland} G.~J.,  {Korista} K.~T.,  {Verner} D.~A.,  {Ferguson} J.~W.,
  {Kingdon} J.~B.,   {Verner} E.~M.,  1998, \mn@doi [\pasp] {10.1086/316190},
  \href {https://ui.adsabs.harvard.edu/abs/1998PASP..110..761F} {110, 761}

\bibitem[\protect\citeauthoryear{{Ferland} et~al.,}{{Ferland}
  et~al.}{2017}]{ferland2017}
{Ferland} G.~J.,  et~al., 2017, \rmxaa, \href
  {https://ui.adsabs.harvard.edu/abs/2017RMxAA..53..385F} {53, 385}

\bibitem[\protect\citeauthoryear{{Ferrara}, {Vallini}, {Pallottini},
  {Gallerani}, {Carniani}, {Kohandel}, {Decataldo}  \& {Behrens}}{{Ferrara}
  et~al.}{2019}]{ferrara2019}
{Ferrara} A.,  {Vallini} L.,  {Pallottini} A.,  {Gallerani} S.,  {Carniani} S.,
   {Kohandel} M.,  {Decataldo} D.,   {Behrens} C.,  2019, \mn@doi [\mnras]
  {10.1093/mnras/stz2031}, \href
  {https://ui.adsabs.harvard.edu/abs/2019MNRAS.489....1F} {489, 1}

\bibitem[\protect\citeauthoryear{{Ferrara} et~al.,}{{Ferrara}
  et~al.}{2022}]{ferrara2022}
{Ferrara} A.,  et~al., 2022, \mn@doi [\mnras] {10.1093/mnras/stac460}, \href
  {https://ui.adsabs.harvard.edu/abs/2022MNRAS.512...58F} {512, 58}

\bibitem[\protect\citeauthoryear{{Finkelstein} et~al.,}{{Finkelstein}
  et~al.}{2015}]{finkelstein2015}
{Finkelstein} S.~L.,  et~al., 2015, \mn@doi [\apj]
  {10.1088/0004-637X/810/1/71}, \href
  {https://ui.adsabs.harvard.edu/abs/2015ApJ...810...71F} {810, 71}

\bibitem[\protect\citeauthoryear{{Foreman-Mackey}, {Hogg}, {Lang}  \&
  {Goodman}}{{Foreman-Mackey} et~al.}{2013}]{foreman-mackey2013}
{Foreman-Mackey} D.,  {Hogg} D.~W.,  {Lang} D.,   {Goodman} J.,  2013, \mn@doi
  [\pasp] {10.1086/670067}, \href
  {https://ui.adsabs.harvard.edu/abs/2013PASP..125..306F} {125, 306}

\bibitem[\protect\citeauthoryear{{Fudamoto} et~al.,}{{Fudamoto}
  et~al.}{2021}]{fudamoto2021}
{Fudamoto} Y.,  et~al., 2021, \mn@doi [\nat] {10.1038/s41586-021-03846-z},
  \href {https://ui.adsabs.harvard.edu/abs/2021Natur.597..489F} {597, 489}

\bibitem[\protect\citeauthoryear{{Fudamoto}, {Inoue}  \& {Sugahara}}{{Fudamoto}
  et~al.}{2022a}]{fudamoto2022b}
{Fudamoto} Y.,  {Inoue} A.~K.,   {Sugahara} Y.,  2022a, arXiv e-prints, \href
  {https://ui.adsabs.harvard.edu/abs/2022arXiv220601879F} {p. arXiv:2206.01879}

\bibitem[\protect\citeauthoryear{{Fudamoto} et~al.,}{{Fudamoto}
  et~al.}{2022b}]{fudamoto2022}
{Fudamoto} Y.,  et~al., 2022b, \mn@doi [\apj] {10.3847/1538-4357/ac7a47}, \href
  {https://ui.adsabs.harvard.edu/abs/2022ApJ...934..144F} {934, 144}

\bibitem[\protect\citeauthoryear{{Fujimoto} et~al.,}{{Fujimoto}
  et~al.}{2020}]{fujimoto2020}
{Fujimoto} S.,  et~al., 2020, \mn@doi [\apj] {10.3847/1538-4357/ab94b3}, \href
  {https://ui.adsabs.harvard.edu/abs/2020ApJ...900....1F} {900, 1}

\bibitem[\protect\citeauthoryear{{Ginsburg} et~al.,}{{Ginsburg}
  et~al.}{2019}]{ginsburg2019}
{Ginsburg} A.,  et~al., 2019, {radio-astro-tools/spectral-cube: Release
  v0.4.5}, Zenodo, \mn@doi{10.5281/zenodo.3558614}

\bibitem[\protect\citeauthoryear{{Glatzle}, {Ciardi}  \& {Graziani}}{{Glatzle}
  et~al.}{2019}]{glatzle2019}
{Glatzle} M.,  {Ciardi} B.,   {Graziani} L.,  2019, \mn@doi [\mnras]
  {10.1093/mnras/sty2514}, \href
  {https://ui.adsabs.harvard.edu/abs/2019MNRAS.482..321G} {482, 321}

\bibitem[\protect\citeauthoryear{{Gould} \& {Salpeter}}{{Gould} \&
  {Salpeter}}{1963}]{gould1963}
{Gould} R.~J.,  {Salpeter} E.~E.,  1963, \mn@doi [\apj] {10.1086/147654}, \href
  {https://ui.adsabs.harvard.edu/abs/1963ApJ...138..393G} {138, 393}

\bibitem[\protect\citeauthoryear{{Graziani}, {Schneider}, {Ginolfi}, {Hunt},
  {Maio}, {Glatzle}  \& {Ciardi}}{{Graziani} et~al.}{2020}]{graziani2020}
{Graziani} L.,  {Schneider} R.,  {Ginolfi} M.,  {Hunt} L.~K.,  {Maio} U.,
  {Glatzle} M.,   {Ciardi} B.,  2020, \mn@doi [\mnras] {10.1093/mnras/staa796},
  \href {https://ui.adsabs.harvard.edu/abs/2020MNRAS.494.1071G} {494, 1071}

\bibitem[\protect\citeauthoryear{{Greve} et~al.,}{{Greve}
  et~al.}{2012}]{greve2012}
{Greve} T.~R.,  et~al., 2012, \mn@doi [\apj] {10.1088/0004-637X/756/1/101},
  \href {https://ui.adsabs.harvard.edu/abs/2012ApJ...756..101G} {756, 101}

\bibitem[\protect\citeauthoryear{{Gullberg} et~al.,}{{Gullberg}
  et~al.}{2015}]{gullberg2015}
{Gullberg} B.,  et~al., 2015, \mn@doi [\mnras] {10.1093/mnras/stv372}, \href
  {https://ui.adsabs.harvard.edu/abs/2015MNRAS.449.2883G} {449, 2883}

\bibitem[\protect\citeauthoryear{{Harikane} et~al.,}{{Harikane}
  et~al.}{2020}]{harikane2020}
{Harikane} Y.,  et~al., 2020, \mn@doi [\apj] {10.3847/1538-4357/ab94bd}, \href
  {https://ui.adsabs.harvard.edu/abs/2020ApJ...896...93H} {896, 93}

\bibitem[\protect\citeauthoryear{{Harikane} et~al.,}{{Harikane}
  et~al.}{2022}]{harikane2022}
{Harikane} Y.,  et~al., 2022, arXiv e-prints, \href
  {https://ui.adsabs.harvard.edu/abs/2022arXiv220801612H} {p. arXiv:2208.01612}

\bibitem[\protect\citeauthoryear{{Hashimoto} et~al.,}{{Hashimoto}
  et~al.}{2018}]{hashimoto2018}
{Hashimoto} T.,  et~al., 2018, \mn@doi [\nat] {10.1038/s41586-018-0117-z},
  \href {https://ui.adsabs.harvard.edu/abs/2018Natur.557..392H} {557, 392}

\bibitem[\protect\citeauthoryear{{Hashimoto} et~al.,}{{Hashimoto}
  et~al.}{2019}]{hashimoto2019}
{Hashimoto} T.,  et~al., 2019, \mn@doi [\pasj] {10.1093/pasj/psz049}, \href
  {https://ui.adsabs.harvard.edu/abs/2019PASJ...71...71H} {71, 71}

\bibitem[\protect\citeauthoryear{{Hayes}, {Schaerer}, {{\"O}stlin},
  {Mas-Hesse}, {Atek}  \& {Kunth}}{{Hayes} et~al.}{2011}]{hayes2011}
{Hayes} M.,  {Schaerer} D.,  {{\"O}stlin} G.,  {Mas-Hesse} J.~M.,  {Atek} H.,
  {Kunth} D.,  2011, \mn@doi [\apj] {10.1088/0004-637X/730/1/8}, \href
  {https://ui.adsabs.harvard.edu/abs/2011ApJ...730....8H} {730, 8}

\bibitem[\protect\citeauthoryear{{Hildebrand}}{{Hildebrand}}{1983}]{hildebrand1983}
{Hildebrand} R.~H.,  1983, \qjras, \href
  {https://ui.adsabs.harvard.edu/abs/1983QJRAS..24..267H} {24, 267}

\bibitem[\protect\citeauthoryear{{Hirashita} \& {Ferrara}}{{Hirashita} \&
  {Ferrara}}{2002}]{hirashita2002}
{Hirashita} H.,  {Ferrara} A.,  2002, \mn@doi [\mnras]
  {10.1046/j.1365-8711.2002.05968.x}, \href
  {https://ui.adsabs.harvard.edu/abs/2002MNRAS.337..921H} {337, 921}

\bibitem[\protect\citeauthoryear{{Hodge} \& {da Cunha}}{{Hodge} \& {da
  Cunha}}{2020}]{hodge2020}
{Hodge} J.~A.,  {da Cunha} E.,  2020, \mn@doi [Royal Society Open Science]
  {10.1098/rsos.200556}, \href
  {https://ui.adsabs.harvard.edu/abs/2020RSOS....700556H} {7, 200556}

\bibitem[\protect\citeauthoryear{{Hogg} \& {Foreman-Mackey}}{{Hogg} \&
  {Foreman-Mackey}}{2018}]{hogg2018}
{Hogg} D.~W.,  {Foreman-Mackey} D.,  2018, \mn@doi [\apjs]
  {10.3847/1538-4365/aab76e}, \href
  {https://ui.adsabs.harvard.edu/abs/2018ApJS..236...11H} {236, 11}

\bibitem[\protect\citeauthoryear{{Hygate et al.}}{{Hygate et
  al.}}{2022}]{hygate2022}
{Hygate et al.} A.,  2022, \mnras, submitted

\bibitem[\protect\citeauthoryear{{Inami} et~al.,}{{Inami}
  et~al.}{2022}]{inami2022}
{Inami} H.,  et~al., 2022, \mn@doi [\mnras] {10.1093/mnras/stac1779}, \href
  {https://ui.adsabs.harvard.edu/abs/2022MNRAS.515.3126I} {515, 3126}

\bibitem[\protect\citeauthoryear{{Inoue} et~al.,}{{Inoue}
  et~al.}{2016}]{inoue2016}
{Inoue} A.~K.,  et~al., 2016, \mn@doi [Science] {10.1126/science.aaf0714},
  \href {https://ui.adsabs.harvard.edu/abs/2016Sci...352.1559I} {352, 1559}

\bibitem[\protect\citeauthoryear{{Inoue}, {Hashimoto}, {Chihara}  \&
  {Koike}}{{Inoue} et~al.}{2020}]{inoue2020}
{Inoue} A.~K.,  {Hashimoto} T.,  {Chihara} H.,   {Koike} C.,  2020, \mn@doi
  [\mnras] {10.1093/mnras/staa1203}, \href
  {https://ui.adsabs.harvard.edu/abs/2020MNRAS.495.1577I} {495, 1577}

\bibitem[\protect\citeauthoryear{{Jarvis} et~al.,}{{Jarvis}
  et~al.}{2013}]{jarvis2013}
{Jarvis} M.~J.,  et~al., 2013, \mn@doi [\mnras] {10.1093/mnras/sts118}, \href
  {https://ui.adsabs.harvard.edu/abs/2013MNRAS.428.1281J} {428, 1281}

\bibitem[\protect\citeauthoryear{{Johnson}, {Leja}, {Conroy}  \&
  {Speagle}}{{Johnson} et~al.}{2021}]{johnson2021}
{Johnson} B.~D.,  {Leja} J.,  {Conroy} C.,   {Speagle} J.~S.,  2021, \mn@doi
  [\apjs] {10.3847/1538-4365/abef67}, \href
  {https://ui.adsabs.harvard.edu/abs/2021ApJS..254...22J} {254, 22}

\bibitem[\protect\citeauthoryear{{Jones}, {Maiolino}, {Caselli}  \&
  {Carniani}}{{Jones} et~al.}{2020}]{jones2020}
{Jones} G.~C.,  {Maiolino} R.,  {Caselli} P.,   {Carniani} S.,  2020, \mn@doi
  [\mnras] {10.1093/mnras/staa2689}, \href
  {https://ui.adsabs.harvard.edu/abs/2020MNRAS.498.4109J} {498, 4109}

\bibitem[\protect\citeauthoryear{{Kashino} et~al.,}{{Kashino}
  et~al.}{2017}]{kashino2017}
{Kashino} D.,  et~al., 2017, \mn@doi [\apj] {10.3847/1538-4357/835/1/88}, \href
  {https://ui.adsabs.harvard.edu/abs/2017ApJ...835...88K} {835, 88}

\bibitem[\protect\citeauthoryear{{Katz}, {Kimm}, {Sijacki}  \&
  {Haehnelt}}{{Katz} et~al.}{2017}]{katz2017}
{Katz} H.,  {Kimm} T.,  {Sijacki} D.,   {Haehnelt} M.~G.,  2017, \mn@doi
  [\mnras] {10.1093/mnras/stx608}, \href
  {https://ui.adsabs.harvard.edu/abs/2017MNRAS.468.4831K} {468, 4831}

\bibitem[\protect\citeauthoryear{{Katz} et~al.,}{{Katz}
  et~al.}{2022}]{katz2022}
{Katz} H.,  et~al., 2022, \mn@doi [\mnras] {10.1093/mnras/stac028}, \href
  {https://ui.adsabs.harvard.edu/abs/2022MNRAS.510.5603K} {510, 5603}

\bibitem[\protect\citeauthoryear{{Kennicutt} \& {Evans}}{{Kennicutt} \&
  {Evans}}{2012}]{kennicutt2012}
{Kennicutt} R.~C.,  {Evans} N.~J.,  2012, \mn@doi [\araa]
  {10.1146/annurev-astro-081811-125610}, \href
  {https://ui.adsabs.harvard.edu/abs/2012ARA&A..50..531K} {50, 531}

\bibitem[\protect\citeauthoryear{{Koch} et~al.,}{{Koch}
  et~al.}{2021}]{koch2021}
{Koch} E.,  et~al., 2021, {radio-astro-tools/radio-beam: v0.3.3}, Zenodo,
  \mn@doi{10.5281/zenodo.4623788}

\bibitem[\protect\citeauthoryear{{Kohandel}, {Pallottini}, {Ferrara},
  {Zanella}, {Behrens}, {Carniani}, {Gallerani}  \& {Vallini}}{{Kohandel}
  et~al.}{2019}]{kohandel2019}
{Kohandel} M.,  {Pallottini} A.,  {Ferrara} A.,  {Zanella} A.,  {Behrens} C.,
  {Carniani} S.,  {Gallerani} S.,   {Vallini} L.,  2019, \mn@doi [\mnras]
  {10.1093/mnras/stz1486}, \href
  {https://ui.adsabs.harvard.edu/abs/2019MNRAS.487.3007K} {487, 3007}

\bibitem[\protect\citeauthoryear{{Lagache}, {Cousin}  \& {Chatzikos}}{{Lagache}
  et~al.}{2018}]{lagache2018}
{Lagache} G.,  {Cousin} M.,   {Chatzikos} M.,  2018, \mn@doi [\aap]
  {10.1051/0004-6361/201732019}, \href
  {https://ui.adsabs.harvard.edu/abs/2018A&A...609A.130L} {609, A130}

\bibitem[\protect\citeauthoryear{{Laporte} et~al.,}{{Laporte}
  et~al.}{2017}]{laporte2017}
{Laporte} N.,  et~al., 2017, \mn@doi [\apjl] {10.3847/2041-8213/aa62aa}, \href
  {https://ui.adsabs.harvard.edu/abs/2017ApJ...837L..21L} {837, L21}

\bibitem[\protect\citeauthoryear{{Laporte} et~al.,}{{Laporte}
  et~al.}{2019}]{laporte2019}
{Laporte} N.,  et~al., 2019, \mn@doi [\mnras] {10.1093/mnrasl/slz094}, \href
  {https://ui.adsabs.harvard.edu/abs/2019MNRAS.487L..81L} {487, L81}

\bibitem[\protect\citeauthoryear{{Leja}, {Carnall}, {Johnson}, {Conroy}  \&
  {Speagle}}{{Leja} et~al.}{2019}]{leja2019}
{Leja} J.,  {Carnall} A.~C.,  {Johnson} B.~D.,  {Conroy} C.,   {Speagle} J.~S.,
   2019, \mn@doi [\apj] {10.3847/1538-4357/ab133c}, \href
  {https://ui.adsabs.harvard.edu/abs/2019ApJ...876....3L} {876, 3}

\bibitem[\protect\citeauthoryear{{Leja}, {Speagle}, {Johnson}, {Conroy}, {van
  Dokkum}  \& {Franx}}{{Leja} et~al.}{2020}]{leja2020}
{Leja} J.,  {Speagle} J.~S.,  {Johnson} B.~D.,  {Conroy} C.,  {van Dokkum} P.,
   {Franx} M.,  2020, \mn@doi [\apj] {10.3847/1538-4357/ab7e27}, \href
  {https://ui.adsabs.harvard.edu/abs/2020ApJ...893..111L} {893, 111}

\bibitem[\protect\citeauthoryear{{Liang} et~al.,}{{Liang}
  et~al.}{2019}]{liang2019}
{Liang} L.,  et~al., 2019, \mn@doi [\mnras] {10.1093/mnras/stz2134}, \href
  {https://ui.adsabs.harvard.edu/abs/2019MNRAS.489.1397L} {489, 1397}

\bibitem[\protect\citeauthoryear{{Lupi} \& {Bovino}}{{Lupi} \&
  {Bovino}}{2020}]{lupi2020}
{Lupi} A.,  {Bovino} S.,  2020, \mn@doi [\mnras] {10.1093/mnras/staa048}, \href
  {https://ui.adsabs.harvard.edu/abs/2020MNRAS.492.2818L} {492, 2818}

\bibitem[\protect\citeauthoryear{{Lutz} et~al.,}{{Lutz}
  et~al.}{2016}]{lutz2016}
{Lutz} D.,  et~al., 2016, \mn@doi [\aap] {10.1051/0004-6361/201527706}, \href
  {https://ui.adsabs.harvard.edu/abs/2016A&A...591A.136L} {591, A136}

\bibitem[\protect\citeauthoryear{{Ma} et~al.,}{{Ma} et~al.}{2018}]{ma2018}
{Ma} X.,  et~al., 2018, \mn@doi [\mnras] {10.1093/mnras/sty1024}, \href
  {https://ui.adsabs.harvard.edu/abs/2018MNRAS.478.1694M} {478, 1694}

\bibitem[\protect\citeauthoryear{{Mancini}, {Schneider}, {Graziani},
  {Valiante}, {Dayal}, {Maio}, {Ciardi}  \& {Hunt}}{{Mancini}
  et~al.}{2015}]{mancini2015}
{Mancini} M.,  {Schneider} R.,  {Graziani} L.,  {Valiante} R.,  {Dayal} P.,
  {Maio} U.,  {Ciardi} B.,   {Hunt} L.~K.,  2015, \mn@doi [\mnras]
  {10.1093/mnrasl/slv070}, \href
  {https://ui.adsabs.harvard.edu/abs/2015MNRAS.451L..70M} {451, L70}

\bibitem[\protect\citeauthoryear{{McCracken} et~al.,}{{McCracken}
  et~al.}{2012}]{mccracken2012}
{McCracken} H.~J.,  et~al., 2012, \mn@doi [\aap] {10.1051/0004-6361/201219507},
  \href {https://ui.adsabs.harvard.edu/abs/2012A&A...544A.156M} {544, A156}

\bibitem[\protect\citeauthoryear{{McLure} et~al.,}{{McLure}
  et~al.}{2013}]{mclure2013}
{McLure} R.~J.,  et~al., 2013, \mn@doi [\mnras] {10.1093/mnras/stt627}, \href
  {https://ui.adsabs.harvard.edu/abs/2013MNRAS.432.2696M} {432, 2696}

\bibitem[\protect\citeauthoryear{{Micha{\l}owski}}{{Micha{\l}owski}}{2015}]{michalowski2015}
{Micha{\l}owski} M.~J.,  2015, \mn@doi [\aap] {10.1051/0004-6361/201525644},
  \href {https://ui.adsabs.harvard.edu/abs/2015A&A...577A..80M} {577, A80}

\bibitem[\protect\citeauthoryear{{Mohan} \& {Rafferty}}{{Mohan} \&
  {Rafferty}}{2015}]{mohanrafferty2015}
{Mohan} N.,  {Rafferty} D.,  2015, {PyBDSF: Python Blob Detection and Source
  Finder}, Astrophysics Source Code Library (\mn@eprint {ascl} {1502.007})

\bibitem[\protect\citeauthoryear{{Morishita} et~al.,}{{Morishita}
  et~al.}{2022}]{morishita2022}
{Morishita} T.,  et~al., 2022, arXiv e-prints, \href
  {https://ui.adsabs.harvard.edu/abs/2022arXiv221109097M} {p. arXiv:2211.09097}

\bibitem[\protect\citeauthoryear{{Naidu} et~al.,}{{Naidu}
  et~al.}{2022}]{naidu2022}
{Naidu} R.~P.,  et~al., 2022, arXiv e-prints, \href
  {https://ui.adsabs.harvard.edu/abs/2022arXiv220709434N} {p. arXiv:2207.09434}

\bibitem[\protect\citeauthoryear{{Oesch}, {Bouwens}, {Illingworth}, {Labb{\'e}}
   \& {Stefanon}}{{Oesch} et~al.}{2018}]{oesch2018}
{Oesch} P.~A.,  {Bouwens} R.~J.,  {Illingworth} G.~D.,  {Labb{\'e}} I.,
  {Stefanon} M.,  2018, \mn@doi [\apj] {10.3847/1538-4357/aab03f}, \href
  {https://ui.adsabs.harvard.edu/abs/2018ApJ...855..105O} {855, 105}

\bibitem[\protect\citeauthoryear{{Pallottini} et~al.,}{{Pallottini}
  et~al.}{2022}]{pallottini2022}
{Pallottini} A.,  et~al., 2022, \mn@doi [\mnras] {10.1093/mnras/stac1281},
  \href {https://ui.adsabs.harvard.edu/abs/2022MNRAS.513.5621P} {513, 5621}

\bibitem[\protect\citeauthoryear{{Popping}, {Somerville}  \&
  {Galametz}}{{Popping} et~al.}{2017}]{popping2017}
{Popping} G.,  {Somerville} R.~S.,   {Galametz} M.,  2017, \mn@doi [\mnras]
  {10.1093/mnras/stx1545}, \href
  {https://ui.adsabs.harvard.edu/abs/2017MNRAS.471.3152P} {471, 3152}

\bibitem[\protect\citeauthoryear{{R{\'e}my-Ruyer} et~al.,}{{R{\'e}my-Ruyer}
  et~al.}{2014}]{remyruyer2014}
{R{\'e}my-Ruyer} A.,  et~al., 2014, \mn@doi [\aap]
  {10.1051/0004-6361/201322803}, \href
  {https://ui.adsabs.harvard.edu/abs/2014A&A...563A..31R} {563, A31}

\bibitem[\protect\citeauthoryear{{Roberts-Borsani} et~al.,}{{Roberts-Borsani}
  et~al.}{2016}]{roberts-borsani2016}
{Roberts-Borsani} G.~W.,  et~al., 2016, \mn@doi [\apj]
  {10.3847/0004-637X/823/2/143}, \href
  {https://ui.adsabs.harvard.edu/abs/2016ApJ...823..143R} {823, 143}

\bibitem[\protect\citeauthoryear{{Sawicki}}{{Sawicki}}{2012}]{sawicki2012}
{Sawicki} M.,  2012, \mn@doi [\pasp] {10.1086/668636}, \href
  {https://ui.adsabs.harvard.edu/abs/2012PASP..124.1208S} {124, 1208}

\bibitem[\protect\citeauthoryear{{Schouws} et~al.,}{{Schouws}
  et~al.}{2022a}]{schouws2022b}
{Schouws} S.,  et~al., 2022a, arXiv e-prints, \href
  {https://ui.adsabs.harvard.edu/abs/2022arXiv220204080S} {p. arXiv:2202.04080}

\bibitem[\protect\citeauthoryear{{Schouws} et~al.,}{{Schouws}
  et~al.}{2022b}]{schouws2022a}
{Schouws} S.,  et~al., 2022b, \mn@doi [\apj] {10.3847/1538-4357/ac4605}, \href
  {https://ui.adsabs.harvard.edu/abs/2022ApJ...928...31S} {928, 31}

\bibitem[\protect\citeauthoryear{{Schreiber}, {Elbaz}, {Pannella}, {Ciesla},
  {Wang}  \& {Franco}}{{Schreiber} et~al.}{2018}]{schreiber2018}
{Schreiber} C.,  {Elbaz} D.,  {Pannella} M.,  {Ciesla} L.,  {Wang} T.,
  {Franco} M.,  2018, \mn@doi [\aap] {10.1051/0004-6361/201731506}, \href
  {https://ui.adsabs.harvard.edu/abs/2018A&A...609A..30S} {609, A30}

\bibitem[\protect\citeauthoryear{{Scoville} et~al.,}{{Scoville}
  et~al.}{2007}]{scoville2007}
{Scoville} N.,  et~al., 2007, \mn@doi [\apjs] {10.1086/516585}, \href
  {https://ui.adsabs.harvard.edu/abs/2007ApJS..172....1S} {172, 1}

\bibitem[\protect\citeauthoryear{{Simpson} et~al.,}{{Simpson}
  et~al.}{2017}]{simpson2017}
{Simpson} J.~M.,  et~al., 2017, \mn@doi [\apj] {10.3847/1538-4357/aa65d0},
  \href {https://ui.adsabs.harvard.edu/abs/2017ApJ...839...58S} {839, 58}

\bibitem[\protect\citeauthoryear{{Smit} et~al.,}{{Smit}
  et~al.}{2014}]{smit2014}
{Smit} R.,  et~al., 2014, \mn@doi [\apj] {10.1088/0004-637X/784/1/58}, \href
  {https://ui.adsabs.harvard.edu/abs/2014ApJ...784...58S} {784, 58}

\bibitem[\protect\citeauthoryear{{Smit} et~al.,}{{Smit}
  et~al.}{2015}]{smit2015}
{Smit} R.,  et~al., 2015, \mn@doi [\apj] {10.1088/0004-637X/801/2/122}, \href
  {https://ui.adsabs.harvard.edu/abs/2015ApJ...801..122S} {801, 122}

\bibitem[\protect\citeauthoryear{{Sommovigo}, {Ferrara}, {Pallottini},
  {Carniani}, {Gallerani}  \& {Decataldo}}{{Sommovigo}
  et~al.}{2020}]{sommovigo2020}
{Sommovigo} L.,  {Ferrara} A.,  {Pallottini} A.,  {Carniani} S.,  {Gallerani}
  S.,   {Decataldo} D.,  2020, \mn@doi [\mnras] {10.1093/mnras/staa1959}, \href
  {https://ui.adsabs.harvard.edu/abs/2020MNRAS.497..956S} {497, 956}

\bibitem[\protect\citeauthoryear{{Sommovigo}, {Ferrara}, {Carniani}, {Zanella},
  {Pallottini}, {Gallerani}  \& {Vallini}}{{Sommovigo}
  et~al.}{2021}]{sommovigo2021}
{Sommovigo} L.,  {Ferrara} A.,  {Carniani} S.,  {Zanella} A.,  {Pallottini} A.,
   {Gallerani} S.,   {Vallini} L.,  2021, \mn@doi [\mnras]
  {10.1093/mnras/stab720}, \href
  {https://ui.adsabs.harvard.edu/abs/2021MNRAS.503.4878S} {503, 4878}

\bibitem[\protect\citeauthoryear{{Sommovigo} et~al.,}{{Sommovigo}
  et~al.}{2022a}]{sommovigo2022}
{Sommovigo} L.,  et~al., 2022a, \mn@doi [\mnras] {10.1093/mnras/stac302}, \href
  {https://ui.adsabs.harvard.edu/abs/2022MNRAS.513.3122S} {513, 3122}

\bibitem[\protect\citeauthoryear{{Sommovigo} et~al.,}{{Sommovigo}
  et~al.}{2022b}]{sommovigo2022b}
{Sommovigo} L.,  et~al., 2022b, \mn@doi [\mnras] {10.1093/mnras/stac2997},
  \href {https://ui.adsabs.harvard.edu/abs/2022MNRAS.517.5930S} {517, 5930}

\bibitem[\protect\citeauthoryear{{Stacey}, {Hailey-Dunsheath}, {Ferkinhoff},
  {Nikola}, {Parshley}, {Benford}, {Staguhn}  \& {Fiolet}}{{Stacey}
  et~al.}{2010}]{stacey2010}
{Stacey} G.~J.,  {Hailey-Dunsheath} S.,  {Ferkinhoff} C.,  {Nikola} T.,
  {Parshley} S.~C.,  {Benford} D.~J.,  {Staguhn} J.~G.,   {Fiolet} N.,  2010,
  \mn@doi [\apj] {10.1088/0004-637X/724/2/957}, \href
  {https://ui.adsabs.harvard.edu/abs/2010ApJ...724..957S} {724, 957}

\bibitem[\protect\citeauthoryear{{Stark}}{{Stark}}{2016}]{stark2016}
{Stark} D.~P.,  2016, \mn@doi [\araa] {10.1146/annurev-astro-081915-023417},
  \href {https://ui.adsabs.harvard.edu/abs/2016ARA&A..54..761S} {54, 761}

\bibitem[\protect\citeauthoryear{{Stefanon} et~al.,}{{Stefanon}
  et~al.}{2019}]{stefanon2019}
{Stefanon} M.,  et~al., 2019, \mn@doi [\apj] {10.3847/1538-4357/ab3792}, \href
  {https://ui.adsabs.harvard.edu/abs/2019ApJ...883...99S} {883, 99}

\bibitem[\protect\citeauthoryear{{Steidel}, {Strom}, {Pettini}, {Rudie},
  {Reddy}  \& {Trainor}}{{Steidel} et~al.}{2016}]{steidel2016}
{Steidel} C.~C.,  {Strom} A.~L.,  {Pettini} M.,  {Rudie} G.~C.,  {Reddy} N.~A.,
    {Trainor} R.~F.,  2016, \mn@doi [\apj] {10.3847/0004-637X/826/2/159}, \href
  {https://ui.adsabs.harvard.edu/abs/2016ApJ...826..159S} {826, 159}

\bibitem[\protect\citeauthoryear{{Sugahara} et~al.,}{{Sugahara}
  et~al.}{2021}]{sugahara2021}
{Sugahara} Y.,  et~al., 2021, \mn@doi [\apj] {10.3847/1538-4357/ac2a36}, \href
  {https://ui.adsabs.harvard.edu/abs/2021ApJ...923....5S} {923, 5}

\bibitem[\protect\citeauthoryear{{Sugahara}, {Inoue}, {Fudamoto}, {Hashimoto},
  {Harikane}  \& {Yamanaka}}{{Sugahara} et~al.}{2022}]{sugahara2022}
{Sugahara} Y.,  {Inoue} A.~K.,  {Fudamoto} Y.,  {Hashimoto} T.,  {Harikane} Y.,
    {Yamanaka} S.,  2022, \mn@doi [\apj] {10.3847/1538-4357/ac7fed}, \href
  {https://ui.adsabs.harvard.edu/abs/2022ApJ...935..119S} {935, 119}

\bibitem[\protect\citeauthoryear{{Tacconi} et~al.,}{{Tacconi}
  et~al.}{2018}]{tacconi2018}
{Tacconi} L.~J.,  et~al., 2018, \mn@doi [\apj] {10.3847/1538-4357/aaa4b4},
  \href {https://ui.adsabs.harvard.edu/abs/2018ApJ...853..179T} {853, 179}

\bibitem[\protect\citeauthoryear{{Tacconi}, {Genzel}  \& {Sternberg}}{{Tacconi}
  et~al.}{2020}]{tacconi2020}
{Tacconi} L.~J.,  {Genzel} R.,   {Sternberg} A.,  2020, \mn@doi [\araa]
  {10.1146/annurev-astro-082812-141034}, \href
  {https://ui.adsabs.harvard.edu/abs/2020ARA&A..58..157T} {58, 157}

\bibitem[\protect\citeauthoryear{{Tamura} et~al.,}{{Tamura}
  et~al.}{2019}]{tamura2019}
{Tamura} Y.,  et~al., 2019, \mn@doi [\apj] {10.3847/1538-4357/ab0374}, \href
  {https://ui.adsabs.harvard.edu/abs/2019ApJ...874...27T} {874, 27}

\bibitem[\protect\citeauthoryear{{Todini} \& {Ferrara}}{{Todini} \&
  {Ferrara}}{2001}]{todini2001}
{Todini} P.,  {Ferrara} A.,  2001, \mn@doi [\mnras]
  {10.1046/j.1365-8711.2001.04486.x}, \href
  {https://ui.adsabs.harvard.edu/abs/2001MNRAS.325..726T} {325, 726}

\bibitem[\protect\citeauthoryear{{Topping} et~al.,}{{Topping}
  et~al.}{2022}]{topping2022}
{Topping} M.~W.,  et~al., 2022, \mn@doi [\mnras] {10.1093/mnras/stac2291},
  \href {https://ui.adsabs.harvard.edu/abs/2022MNRAS.516..975T} {516, 975}

\bibitem[\protect\citeauthoryear{{Vallini}, {Gallerani}, {Ferrara},
  {Pallottini}  \& {Yue}}{{Vallini} et~al.}{2015}]{vallini2015}
{Vallini} L.,  {Gallerani} S.,  {Ferrara} A.,  {Pallottini} A.,   {Yue} B.,
  2015, \mn@doi [\apj] {10.1088/0004-637X/813/1/36}, \href
  {https://ui.adsabs.harvard.edu/abs/2015ApJ...813...36V} {813, 36}

\bibitem[\protect\citeauthoryear{{Vallini}, {Ferrara}, {Pallottini}  \&
  {Gallerani}}{{Vallini} et~al.}{2017}]{vallini2017}
{Vallini} L.,  {Ferrara} A.,  {Pallottini} A.,   {Gallerani} S.,  2017, \mn@doi
  [\mnras] {10.1093/mnras/stx180}, \href
  {https://ui.adsabs.harvard.edu/abs/2017MNRAS.467.1300V} {467, 1300}

\bibitem[\protect\citeauthoryear{{Vallini}, {Ferrara}, {Pallottini}, {Carniani}
   \& {Gallerani}}{{Vallini} et~al.}{2021}]{vallini2021}
{Vallini} L.,  {Ferrara} A.,  {Pallottini} A.,  {Carniani} S.,   {Gallerani}
  S.,  2021, \mn@doi [\mnras] {10.1093/mnras/stab1674}, \href
  {https://ui.adsabs.harvard.edu/abs/2021MNRAS.505.5543V} {505, 5543}

\bibitem[\protect\citeauthoryear{{Van der Wel} et~al.,}{{Van der Wel}
  et~al.}{2014}]{vanderwel2014}
{Van der Wel} A.,  et~al., 2014, \mn@doi [\apj] {10.1088/0004-637X/788/1/28},
  \href {https://ui.adsabs.harvard.edu/abs/2014ApJ...788...28V} {788, 28}

\bibitem[\protect\citeauthoryear{{Viero}, {Sun}, {Chung}, {Moncelsi}  \&
  {Condon}}{{Viero} et~al.}{2022}]{viero2022}
{Viero} M.~P.,  {Sun} G.,  {Chung} D.~T.,  {Moncelsi} L.,   {Condon} S.~S.,
  2022, \mn@doi [\mnras] {10.1093/mnrasl/slac075}, \href
  {https://ui.adsabs.harvard.edu/abs/2022MNRAS.516L..30V} {516, L30}

\bibitem[\protect\citeauthoryear{{Vijayan}, {Clay}, {Thomas}, {Yates},
  {Wilkins}  \& {Henriques}}{{Vijayan} et~al.}{2019}]{vijayan2019}
{Vijayan} A.~P.,  {Clay} S.~J.,  {Thomas} P.~A.,  {Yates} R.~M.,  {Wilkins}
  S.~M.,   {Henriques} B.~M.,  2019, \mn@doi [\mnras] {10.1093/mnras/stz1948},
  \href {https://ui.adsabs.harvard.edu/abs/2019MNRAS.489.4072V} {489, 4072}

\bibitem[\protect\citeauthoryear{{Watson}, {Christensen}, {Knudsen}, {Richard},
  {Gallazzi}  \& {Micha{\l}owski}}{{Watson} et~al.}{2015}]{watson2015}
{Watson} D.,  {Christensen} L.,  {Knudsen} K.~K.,  {Richard} J.,  {Gallazzi}
  A.,   {Micha{\l}owski} M.~J.,  2015, \mn@doi [\nat] {10.1038/nature14164},
  \href {https://ui.adsabs.harvard.edu/abs/2015Natur.519..327W} {519, 327}

\bibitem[\protect\citeauthoryear{{Weingartner} \& {Draine}}{{Weingartner} \&
  {Draine}}{2001}]{weingartner2001}
{Weingartner} J.~C.,  {Draine} B.~T.,  2001, \mn@doi [\apj] {10.1086/318651},
  \href {https://ui.adsabs.harvard.edu/abs/2001ApJ...548..296W} {548, 296}

\bibitem[\protect\citeauthoryear{{Whitler}, {Stark}, {Endsley}, {Leja},
  {Charlot}  \& {Chevallard}}{{Whitler} et~al.}{2022}]{whitler2022}
{Whitler} L.,  {Stark} D.~P.,  {Endsley} R.,  {Leja} J.,  {Charlot} S.,
  {Chevallard} J.,  2022, arXiv e-prints, \href
  {https://ui.adsabs.harvard.edu/abs/2022arXiv220605315W} {p. arXiv:2206.05315}

\bibitem[\protect\citeauthoryear{{Witstok} et~al.,}{{Witstok}
  et~al.}{2022}]{witstok2022}
{Witstok} J.,  et~al., 2022, \mn@doi [\mnras] {10.1093/mnras/stac1905}, \href
  {https://ui.adsabs.harvard.edu/abs/2022MNRAS.515.1751W} {515, 1751}

\bibitem[\protect\citeauthoryear{{Yan}, {Ma}, {Ling}, {Cheng}, {Huang}  \&
  {Zitrin}}{{Yan} et~al.}{2022}]{yan2022}
{Yan} H.,  {Ma} Z.,  {Ling} C.,  {Cheng} C.,  {Huang} J.-s.,   {Zitrin} A.,
  2022, arXiv e-prints, \href
  {https://ui.adsabs.harvard.edu/abs/2022arXiv220711558Y} {p. arXiv:2207.11558}

\bibitem[\protect\citeauthoryear{{Zanella} et~al.,}{{Zanella}
  et~al.}{2018}]{zanella2018}
{Zanella} A.,  et~al., 2018, \mn@doi [\mnras] {10.1093/mnras/sty2394}, \href
  {https://ui.adsabs.harvard.edu/abs/2018MNRAS.481.1976Z} {481, 1976}

\makeatother
\end{thebibliography}

\appendix

\section{The Dust Morphology of REBELS-38}
\label{app:rebels_38morphology}

In the moderate resolution Band 8 imaging of $0\farcs61\times0\farcs48$, REBELS-38 is resolved into two components detected at a peak significance of $4.3$ and $4.1\sigma$. While unresolved in the original REBELS Band 6 observations, \citet{inami2022} noted an offset between the peaks of its rest-frame UV and dust continuum emission of $0\farcs57\pm0\farcs25$, corresponding to a physical offset of $3.1\pm 1.4\,$kpc. In the higher resolution Band 8 data, we observe that the fainter clump (B) is nearly co-spatial with the UV emission (offset of $0.9\pm 0.4\,$kpc), while clump A is offset to the east by $4.0\pm 0.4\,$kpc. The separation between the two clumps themselves equals $0\farcs73\pm 0\farcs11$ ($4.0 \pm 0.6\,$kpc), and both are spatially unresolved individually ($r_{1/2} < 1.5\,$kpc). Given this small separation, it is likely that the two dusty clumps are individual components of a singular galaxy, although we note the possibility that REBELS-38 could be a late-stage merger.

\section{Dust Temperature Posteriors}
\label{app:posteriors}

\begin{figure*}
    \centering
    \includegraphics[width=0.8\textwidth]{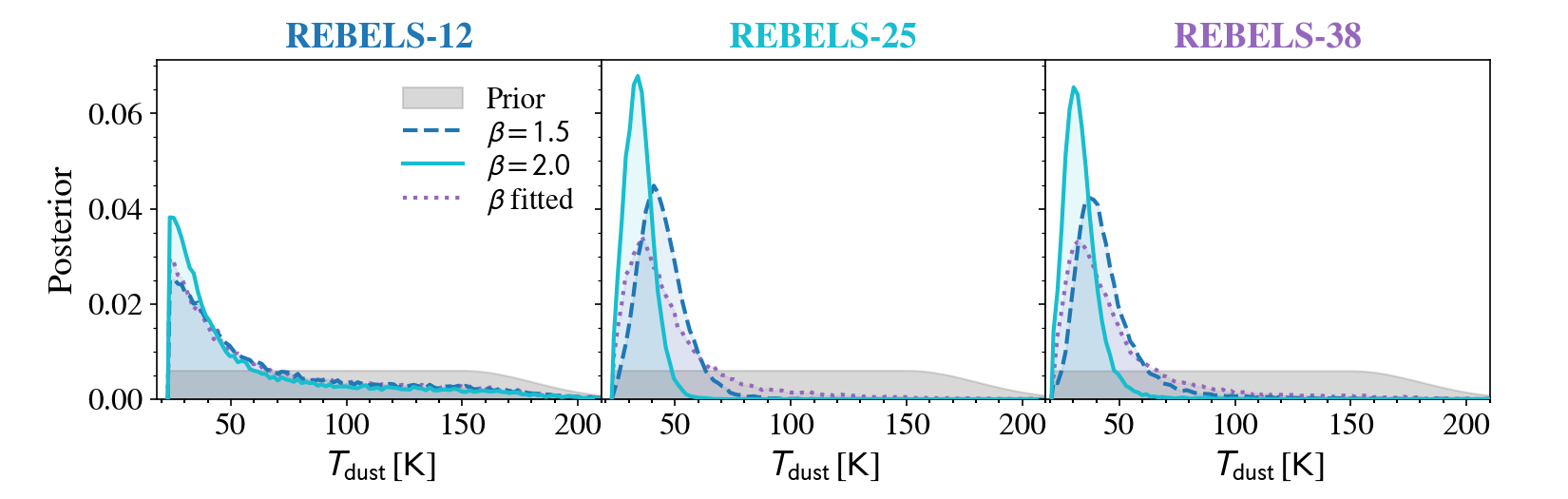}
    \includegraphics[width=0.8\textwidth]{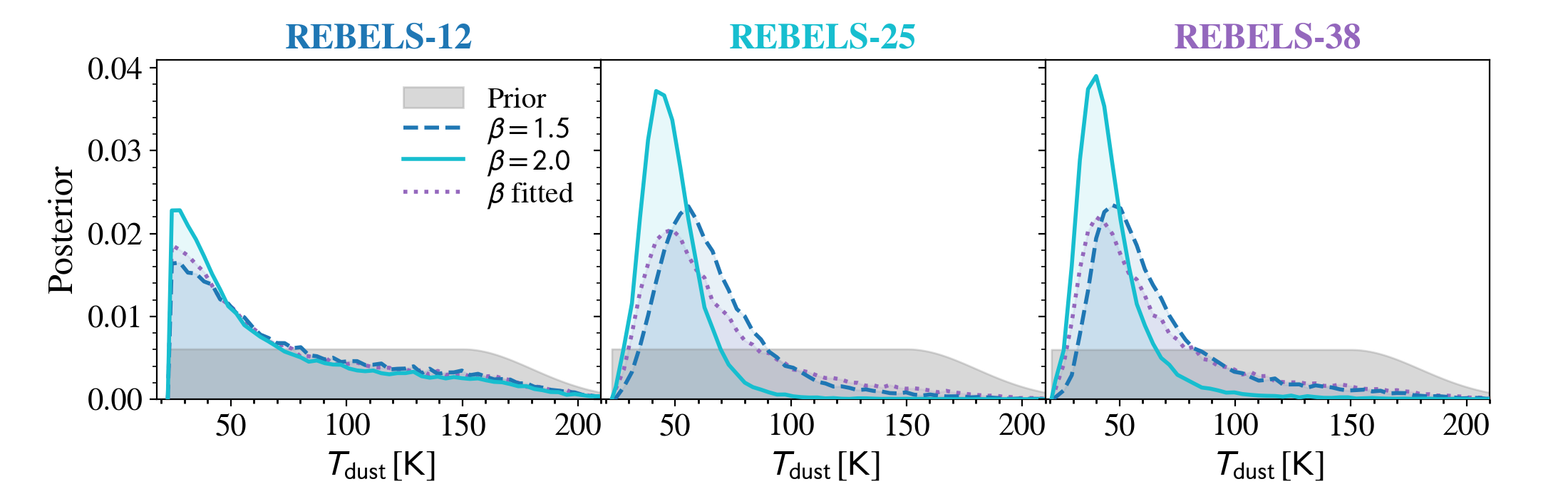}
    \caption{Posterior dust temperature distributions for REBELS-12, REBELS-25 and REBELS-38, assuming optically thin (\textbf{top row}) and thick dust ($\lambda_\mathrm{thick}=100\,\mu$m; \textbf{bottom row}). The distributions are shown both for a fixed dust emissivity ($\beta=1.5, 2.0$; dashed and solid lines respectively), as well as for $\beta$ included in the fit (dotted). The prior on the dust temperature is shown in grey, and is flat for $T_{\mathrm{CMB},z} < T_{\mathrm{dust},z} < 150\,$K, after which it smoothly decreases by a Gaussian with $\sigma=30\,$K. For REBELS-25 and REBELS-38, which have robust detections in both ALMA Bands 6 and 8, the dust temperatures are well constrained, and are independent of the precise shape of the assumed prior.}
    \label{fig:Tdust_posterior}
\end{figure*}

We show the posterior distributions of the inferred dust temperatures for REBELS-12, REBELS-25 and REBELS-38 in Figure \ref{fig:Tdust_posterior}, assuming optically thin (top row) and thick dust (bottom row; $\lambda_\mathrm{thick}=100\,\mu$m). For REBELS-12, a long tail towards higher temperatures remains, owing to its non-detection at rest-frame $\sim90\,\mu$m. The posteriors for REBELS-25 and REBELS-38, on the other hand, are narrow and single-peaked.

\section{The [CII], [OIII] and Dust Morphology of REBELS-12}
\label{app:rebels12_cii}

\begin{figure}
    \centering
    \includegraphics[width=0.49\textwidth]{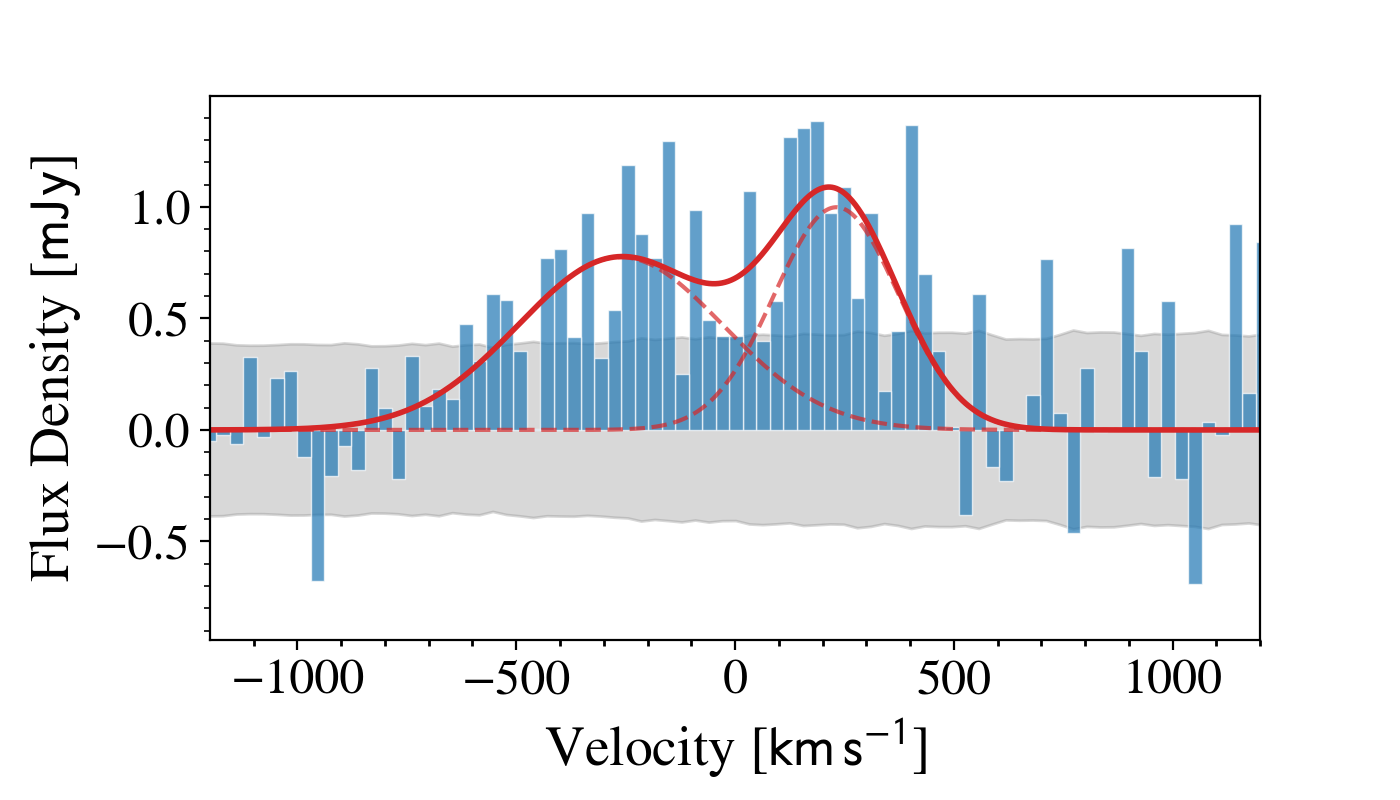}
    \caption{Double Gaussian fit to the [CII] emission in REBELS-12. The redder Gaussian covers a similar velocity range as the [OIII] emission in this source (c.f., Figure \ref{fig:OIII}), and contains approximately half of the total [CII] flux. On the other hand, no counterpart in [OIII] emission is seen for the bluer Gaussian component.}
    \label{fig:R12_CII_spec}
\end{figure}

\begin{figure}
    \centering
    \includegraphics[width=0.35\textwidth]{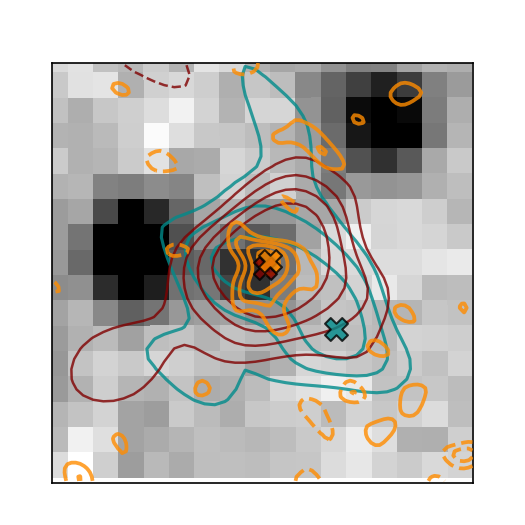}
    \caption{Moment-0 maps of the blue and red Gaussians identified in the 1D [CII] spectrum of REBELS-12 (Figure \ref{fig:R12_CII_spec}), on top of stacked, astrometry-corrected JHK-imaging. The orange contours show the [OIII] emission. Contours of both emission lines start at $2\sigma$ and increase in steps of $1\sigma$. The peak of the moment-0 maps are indicated by crosses of the same color. The red Gaussian is co-spatial with the rest-frame UV and [OIII] emission, while the blue Gaussian appears extended, being both spectrally and spatially offset from the red component.}
    \label{fig:R12_CII_JHK}
\end{figure}

As outlined in Section \ref{sec:results_OIII}, REBELS-12 shows a broad [CII] line that appears to consist of two Gaussian components. Its [OIII] emission, on the other hand, is significantly narrower and coincides only with the redder [CII] component. In Figure \ref{fig:R12_CII_spec}, we fit the [CII] emission with two Gaussians, which contain a flux of $S_\mathrm{[CII]}^\mathrm{blue} = 443_{-130}^{+237}\,\mathrm{mJy}\,\mathrm{km}\,\mathrm{s}^{-1}$ and $S_\mathrm{[CII]}^\mathrm{red} = 358_{-186}^{+141}\,\mathrm{mJy}\,\mathrm{km}\,\mathrm{s}^{-1}$. The total flux is consistent with that obtained from a single-Gaussian fit (Table \ref{tab:emission_line_parameters}).

We image the channels corresponding to the FWHMs of the blue and red Gaussians separately, and plot the contours on top of stacked JHK-imaging and [OIII] emission in Figure \ref{fig:R12_CII_JHK}. The redder Gaussian component is co-spatial with both the UV and rest-frame UV emission, while the peak of the blue component is offset to the southwest by $\sim1''$. This may indicate that REBELS-12 is a merger, consisting of a bright, star-forming component emitting in [OIII] and a low-SFR galaxy with no discernible UV or dust continuum emission. This scenario is particularly appealing given that a serendipitous, UV-dark galaxy was previously found $\sim11\farcs5$ away from REBELS-12 by \citet{fudamoto2021}, implying that REBELS-12 may be part of a larger cosmic structure at $z\approx7.3$.

Assuming REBELS-12 is indeed a merging galaxy, we compute an upper limit on the [OIII]/[CII] ratio of the blue component. We assume the [OIII] FWHM is equal to that of the observed red component and adopt peak line flux of $3\times$ the RMS across the FWHM (corresponding to the RMS in a $130\,\mathrm{km}\,\mathrm{s}^{-1}$ channel). We then infer an upper limit of $S_\mathrm{[OIII]}^\mathrm{blue} \lesssim 280\,\mathrm{mJy}\,\mathrm{km}\,\mathrm{s}^{-1}$, and a line ratio of $\mathrm{[OIII]}/\mathrm{[CII]} < 1.1$ for the blue component. For the red component, on the other hand, we infer $\mathrm{[OIII]}/\mathrm{[CII]} = 3.4_{-1.0}^{+3.6}$ (c.f., $\mathrm{[OIII]}/\mathrm{[CII]} = 1.5_{-0.5}^{+1.0}$ when treating the system as a single component; Table \ref{tab:emission_line_parameters}).

Finally, we note that REBELS-12 consists of two dust components, only one of which is co-spatial with the [OIII], UV and (red) [CII] emission (c.f., Figure \ref{fig:moment0}). If we include only the rest-frame $160\,\mu$m flux density of the co-spatial dust emission in the MBB fitting ($S_\nu = 39 \pm 12\,\mu$Jy), we obtain a dust temperature of $T_{\mathrm{dust},z} = 78_{-41}^{+68}\,$K for our fiducial model of $\beta=2.0$. As expected, a lower Band 6 flux density allows for a wider range of solutions with hot dust. However, irrespective of whether the full Band 6 flux density, or the flux in the single, co-spatial component is adopted, the dust temperature of REBELS-12 cannot accurately be constrained.

\section{Comparison to Single-band Dust Temperature Modelling}
\label{app:singleband_Tdust}

\begin{figure}
    \centering
    \includegraphics[width=0.5\textwidth]{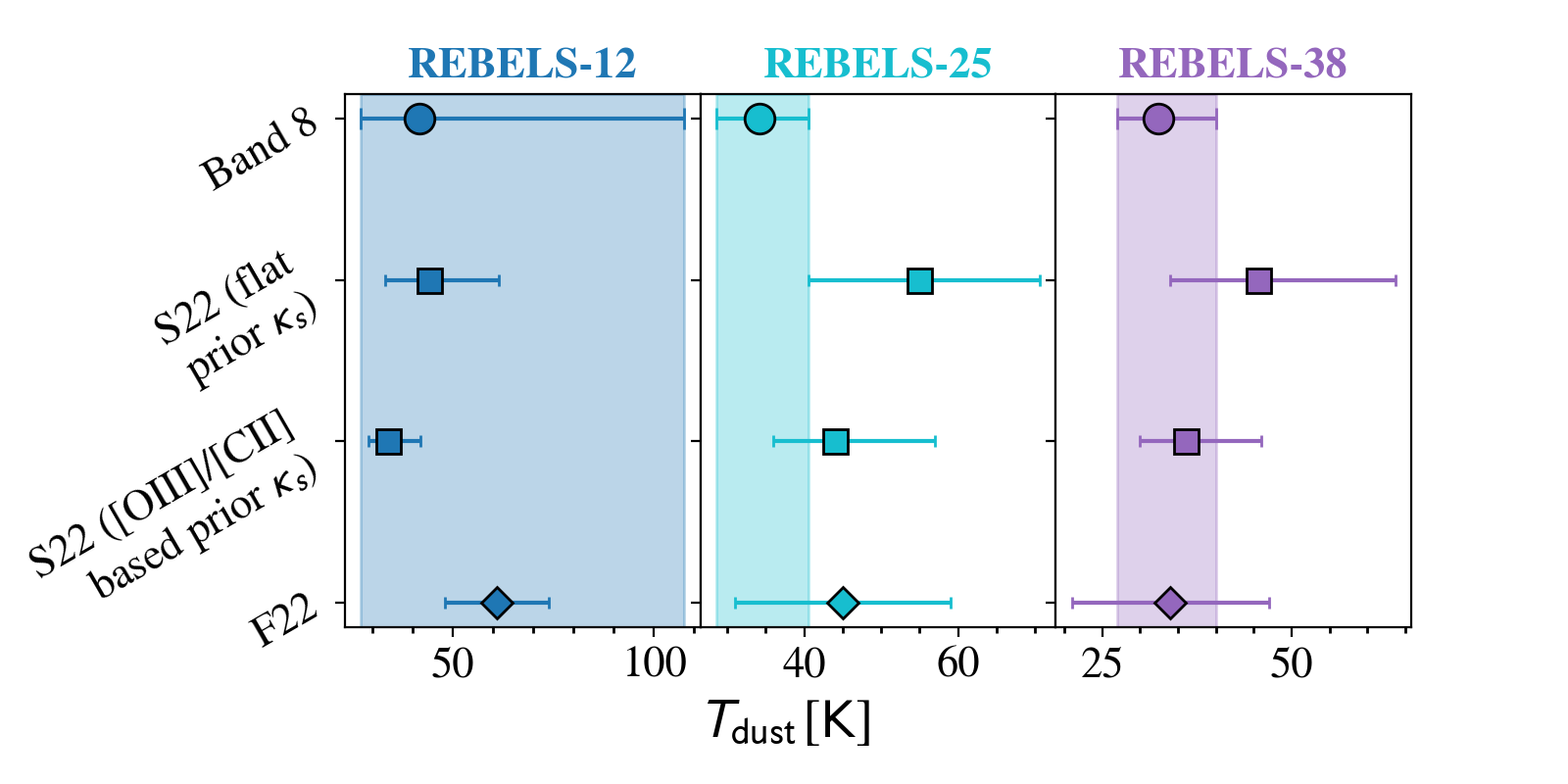}
    \caption{A comparison of different dust temperature measurements for the three REBELS targets discussed in this work. The top row shows $T_\mathrm{dust}$ obtained from MBB fitting (assuming $\beta=2.0$), while the middle two rows show the single-band dust temperature predictions from \citet[S22]{sommovigo2022}. We show the inferred dust temperature for two assumed values of the burstiness $\kappa_{s}$ in their models; the original value assumed by \citet{sommovigo2022} and the updated values based on the measured [OIII]/[CII] ratios as discussed in this work ($\kappa_{s} \sim 10 - 15$). In addition, we show the dust temperature predictions from \citet[F22]{fudamoto2022b} in the bottom row. While a larger sample size is required for a more detailed comparison, the single-band dust temperature models successfully predict the dust temperatures measured from the two-band fit for REBELS-25 and REBELS-38 under careful assumptions on the burstiness parameter and/or infrared sizes.}
    \label{fig:sommovigo_comparison}
\end{figure}

In recent years, there has been significant interest in estimating dust temperatures of high-redshift galaxies without having to resort to expensive multi-band ALMA photometry. We here compare our Band 8 dust temperature determinations to two models providing such estimates, described in detail in \citet{sommovigo2021} and \citet{fudamoto2022b}.

\subsubsection*{Comparison to \citet{sommovigo2022}}

\citet{sommovigo2021} introduce a model to infer dust temperatures and masses through the combined detection of the [CII] emission line and underlying $158\,\mu$m continuum. In short, their models utilize the fact that both the [CII] and far-infrared luminosities of a galaxy trace its star formation rate, while the [CII] emission is also a proxy for gas mass and therefore by extension dust mass. As a result, a single dust continuum detection combined with an indirect measurement of the infrared luminosity through the [CII]-based SFR provides insight into the dust SED, and hence the dust temperature. A recent application of their models to the $z=7.13$ lensed galaxy A1689-zD1 by \citet{bakx2021} demonstrates good agreement between the dust temperature obtained from four-band MBB fitting and the single-band temperature prediction. 

In a recent work, \citet{sommovigo2022} use their model to estimate the dust temperatures of the thirteen REBELS targets for which both a dust continuum and [CII] detection are available, which includes the three targets analyzed in this work. To be consistent with the \citet{sommovigo2022} models, who assume Milky Way-like dust, we compare the dust temperatures obtained when $\beta=2.0$ is adopted. The results of this comparison are shown in Figure \ref{fig:sommovigo_comparison} (first two rows). We find that the model predictions overestimate the dust temperatures of REBELS-25 and REBELS-38, for which Band 8 detections are available, by $\Delta T_\mathrm{dust}\approx15\,$K. For REBELS-12, on the other hand, a comparison of the fitted and predicted dust temperatures provides only limited insight, given the large uncertainty on its MBB-fitted dust temperature.

We can explain the observed difference for REBELS-25 and REBELS-38 by returning to the burstiness parameter $\kappa_s$. In their models, \citet{sommovigo2022} assumed a uniform distribution for the burstiness for the REBELS sources between $\kappa_s = 1 - 50$, motivated by the literature sample analyzed in \citet{vallini2021}. Given that the burstiness cannot be constrained with just [CII] information, their inferred posterior on $\kappa_\mathrm{s}$ resembles the prior, which hence corresponds to a typical assumed burstiness of $\kappa_\mathrm{s} \sim 25 \pm 15$.

However, using the new [OIII] observations for REBELS-12 and REBELS-25, we infer a moderate $\kappa_\mathrm{s} \approx 10 - 15$ for our sample (Section \ref{sec:discussion_OIII}). We therefore re-run the \citet{sommovigo2022} models with these lower burstiness parameters, and include the updated dust temperature predictions in Figure \ref{fig:sommovigo_comparison} (third row). We also re-run their models for REBELS-38, for which no [OIII] data is available, under the assumption that it has the same burstiness as REBELS-12 and REBELS-25. Given that REBELS-38, like REBELS-25, is a massive ($M_\star \approx 10^{10.4}\,M_\odot$; Table \ref{tab:source_parameters}) galaxy with a low dust temperature, we expect it to similarly have a low [OIII]/[CII] ratio.

By assuming these physically motivated lower values for $\kappa_\mathrm{s}$, the predicted $T_\mathrm{dust}$ decreases by an average $\sim 15\,$K, thereby improving the agreement with the temperatures inferred through modified blackbody fitting. As such, care must be taken to adopt suitable priors when adopting single-band dust temperatures.

\subsubsection*{Comparison to \citet{fudamoto2022b}}
We next compare to the single-band dust temperature model from \citet{fudamoto2022b}, which builds upon the framework by \citet{inoue2020}. In short, their model approximates a galaxy as spherical object of size $r_\mathrm{IR}$ consisting of uniformly distributed dust clumps within a homogeneous interstellar medium. The galaxy as a whole is assumed to be in radiative equilibrium, which implies that a fraction of the UV emission associated with massive star formation is absorbed by dust and subsequently re-emitted in the infrared regime. Given, then, a measured UV luminosity, infrared size and dust continuum flux density at any single wavelength, a galaxy's dust temperature (and mass) can be inferred under the assumption of a clumpiness parameter (see \citealt{inoue2020} for details). We here assume the average clumpiness parameter of $\log_{10}(\xi) = -1.02 \pm 0.41$ measured across six $z\gtrsim5$ galaxies with 3 or more ALMA continuum detections by \citet{fudamoto2022b}.

We note that some of the REBELS targets -- including REBELS-12 and REBELS-38 analyzed in this work -- show evidence for spatial offsets between their dust and UV emission \citep{inami2022}, which introduces additional uncertainty in the applicability of dust temperature models based upon radiative equilibrium (see also \citealt{ferrara2022}). However, in the absence of available high-resolution observations for the bulk of the high-redshift dust-detected population, such models may still provide a useful first-look into their dust properties. As such, we apply the \citet{fudamoto2022b} models here while keeping this caveat in mind. \\

We adopt the UV luminosities of our three REBELS targets from Stefanon et al. (in prep), and utilize the IR sizes determined in Section \ref{sec:discussion_OIII}. For REBELS-12, we investigate two scenarios, whereby we adopt either the total Band 6 flux density across its two dust components, or the flux density in the co-spatial dust component only (Section \ref{sec:discussion_OIII}). However, we find that the resulting predicted temperatures agree to within $<2\,$K. For REBELS-38, we model the two dust components seen in the Band 8 data separately (assuming $r_\mathrm{IR} = 1.0 \pm 0.3\,$kpc for both clumps A\&B), but also find that they are characterized by a near-identical dust temperature. 

We show the predicted dust temperatures using the \citet{fudamoto2022b} models in Figure \ref{fig:sommovigo_comparison} (bottom row). We assume a conservative systematic $11\,$K uncertainty -- resulting from the assumption of a fixed clumpiness -- and add it in quadrature to the fitting uncertainties, following the discussion in \citet{fudamoto2022b}. We find that, despite the aforementioned uncertainties and caveats, the predicted single-band dust temperatures are in good agreement with the values derived from MBB fitting. Interestingly, a relatively large temperature of $T_\mathrm{dust} = 61 \pm 13\,$K is predicted for REBELS-12, driven predominantly by its large UV luminosity ($L_\mathrm{UV} \approx 2\times10^{11} L_\odot$). Deeper ALMA observations of REBELS-12 at either Band 6 or 8 are required to verify if it indeed hosts warmer dust than REBELS-25 and REBELS-38.

\section{MBB Fitting of Literature Sample}
\label{app:literature}

In Section \ref{sec:discussion_Tdust}, we consistently analyze the dust SEDs of all nine $z > 6$ literature sources targeted at two or more ALMA bands, at least one of which provides a continuum detection. We list the parameters obtained from modified blackbody fitting in Table \ref{tab:literature}. We note that for sources with poor photometry (e.g., a single, modest-S/N dust detection and one or multiple upper limits), the inferred dust temperatures -- and therefore the inferred dust masses and IR luminosities -- depend strongly on the adopted prior on $T_\mathrm{dust}$.

\begin{landscape}
\begin{table}
    
    \def\arraystretch{1.2}
    
    \centering
    
	\caption{Dust parameters derived from optically thin modified blackbody fitting to nine literature sources, organized in the same manner as Table \ref{tab:MBB_parameters}. \\
	\textbf{References:} B20: \citet{bakx2020}, B21: \citet{bakx2021}, H20: \citet{harikane2020}, L17: \citet{laporte2017}, L19: \citet{laporte2019}, M22: \citet{morishita2022}, S21: \citet{sugahara2021}, W22: \citet{witstok2022}.}
	
	\begin{tabular}{llllllllll}

    \hline 
    
	& \textbf{A1689-zD1} & \textbf{A2744-YD4} & \textbf{Big Three Dragons} & \textbf{COS-3018555981} & \textbf{J0217-0208} & \textbf{J1211-0118} & \textbf{MACS0416-Y1} & \textbf{UVISTA-Z-001} & \textbf{UVISTA-Z-019} \\
	
	\hline
	\textbf{Reference} & B21 & L17, L19, M22 & H19, S21 & W22 & H20 & H20 & B20 & W22 & W22 \\ 
    
    \hline 
    $\beta$ & $1.5$ & $1.5$ & $1.5$ & $1.5$ & $1.5$ & $1.5$ & $1.5$ & $1.5$ & $1.5$ \\
$\log_{10}(\mathrm{M}_\mathrm{dust}/M_\odot)$ &$7.13_{-0.13}^{+0.14}$ & $5.51_{-0.20}^{+0.25}$ & $6.74_{-0.38}^{+0.53}$ & $7.40_{-0.57}^{+0.68}$ & $7.32_{-0.77}^{+0.95}$ & $7.12_{-0.56}^{+0.70}$ & $5.82_{-0.17}^{+0.26}$ & $6.73_{-0.68}^{+0.88}$ & $6.50_{-0.34}^{+0.71}$ \\
$T_\mathrm{dust}\,[\mathrm{K}]$ &$45_{-4}^{+4}$ & $143_{-44}^{+36}$ & $87_{-37}^{+60}$ & $37_{-11}^{+22}$ & $50_{-23}^{+67}$ & $65_{-29}^{+65}$ & $143_{-43}^{+35}$ & $65_{-30}^{+71}$ & $100_{-55}^{+59}$ \\
$\log_{10}(\mathrm{L}_\mathrm{IR}/L_\odot)$ &$11.45_{-0.08}^{+0.08}$ & $12.56_{-0.64}^{+0.41}$ & $12.64_{-0.77}^{+0.87}$ & $11.32_{-0.21}^{+0.54}$ & $11.92_{-0.45}^{+1.22}$ & $12.31_{-0.71}^{+1.10}$ & $12.88_{-0.59}^{+0.39}$ & $11.91_{-0.59}^{+1.07}$ & $12.70_{-1.15}^{+0.82}$ \\
\hline
$\beta$ & $2.0$ & $2.0$ & $2.0$ & $2.0$ & $2.0$ & $2.0$ & $2.0$ & $2.0$ & $2.0$ \\
$\log_{10}(\mathrm{M}_\mathrm{dust}/M_\odot)$ &$7.28_{-0.13}^{+0.15}$ & $5.48_{-0.20}^{+0.29}$ & $7.01_{-0.55}^{+0.62}$ & $7.71_{-0.51}^{+0.64}$ & $7.58_{-0.91}^{+0.92}$ & $7.47_{-0.73}^{+0.72}$ & $5.78_{-0.20}^{+0.31}$ & $6.90_{-0.94}^{+0.95}$ & $6.49_{-0.36}^{+0.73}$ \\
$T_\mathrm{dust}\,[\mathrm{K}]$ &$39_{-3}^{+3}$ & $138_{-45}^{+37}$ & $61_{-23}^{+54}$ & $31_{-7}^{+12}$ & $39_{-14}^{+51}$ & $45_{-16}^{+48}$ & $135_{-45}^{+39}$ & $52_{-22}^{+74}$ & $96_{-52}^{+61}$ \\
$\log_{10}(\mathrm{L}_\mathrm{IR}/L_\odot)$ &$11.33_{-0.07}^{+0.07}$ & $12.80_{-0.74}^{+0.47}$ & $12.24_{-0.64}^{+1.09}$ & $11.23_{-0.14}^{+0.33}$ & $11.70_{-0.31}^{+1.20}$ & $11.89_{-0.44}^{+1.16}$ & $13.06_{-0.73}^{+0.47}$ & $11.71_{-0.46}^{+1.31}$ & $12.85_{-1.28}^{+0.97}$ \\
\hline
$\beta$ & $1.7_{-0.4}^{+0.5}$ & $2.1_{-0.4}^{+0.4}$ & $1.5_{-0.4}^{+0.5}$ & $1.5_{-0.5}^{+0.5}$ & $1.6_{-0.5}^{+0.5}$ & $1.6_{-0.5}^{+0.5}$ & $2.1_{-0.4}^{+0.4}$ & $1.6_{-0.5}^{+0.5}$ & $1.8_{-0.5}^{+0.5}$ \\
$\log_{10}(\mathrm{M}_\mathrm{dust}/M_\odot)$ &$7.19_{-0.18}^{+0.20}$ & $5.47_{-0.23}^{+0.31}$ & $6.76_{-0.38}^{+0.58}$ & $7.43_{-0.69}^{+0.72}$ & $7.34_{-0.78}^{+0.95}$ & $7.17_{-0.60}^{+0.76}$ & $5.77_{-0.22}^{+0.32}$ & $6.74_{-0.71}^{+0.93}$ & $6.50_{-0.36}^{+0.73}$ \\
$T_\mathrm{dust}\,[\mathrm{K}]$ &$42_{-6}^{+7}$ & $135_{-45}^{+39}$ & $85_{-39}^{+62}$ & $36_{-11}^{+30}$ & $48_{-21}^{+67}$ & $60_{-28}^{+70}$ & $134_{-46}^{+40}$ & $63_{-30}^{+72}$ & $97_{-53}^{+60}$ \\
$\log_{10}(\mathrm{L}_\mathrm{IR}/L_\odot)$ &$11.40_{-0.13}^{+0.14}$ & $12.79_{-0.77}^{+0.58}$ & $12.59_{-0.79}^{+0.83}$ & $11.33_{-0.21}^{+0.63}$ & $11.89_{-0.44}^{+1.18}$ & $12.22_{-0.69}^{+1.11}$ & $13.05_{-0.75}^{+0.56}$ & $11.88_{-0.59}^{+1.09}$ & $12.74_{-1.19}^{+0.92}$ \\
\hline

    \def\arraystretch{1.0}
    
    \label{tab:literature}
    \end{tabular}
    
\end{table}
\end{landscape}

\let\clearpage\relax


\bsp	
\label{lastpage}
\end{document}